UNIL | Université de Lausanne

Faculté des géosciences et de l'environnement
Institut des Sciences de la Terre

# Development of Massively Parallel Near Peak Performance Solvers for Three-Dimensional Geodynamic Modelling

## Thèse de doctorat

présentée à la Faculté des géosciences et de l'environnement de l'Université de Lausanne par

**Samuel Omlin**

Master en Informatique et Méthodes Mathématiques, Université de Lausanne, 2010

pour l'obtention du grade de Docteur en Sciences de la Terre
Université de Lausanne

**Jury**

| | |
|---|---|
| M. Le Professeur Yury Podladchikov | Directeur de thèse |
| M. Le Professeur Stefan Schmalholz | Expert interne |
| M. Le Professeur Taras Gerya | Expert externe |

sous la présidence de M. Le Professeur Michel Jaboyedoff

UNIL | Université de Lausanne
Décanat Géosciences et de l'Environnement
bâtiment Géopolis
CH-1015 Lausanne

# IMPRIMATUR

Vu le rapport présenté par le jury d'examen, composé de

| | |
|---|---|
| Président de la séance publique : | M. le Professeur Michel Jaboyedoff |
| Président du colloque : | M. le Professeur Michel Jaboyedoff |
| Directeur de thèse : | M. le Professeur Yury Podladchikov |
| Expert interne: | M. le Professeur Stefan Schmalholz |
| Expert externe : | M. le Professeur Taras Gerya |

Le Doyen de la Faculté des géosciences et de l'environnement autorise l'impression de la thèse de

**Monsieur Samuel OMLIN**

Titulaire d'une
*Maîtrise universitaire ès Lettres,*
*Informatique et méthodes mathématiques*
*de l'Université de Lausanne*

intitulée

**DEVELOPMENT OF MASSIVELY PARALLEL NEAR PEAK PERFORMANCE SOLVERS FOR THREE-DIMENSIONAL GEODYNAMIC MODELLING**

Lausanne, le 26 janvier 2017

Pour le Doyen de la Faculté des géosciences et de l'environnement

Professeur Michel Jaboyedoff

*To my wife and friend Rossi*

## *Abstract*

Academia and industry have not yet been able to adapt software and numerical algorithms to the rapid, fundamental changes in hardware evolution since the beginning of the 21st century. Many of today's applications and algorithms are as a result not well suited for the available hardware, what implies performances far below hardware's peak. We address in this thesis the current need to design new parallel algorithms and tools that ease the development of applications that are suited for present and future hardware. We present (1) the MATLAB HPC compiler *HPC*$^{.m}$, which greatly simplifies the building of parallel high performance applications and (2) parallel algorithms for the 3D simulation of strongly nonlinear processes as mechanical and reactive porosity waves.

To simulate mechanical porosity waves we employ a simple numerical algorithm that permits to resolve the deformation of fluid-filled viscoelastic porous media in 3D. The utilized mathematical model is based on Biot's poroelastic theory, extended to account for viscous deformation and plastic yielding. The algorithm is designed for massively parallel high performance computing. It employs finite difference stencil calculations on a staggered grid to approximate spatial derivatives. Pseudo-transient iterations were utilized to formulate an explicit algorithm and Picard iterations to resolve the nonlinearities. The modelling results exhibit the impact of decompaction weakening on the formation of three-dimensional solitary-wave-like moving porosity channels. We evaluate the algorithm's suitability for the building of high performance massively parallel 3D solvers and compare the achievable performance and parallel scalability to the ones of other state-of-the-art algorithms that were developed aiming at large-scale simulations. Our algorithm is found to be better suited for the building of high performance massively parallel 3D solvers than the other algorithms considered in this thesis.

To simulate reactive porosity waves we use a solver for 3D deformation of fluid-filled reactive viscous porous media. The employed algorithm essentially uses the same numerical methods as the algorithm for the simulation of mechanical porosity waves. The Damköhler number (Da) of the simulations is varied in order to estimate the respective roles of viscous deformation (low Da) and reaction (high Da) on wave propagation. 3D waves are found to propagate independently of their source at constant speed by passing through each other for all the investigated Da. Soliton-like wave propagation as a result of metamorphic reaction provides an efficient mechanism for fluid flow in the Earth's crust. It is expected that this mechanism takes place at the meter-scale in the lower crust and at the kilometer-scale in the upper crust providing explanations for both metamorphic veins formation and fluid extraction in subduction zones.

The tool HPC$^{.m}$, developed here, transforms simple MATLAB scripts into parallel high performance applications for GPU-, CPU- and MIC-supercomputers, clusters or workstations. It is designed for stencil-based applications, in particular for iterative PDE solvers that use finite differences and a regular Cartesian grid. Its core is a source-to-source translator. We illustrate in this thesis the great performance and versatility of HPC$^{.m}$ by deploying it to generate solvers for a variety of physical processes across multiple earth science disciplines. All solvers run close to hardware's peak performance and scale linearly on the 80 GPUs of the Octopus cluster, hosted by the Institute of Earth Sciences at University of Lausanne (Lausanne, Switzerland). They achieve moreover a speedup over the fully vectorised MATLAB input script of about 250x to 500x on one GPU, of 1000x to 2000x on one workstation with 4 GPUs and of 17 000x to 35 000x on 80 GPUs. Additionally, we show that our nonlinear poroviscoelastic two-phase flow solver scales also linearly on the 5000 GPUs of the Piz Daint supercomputer at the Swiss National Supercomputing Centre (CSCS, Lugano, Switzerland), achieving a speedup over the fully vectorised MATLAB input script of over 500 000x. We expect a similar scaling for all the ten solvers. The source-to-source translator contained in HPC$^{.m}$ is, to the authors knowledge, the first that can automatically perform all tasks that are required for the generation of a massively parallel near peak performance supercomputing application from a code developed in a classical prototyping environment such as MATLAB.

# Contents

# INTRODUCTION

## 1 The gap between the evolution of hardware and software

The famous law of Intel co-founder Gordon Moore [1] states that the number of transistors per unit of area on a chip[1] doubles every 18-24 months. Moore found this law in 1965 and it is still correct today (see Figure 1). Until the early 2000s, it lead to a doubling of processor clock speeds every 18-24 months and uniprocessor performance increased alongside this [2] (see Figure 1). Since then, it has not anymore been cost effective to increase processor clock speeds – the *power wall* was hit – and aggressive uniprocessor performance scaling has reached its end [2,3] (see Figure 1). The doubling of clock speeds every 18-24 months has been replaced by a doubling of cores, threads or other parallelism mechanisms [4] (see Figure 1). Simply put, the traditional single-core CPUs from the last century have been replaced by multi-core CPUs and many-core architectures such as GPUs[2] and MICs[3].

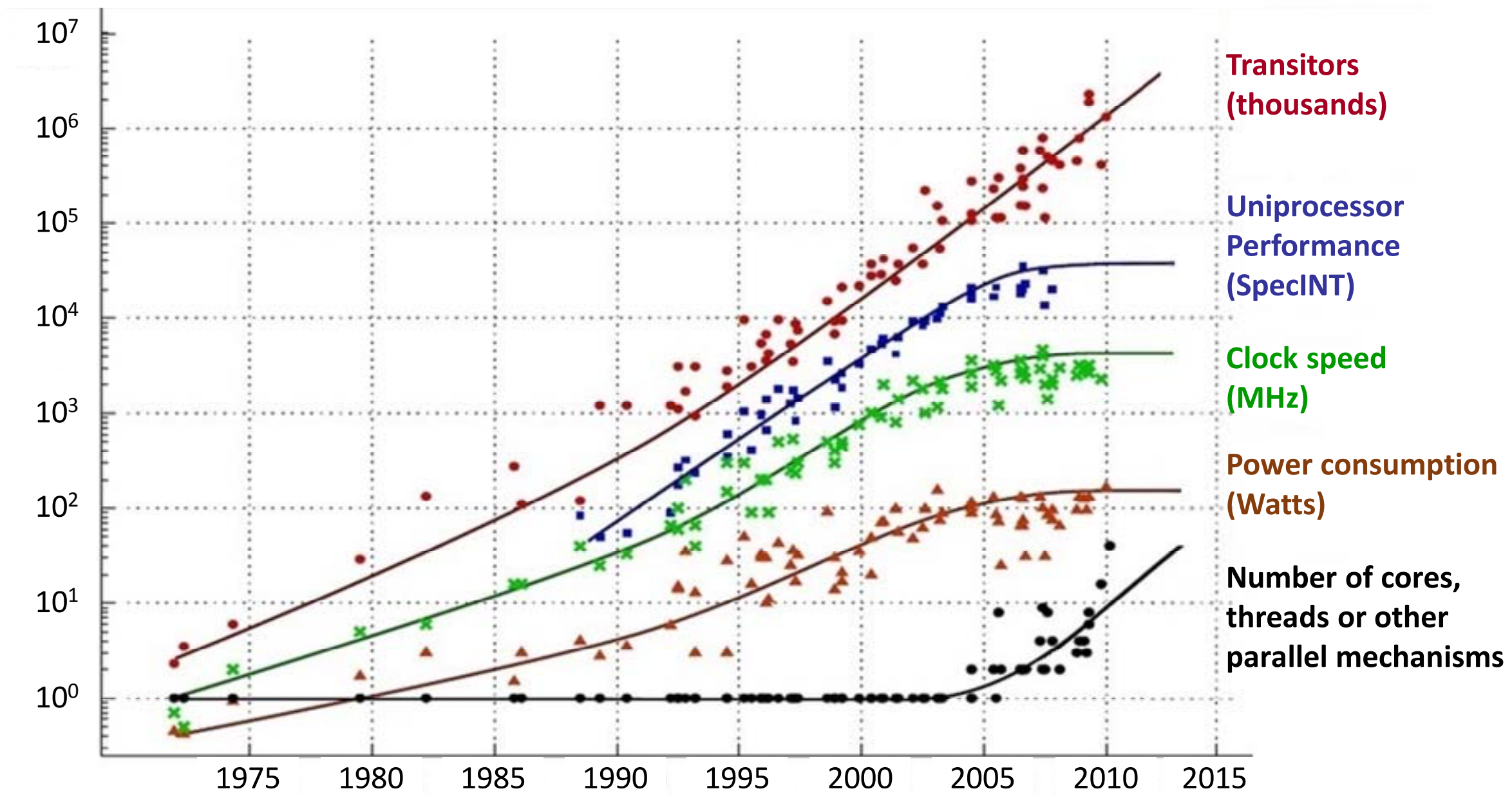

*Figure 1. Moore's law on the increase of transistors per unit of area on a chip in time and its relation to processor clock speed, uniprocessor performance, power consumption and number of cores, threads or other parallelism mechanisms. Source: Samuel Naffziger, AMD (modified).*

[1] also called 'microchip' or 'integrated circuit'
[2] GPU: Graphic Processing Unit
[3] MIC: Many Integrated Core architecture from Intel

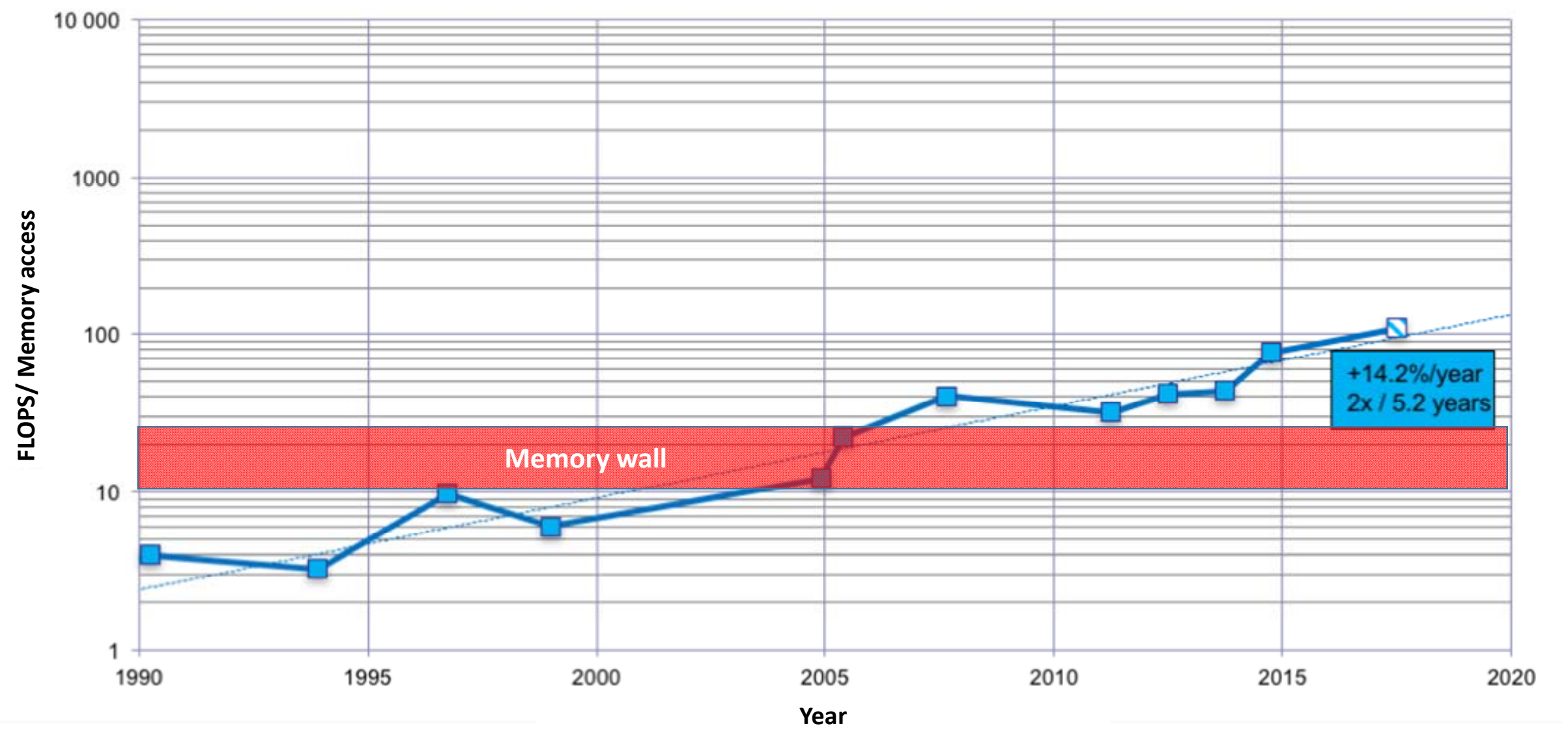

*Figure 2. The evolution of the FLOPS per Memory access ratio and the predicted memory wall. Source: John McCalpin, Texas Advanced Computing Center (modified).*

The improvement of memory access speed has not been comparable to the impressive increase of computation speed over the last decades. The ratio of the performable amount of floating point operations (FLOPS) per floating point number accessed in memory (*FLOPS per Memory access ratio*) has therefore been constantly increasing. The majority of software, however, requires only few operations per accessed number. The achievable performance is thus totally determined by memory access speed [5]. This *memory wall* was predicted in 1995 by Wulf and McKee [6] who noted that from the beginning of the 21$^{st}$ century on, "*system performance is totally determined by memory speed; making the processor faster won't affect the wall-clock time to complete an application*" (Figure 2, shows the evolution of the FLOPS per Memory access ratio and the predicted memory wall).

The fundamental changes in hardware evolution since the beginning of the century have drastic implications for present and future software: (1) applications without proper parallelization and a focus on optimal memory access reach only a small fraction of the performance that current hardware can offer, and (2) the same applications will hardly benefit from the performance increase of future hardware as it will be ever more parallel.

The evolution of software is at present far behind hardware evolution [7]. Many applications therefore run with a performance that is far below hardware's peak. There are three major reasons for this:

- A large amount of today's software and employed algorithms were *developed before the appearance of major parallelism and the memory wall*; thus, neither the software nor the algorithms are typically well suited for current hardware.
- The development of parallel high performance software is very time consuming; *the programmers' productivity is very low* [7].
- Software that may achieve a performance close to the maximum of what present and future hardware can deliver is at present very difficult to develop; the development of such parallel high performance software is normally *a task for computer scientists*.

Many existing sophisticated algorithms are to a major extent serial. Parallelization is therefore only possible to a small degree and the scalability is very limited. Moreover, the data access pattern is often rather random, while regular memory access is fastest. Random memory access can be up to two order of magnitudes slower than regular memory access [8]. Random memory access leads in addition normally to more cache misses, which increase the required amount of accesses to main memory and may cause the processor to frequently wait for data and thus further slow down the application.

We identify two major challenges needed to be addressed in order to bridge the gap between hardware and software evolution:

- New algorithms that are suited for parallelisation and have simple regular data access patterns must be designed [4].
- Tools that increase the programmer's productivity and that permit non-computer scientists to develop parallel high performance software have to be developed.

## 2 Parallel algorithms and high performance software development tools

The challenge of developing new algorithms that are well suited for current and future hardware is approached essentially in two ways. The first consists in modifying algorithms that are sophisticated and state-of-the-art but typically not well suited for current hardware. Efforts in this direction are often of limited success, because such algorithms are in many cases to a major extent inherently serial; to transform an inherently serial algorithm into a fully parallel algorithm is an enormous challenge and in many cases it may result practically impossible. The second way to approach the development of new parallel algorithms consists in the building on simple algorithms that are well suited for parallelism and have simple regular data access patterns, but are typically rather old. This manner of development has been successful in many cases (e.g. [9–12]). Particularly well adapted for current hardware are stencil-based methods[4], which allow, for example, to build solvers for partial differential equations (PDE) [13]. So following this second approach of algorithm development, a PDE solver, for instance, is rather built with iterative methods using finite differences and a regular Cartesian grid, than with direct methods using complex techniques to compute derivatives and an unstructured grid. In fact, finite differences are computed with locally neighbouring grid cells, i.e. the data access pattern is simple and regular. Moreover, a computational domain that is defined by a regular Cartesian grid is very easy to decompose into many small domains for which an iteration or time step can be solved independently, i.e. in parallel.

The lack of software that is well suited for today's hardware gave birth to many projects that aim at increasing the programmer's productivity. This includes the design of new general purpose programming languages [14–16] and domain specific languages [11,17,18], as well as the development of new libraries [19–21], and the building of source-to-source translators that automatize some of the steps in the development of parallel high performance software [11,17,18,22]. All of these projects aim to simplify the programming by trying to separate application-specific programming logic from implementation. It is obvious that domain or method specific approaches require less development time than very general approaches as, for instance, the design of new general-purpose programming languages. They may moreover achieve greater performances because they can employ aggressive domain or method specific performance optimizations and parallelization techniques.

[4] A stencil can be considered as a window on the data. Stencil-based methods move this window over the application data and perform computations with the data contained in the window. Finite difference methods are an example for stencil-based methods.

Source-to-source translation has become a popular approach in High Performance Computing (HPC) because it allows for automatic optimisations and parallelization. Many of these source-to-source translators limit themselves to stencil-based applications [11,17,18,22] to permit aggressive stencil method specific performance optimizations. The input source language for these source-to-source translators is typically a domain specific language, which has been developed on top of a compilable general purpose language like C or C++ [11,18].

In the scope of this thesis, we have developed HPC.m – the MATLAB HPC compiler. It is a tool that responds to the major challenges of nowadays software development (listed in section 1): it increases the programmer's productivity and permits non-computer scientists to develop parallel high performance software. HPC.m transforms simple MATLAB scripts within a few seconds into parallel high performance applications for GPU-, CPU- and MIC-supercomputers, clusters or workstations. It is designed for stencil-based applications, in particular for iterative PDE solvers that use finite differences and a regular Cartesian grid. Its core is a source-to-source translator. To the authors' knowledge, *it is the first source-to-source translator that can automatically perform all tasks that are required for the generation of a parallel near peak performance supercomputing application from a code developed in a classical prototyping environment such as MATLAB*.

## 3 The thesis's objective: spreading of HPC in earth sciences

Spreading of HPC is a major objective in academia around the world and important efforts have been implemented for this purpose. An example is the CADMOS[5] initiative in western Switzerland, a joint project of the Universities of Geneva and Lausanne and of the Swiss Federal Institute of Technology Lausanne (EPFL) that was initiated in 2009 with the goal to expand numerical HPC modelling activities within the three member institutions. Multiple professorship positions were created within the scope of this project and a supercomputer that cost a two digit million dollar amount was acquired. Such projects are of ever increasing importance because (1) numerical simulations play a key role in today's science and technology and have veritably become the third way of research, adding to experiences and theory [23], and (2) HPC opens new doors for numerical modelling, as simulations that were previously too computationally heavy become feasible. Moreover, it is not only important to extend current HPC modelling activities which are at present typically made in physics, chemistry, material sciences, hydrology, biology and geophysics, but also new domains should be reached such as economy and finances, language treatment, large-scale climate modelling, environmental risk analysis, engineering and architecture [23].

[5] Center for Advanced Modeling Science

The principal aim of this thesis is to spread HPC across the disciplines in earth sciences. The $HPC^{.m}$ software developed here is an ideal tool for numerical modellers to conduct simulations in three dimensions at high resolution. This is of particular interest for state-of-the-art problems dealing with physical processes that are very localised in space and time and for which it is a priori unknown where and when they occur (particularly characteristic for nonlinear processes). Resolving such localized processes accurately requires high-resolution in both space and time, which demands scalable high performance solvers. $HPC^{.m}$ makes it easy for earth scientists to generate such solvers, because $HPC^{.m}$ input scripts are developed in MATLAB, a prototyping environment that is familiar to most numerical modellers.

We illustrate in this thesis the great performance and versatility of $HPC^{.m}$ by deploying it to generate solvers for a variety of physical processes across multiple earth science disciplines (article 2, chapter *$HPC^{.m}$: from MATLAB to HPC on GPU, CPU and MIC*): shallow water, glacier flow, convection, scalar and reactive porosity waves, poroviscoelastic two-phase flow, shear heating, seismics, acoustic wave propagation and heat diffusion (a selection of MATLAB input scripts for $HPC^{.m}$ are found in the thesis's appendix). Moreover, we perform state-of-the-art modelling with the help of $HPC^{.m}$: we simulate mechanical and reactive porosity waves in three dimensions (see article 1 and 3, chapter *Massively Parallel Simulation of Three dimensional Viscoelastic Deformation Coupled to Porous Fluid Flow* and chapter *Porous fluids extraction by reactive solitary waves in 3-D*, respectively). To this purpose, we design numerical algorithms that are optimally suited for today's hardware, i.e. algorithms that are appropriate for parallelisation and have simple regular data access patterns. With the development of $HPC^{.m}$ and new parallel algorithms, we try to respond to all identified major challenges (listed in section 1) with the aim to bridge the gap between hardware and software evolution.

## 4 References

[1] G.E. Moore, Cramming More Components Onto Integrated Circuits (Reprinted from Electronics, pg 114-117, April 19, 1965), Proc. IEEE. 86 (1998) 82–85. doi:10.1109/JPROC.1998.658762.

[2] P. Kogge, K. Bergman, S. Borkar, D. Campbell, W. Carson, W. Dally, M. Denneau, P. Franzon, W. Harrod, K. Hill, J. Hiller, S. Karp, S. Keckler, D. Klein, R. Lucas, M. Richards, A. Scarpelli, S. Scott, A. Snavely, T. Sterling, R.S. Williams, K. Yelick, ExaScale Computing Study : Technology Challenges in Achieving Exascale Systems, 2008. doi:10.1.1.165.6676.

[3] S.W. Keckler, W.J. Dally, B. Khailany, M. Garland, D. Glasco, GPUs and the Future of Parallel Computing, IEEE Micro. 31 (2011) 7–17. doi:10.1109/MM.2011.89.

[4] J. Shalf, S. Dosanjh, J. Morrison, Exascale Computing Technology Challenges, in: J.M.L.M. Palma, M. Daydé, O. Marques, J.C. Lopes (Eds.), High Perform. Comput. Comput. Sci. – VECPAR 2010, Springer Berlin Heidelberg, 2011: pp. 1–25. doi:10.1007/978-3-642-19328-6_1.

[5] A. Tate, A. Kamil, A. Dubey, A. Größlinger, B. Chamberlain, B. Goglin, C. Edwards, C.J. Newburn, D. Padua, D. Unat, E. Jeannot, F. Hannig, T. Gysi, H. Ltaief, J. Sexton, J. Labarta, J. Shalf, K. Fürlinger, K. O'brien, L. Linardakis, M. Besta, M.-C. Sawley, M. Abraham, M. Bianco, M. Pericàs, N. Maruyama, P.H.J. Kelly, P. Messmer, R.B. Ross, R. Cledat, S. Matsuoka, T. Schulthess, T. Hoefler, V.J. Leung, Programming Abstractions for Data Locality, PADAL Workshop 2014, April 28--29, Swiss National Supercomputing Center (CSCS), Lugano, Switzerland, Lugano, Switzerland, 2014. hal.inria.fr/hal-01083080/ (accessed October 30, 2016).

[6] W.A. Wulf, S.A. McKee, Hitting the memory wall: implications of the obvious, ACM SIGARCH Comput. Archit. News. 23 (1995) 20–24. doi:10.1145/216585.216588.

[7] R. Stefano, T.S. Hailu, Solve. The Exascale Effect: Benefits of Supercomputing Investment for U.S. Industry, Washington, D.C., 2014. http://www.compete.org/reports/all/2695-solve.

[8] T. Kaldewey, A. Di Blas, J. Hagen, E. Sedlar, S. Brandt, Memory matters, 29th IEEE Real-

Time Syst. Symp. (2008) 1–4. http://www.kaldewey.com/pubs/Memory_Matters__RTSS08.pdf.

[9] K. Datta, S. Williams, V. Volkov, J. Carter, L. Oliker, J. Shalf, K. Yelick, Auto-tuning the 27-point stencil for multicore, in: Proc. iWAPT2009 Fourth Int. Work. Autom. Perform. Tuning, 2009: p. 17. http://citeseerx.ist.psu.edu/viewdoc/download?doi=10.1.1.157.7808&rep=rep1&type=pdf\nhttp://bebop.cs.berkeley.edu/pubs/iwapt2009_datta.pdf.

[10] M. Krotkiewski, M. Dabrowski, Efficient 3D stencil computations using CUDA, Parallel Comput. 39 (2013) 533–548. doi:10.1016/j.parco.2013.08.002.

[11] T. Gysi, C. Osuna, O. Fuhrer, M. Bianco, T.C. Schulthess, STELLA: A Domain-specific Tool for Structured Grid Methods in Weather and Climate Models, in: Proc. Int. Conf. High Perform. Comput. Networking, Storage Anal. - SC ’15, ACM Press, New York, New York, USA, 2015: pp. 1–12. doi:10.1145/2807591.2807627.

[12] P. Micikevicius, 3D Finite Difference Computation on GPUs using CUDA, in: GPGPU-2 Proc. 2nd Work. Gen. Purp. Process. Graph. Process. Units, ACM Press, New York, New York, USA, 2009: pp. 79–84. doi:10.1145/1513895.1513905.

[13] J.C. Strikwerda, Finite Difference Schemes and Partial Differential Equations, Second Edition, Society for Industrial and Applied Mathematics, 2004. doi:10.1137/1.9780898717938.

[14] R.W. Numrich, J. Reid, Co-array Fortran for parallel programming, ACM SIGPLAN Fortran Forum. 17 (1998) 1–31. doi:10.1145/289918.289920.

[15] E. Allen, D. Chase, J. Hallett, V. Luchangco, J.-W. Maessen, S. Ryu, G.L.S. Jr, S. Tobin-Hochstadt, The Fortress Language Specification, 2007. www.eecis.udel.edu/~cavazos/cisc879-spring2008/papers/fortress.pdf (accessed November 14, 2016).

[16] W.W. Carlson, J.M. Draper, D.E. Culler, K. Yelick, E. Brooks, K. Warren, Introduction to UPC and language specification, 1999. www3.uji.es/~aliaga/UPC/upc_intro.pdf (accessed November 14, 2016).

[17] D.A. Orchard, M. Bolingbroke, A. Mycroft, Ypnos: Declarative, Parallel Structured Grid Programming, in: Proc. 5th ACM SIGPLAN Work. Declar. Asp. Multicore Program. - DAMP ’10, ACM Press, New York, New York, USA, 2010: pp. 15–24. doi:10.1145/1708046.1708053.

[18] N. Maruyama, Physis: An Implicitly Parallel Programming Model for Stencil Computations on Large-Scale GPU-Accelerated Supercomputers, in: Proc. 2011 Int. Conf. High Perform. Comput. Networking, Storage Anal. - SC ’11, ACM Press, New York, New York, USA, 2011. doi:10.1145/2063384.2063398.

[19] OpenCFD Ltd (ESI Group), OpenFOAM, (n.d.). www.openfoam.com/documentation/user-guide/ (accessed November 14, 2016).

[20] J. Choi, J. Dongarra, S. Ostrouchov, A. Petitet, D. Walker, R.C. Whaley, A proposal for a set of parallel basic linear algebra subprograms, in: Appl. Parallel Comput. Comput. Physics, Chem. Eng. Sci., Springer Berlin Heidelberg, 1996: pp. 107–114. doi:10.1007/3-540-60902-4_13.

[21] S. Balay, W.D. Gropp, L.C. McInnes, B.F. Smith, Efficient Management of Parallelism in Object-Oriented Numerical Software Libraries, in: Mod. Softw. Tools Sci. Comput., Birkhäuser Boston, Boston, MA, 1997: pp. 163–202. doi:10.1007/978-1-4612-1986-6_8.

[22] D. Unat, X. Cai, S.B. Baden, Mint: Realizing CUDA performance in 3D Stencil Methods with Annotated C, in: Proc. Int. Conf. Supercomput. - ICS ’11, 2011: p. 214. doi:10.1145/1995896.1995932.

[23] G. Margaritondo, Le projet CADMOS, Flash Inform. (2009) 2. http://flashinformatique.epfl.ch/spip.php?article1833.

# Massively Parallel Simulation of Three-dimensional Viscoelastic Deformation Coupled to Porous Fluid Flow

Samuel Omlin[1], Ludovic Räss[1], and Yury Y. Podladchikov[1]

[1]Institute of Earth Sciences, University of Lausanne, Lausanne, Switzerland

## *Abstract*

We present a simple numerical algorithm to simulate the deformation of fluid-filled viscoelastic porous media in 3D. The employed mathematical model is based on Biot's poroelastic theory, extended to account for viscous deformation and plastic yielding. The algorithm is designed for massively parallel high performance computing in three dimensions. It employs finite difference stencil calculations on a staggered grid to approximate spatial derivatives. Pseudo-transient iterations are utilized to formulate an explicit algorithm and Picard iterations to resolve the nonlinearities. Iterative implicit schemes improve the algorithm's stability and convergence and permit to choose reasonable time steps at high spatial resolutions. Damping accelerates the convergence of these implicit iterations additionally. As a numerically challenging example we consider the dynamics of spontaneous channel formation in fluid-filled viscoelastic porous media. The modelling results exhibit the impact of decompaction weakening on the formation of three-dimensional solitary-wave-like moving porosity channels. Our solver runs close to hardware limit and achieves linear weak scaling on the 5000 GPUs of Piz Daint, a Cray XC 30 supercomputer at the Swiss National Supercomputing Center (CSCS, Lugano, Switzerland), cumulating over 10 million CUDA cores. We evaluate the algorithm's suitability for the building of high performance massively parallel 3D solvers and compare the achievable performance and parallel scalability to the ones of other state-of-the-art algorithms that were developed aiming at large-scale simulations. To this purpose we introduce an effective memory throughput metric as a formal evaluation criteria for the algorithms' performance. Our algorithm is found to be better suited for the building of high performance massively parallel 3D solvers than the other algorithms considered in this paper. We attribute this result to a simpler, more regular and more local data access pattern of our algorithm compared to the others.

## *Keywords*

Fluid-filled porous media; Deformation; Large-scale simulations; Massive parallelization; Performance evaluation

# 1 Introduction

Fluid-filled viscoelastic porous media has been studied extensively since the beginning of the twentieth century (e.g. [1–11]). Von Terzaghi [7] proposed the first model of one-dimensional consolidation. Biot [4] derived equations for three-dimensional consolidation first in 1941 and reformulated and extended his theory in the following years [1–3,5]. Similar formulations were suggested by, among others, Verruijt [8] and Rice and Cleary [6]. More details can be found in recent books, e.g. [12–15], and reviews, e.g. [16–19].

Many of today's applications that deal with fluid-filled elastic porous media require large-scale simulations and must be able to handle large contrasts in material properties. Direct solvers are generally the most robust to deal with high material property contrasts, but have a limited parallel scalability, particularly in 3D [20–22]. Iterative algorithms have thus been developed to solve the systems of partial differential equations describing fluid-filled poro(visco-)elastic media. They are less robust but potentially very scalable. State-of-the-art algorithms include the two presented by Haga et al. [23–25] and Chen et al. [26–28]. They both are finite element algorithms that combine iterative methods with sophisticated pre-conditioners. Haga et al. [24,25] precondition a conjugate gradient-type method with algebraic multigrid in order to optimally deal with heterogeneous material parameters, in particular, highly variable permeability and significant jumps in elastic properties. Chen et al. [27,28] precondition the (simplified) quasi-minimal residual (QMR/SQMR) method with variants of successive over-relaxation (SOR); they optimized the algorithm particularly for high contrasts in material stiffness and include non-associated plasticity into the model. Aiming at large scale simulations, Haga et al. [25] and Chen et al. [26] have developed parallel CPU and GPU implementations, respectively.

The DynEarthSol3D solver [29] utilizes nearly the same system of equations as the one that describes fluid-filled (visco-)elastic porous media. DynEarthSol3D does not consider fluid phase pressure but solves for the temperature instead. DynEarthSol3D utilizes momentum conservation equations for the solid phase that are similar to the ones solved by porous media simulators. DynEarthSol3D has been developed primarily to study the long term deformation of earth's lithosphere. It implements a finite element algorithm based on the Fast Lagrangian Analysis of Continua (FLAC) algorithm, uses the EVP material model and does explicit forward time stepping [30]. Aiming at large-scale simulations, Ta et al. [29] have developed an implementation of DynEarthSol3D for Multi-Core CPU, for GPU and for CPU with embedded GPU.

Each of the three state-of-the-art algorithms greatly outperforms the similar methods it compares with. We present in this paper though a simple algorithm which performs in high resolution similarly good as these sophisticated algorithms. We propose it therefore as a benchmark for present and future algorithms that are potential candidates *for the building of high performance massively parallel 3D solvers* for deformation of fluid-filled (visco-)elastic porous media.

The proposed benchmark algorithm is a combination of simple and well-known standard numerical methods to solve systems of partial differential equations. It uses pseudo-transient iterations to find the implicit solution for time derivatives and it computes spatial derivatives from the smallest possible finite difference stencil. This results in a very simple and regular memory access pattern, what permits to build 3D solvers that achieve a performance close to hardware limit [31,32]. Moreover, the method being highly local, it suits optimally for massive parallelization. High parallel efficiency and linear parallel scaling may be achieved on large supercomputers.

Sophisticated algorithms as those in [25,26,29], in contrast, require often a quite random memory access pattern. Random memory access can be up to two orders of magnitude slower than regular memory access [33]. In addition, random memory access leads normally to more cache misses what increases the required amount of accesses to main memory and may cause the processor to frequently wait for data ('stalling') and therefore further slows down the application. Applications with a lot of random memory access achieve normally performances that are orders of magnitude below hardware's peak performance.

To sum up, today's sophisticated algorithms to simulate the deformation of fluid-filled (visco-)elastic porous media typically reduce the number of iterations that are required to reach the solution, but their implementations may face important trade-offs in the achievable performance and in parallel scalability. A lower performance means that each iteration takes longer. This may balance out the reduction of the number of iterations.

The objectives of this work are (1) to evaluate our simple algorithm's suitability for the building of high performance massively parallel 3D solvers for the deformation of fluid-filled (visco-)elastic porous media and (2) to compare the achievable performance and parallel scalability to the ones of other algorithms that were developed aiming at large-scale simulations. To this purpose we measure the performance and parallel scalability of our multi-GPU solver and compare it to the best-known implementation of the other algorithms. We make this simple choice because traditional algorithm complexity analysis does not take into account how efficient an algorithm may be implemented on actual available hardware. It is though crucial for us, given our aim to evaluate the algorithms suitability for building massively parallel 3D solvers that can efficiently use modern supercomputers with thousand or even hundred thousands of compute nodes.

We would like to highlight another feature of our simple algorithm: small time steps are taken, i.e. a high resolution in time is achieved. It allows therefore to resolve processes that happen very localized in time and at a moment and a place unknown in advance. Algorithms designed to take large time steps or to solve directly for steady state cannot capture such processes.

The paper is organized as follows. In section 2, we present the mathematical model, the numerical algorithm and our methodology to measure performance and parallel scaling. In section 3, we present some representative modelling results and report the performance and the parallel scalability of our solver and compare it with the results from [25,26,29]. In section 4, we give concluding remarks. Section 5 is a glossary.

# 2 Methods

## 2.1 Mathematical model

We present here a closed system of equations for deformation of fluid-filled viscoelastic porous media in 3D. The core of our model is the constitutive equations of Biot's poroelastic theory for an isotropic fluid-filled porous medium [4], extended to account for viscous deformation [17]. They are written in time incremental form as

$$\nabla_k V_k^s = -\frac{1}{K_d}\left(\frac{d^s \bar{P}}{dt} - \alpha \frac{d^f P_f}{dt}\right) - \frac{\bar{P} - P_f}{(1-\varphi)\eta_\varphi} \tag{1}$$

and

$$\nabla_k\big(\varphi(V_k^f - V_k^s)\big) = \frac{\alpha}{K_d}\left(\frac{d^s \bar{P}}{dt} - \frac{1}{B}\frac{d^f P_f}{dt}\right) + \frac{\bar{P} - P_f}{(1-\varphi)\eta_\varphi}, \tag{2}$$

where $V_k^s$ is the solid velocity, $V_k^f$ is the fluid velocity, $\bar{P}$ is the total pressure, $P_f$ is the fluid pressure and $\varphi$ is the porosity; $K_d$ is the drained bulk modulus, $\alpha$ is the Biot-Willis coefficient, $B$ is Skempton's coefficient and $\eta_\varphi$ is the effective bulk viscosity; $\frac{d^s}{dt} = \frac{\partial}{\partial t} + \vec{v}_s \cdot \nabla$ and $\frac{d^f}{dt} = \frac{\partial}{\partial t} + \vec{v}_f \cdot \nabla$ denote the material time derivatives with respect to the solid and fluid, respectively. These equations were derived by combining Maxwell's viscoelastic bulk rheology with mass conservation equations for fluid and solid [17].

We utilize additionally usual total momentum conservation equations for both solid and fluid (3), fluid momentum conservation equations combined with fluid's viscous rheology (Darcy's law of viscous fluid flow through porous media) (4) and Maxwell's viscoelastic deviatoric rheology (5) (c.f. [17]):

$$\nabla_j(\bar{\tau}_{ij} - \bar{P}\delta_{ij}) - g_i\bar{\rho} = 0, \tag{3}$$

$$\varphi\big(V_i^f - V_i^s\big) = -\frac{k}{\mu_f}\big(\nabla_i P_f + g_i \rho_f\big) \tag{4}$$

and

$$\nabla_i V_j^s + \nabla_j V_i^s - \frac{2}{3}(\nabla_k V_k^s)\,\delta_{ij} = \frac{1}{G}\frac{d^\nabla \bar{\tau}_{ij}}{dt} + \frac{\bar{\tau}_{ij}}{\mu_s}. \tag{5}$$

The symbols $V_i^s$, $V_i^f$, $\bar{P}$ and $P_f$ are the same as in the Eqns. (1) and (2), $\bar{\tau}_{ij}$ is the total stress deviator; $g_i$ is the gravitational acceleration, $\bar{\rho}$ is the total density, $\rho_f$ is the fluid density, $k$ is the permeability, $\mu_f$ is the fluid shear viscosity, $\mu_s$ is the solid shear viscosity and $G$ is the elastic shear modulus of the solid; $\frac{d^\nabla}{dt}$ denotes an objective time derivative (e.g. Jaumann, for details see for instance [34]) and $\delta_{ij}$ the Kronecker-delta.

The model accounts for tensile micro-fracturing by employing *decompaction weakening* [35] that reduces the solid shear viscosity $\mu_s$ by constant factor $R_0$, if $P_f > \bar{P}$. A closure relation for the porosity evolution (6) and equations for the nonlinear (Carman-Kozeny) porosity dependent permeability (7), for the solid shear viscosity (8), for the decompaction weakening factor, $R$, for the bulk viscosity (10) and the total density (11) close the system of equations [17]:

$$\frac{1}{1-\varphi}\frac{d^s\varphi}{dt} = -\frac{1}{K_\varphi}\left(\frac{\partial \bar{P}}{\partial t} - \frac{\partial P_f}{\partial t}\right) - \frac{\bar{P} - P_f}{\eta_\varphi}, \tag{6}$$

$$k = k_0\varphi^3, \tag{7}$$

$$\mu_s = \frac{\frac{\mu_0}{R}}{1 + \left(\frac{\tau_{II}}{\tau_0}\right)^{(n-1)}}, \tag{8}$$

$$R = \begin{cases} 1, & P_f \leq \bar{P} \\ R_0, & P_f > \bar{P} \end{cases}, \tag{9}$$

$$\eta_\varphi = \eta_0\frac{\mu_s}{\varphi} \tag{10}$$

and

$$\bar{\rho} = \varphi\,\rho_f + (1-\varphi)\rho_s. \tag{11}$$

The symbols introduced here have the following meanings: $K_\varphi$ is the effective bulk modulus, $k_0$ is the reference permeability, $\mu_0$ is the reference solid shear viscosity, $\eta_0$ is the reference bulk viscosity, $\tau_0$ is the characteristic stress at the transition between linear and power law viscous behaviour [36], $\tau_{II}$ is the second invariant of the deviatoric stress tensor, $n$ is the power law viscosity exponent and $\rho_s$ is the solid density.

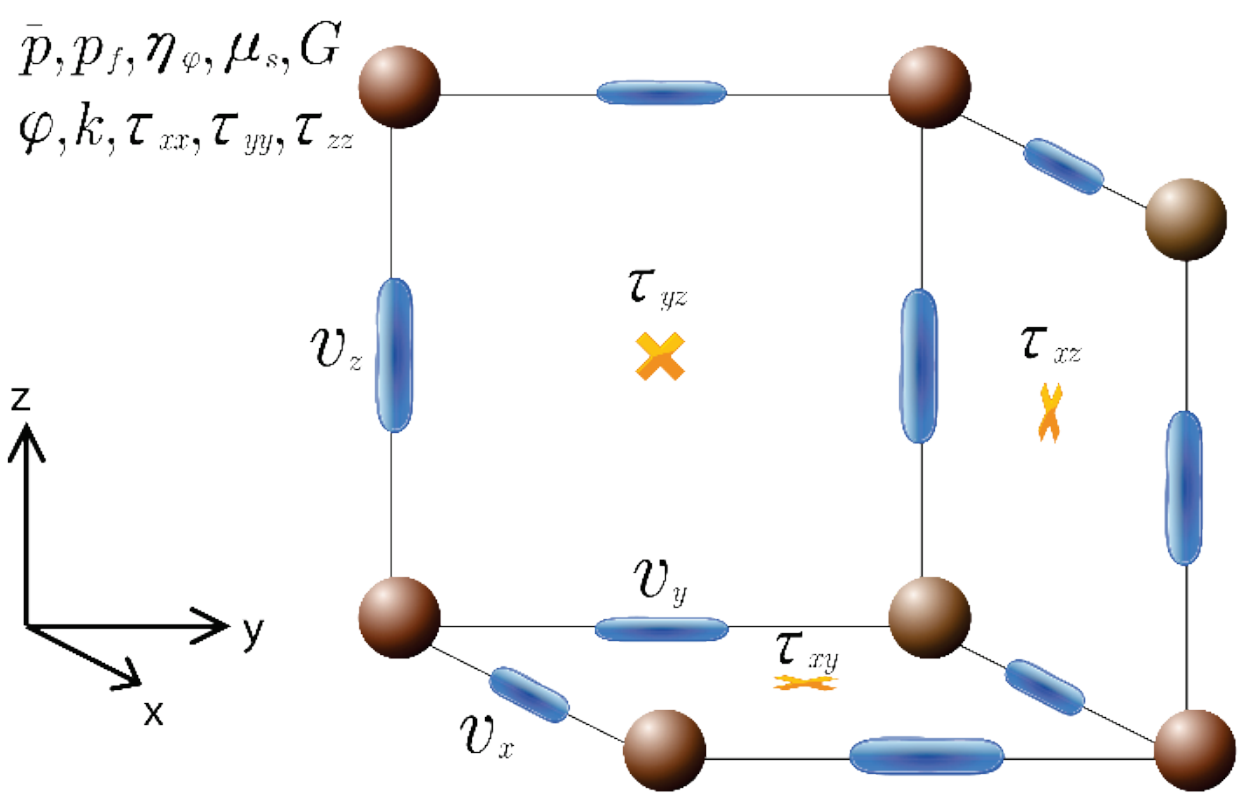

*Figure 1. Location of the fields of the 3D two-phase solver on the staggered grid.*

## 2.2 Numerical algorithm

We describe in the following the numerical algorithm that we employ in our solver for the deformation of fluid-filled viscoelastic porous media in 3D. We refer to this solver as *3D two-phase solver*[1] in the remainder of the article. The algorithm is designed for massively parallel high performance computing in three dimensions.

Our algorithm uses a staggered, regular Cartesian three dimensional grid that is based on the velocity-stress formulation which was used by Madariaga [37] in the simulation of fault-rupture dynamics; the method was later extended to seismic wave propagation in two-dimensional media [38–40] and also applied to three-dimensional media [41–43]. Figure 1 shows how the fields of Eqns. (1) to (11) are located on the staggered grid.

Spatial derivatives are approximated in our algorithm with finite differences using the smallest possible stencil. Pseudo-transient iterations are utilized to formulate an explicit algorithm that includes iterative implicit schemes, which improve the algorithm's stability and convergence and permit to choose sufficiently large time steps at high spatial resolutions. Although the method is iterative, it permits to find the implicit solution for the system of equations, i.e. the spatial derivatives are computed from the fields at time $t + \Delta t$ ($\Delta t$ is the time increment). We explain the method and its simple derivation briefly in the following.

[1] Here, the two phases are the fluid and the solid phases, whereas often the term *two-phase* is assumed to refer to two fluid phases.

We want to formulate an iterative solution algorithm for the differential equation

$$\frac{\partial U}{\partial t} = f(U), \tag{12}$$

where $U$ is an unknown and $f(U)$ a known function of $U$ that may include its spatial derivatives. The implicit discrete form is

$$\frac{U_{t+\Delta \mathrm{t}} - U_{\mathrm{t}}}{\Delta t} = f(U_{t+\Delta \mathrm{t}}), \tag{13}$$

where $U_t$ and $U_{t+\Delta \mathrm{t}}$ refer to $U$ at time $t$ and $t + \Delta \mathrm{t}$. Equation (12) is equivalent to

$$\frac{\partial U}{\partial \tau} = f(U) - \frac{\partial U}{\partial t}, \tag{14}$$

if $\frac{\partial U}{\partial \tau} = 0$ ($\tau$ denotes pseudo-transient time). The implicit discrete form is

$$\frac{U_{\tau+\Delta\tau} - U_{\tau}}{\Delta\tau} = f(U_{t+\Delta \mathrm{t}}) - \frac{U_{t+\Delta \mathrm{t}} - U_{\mathrm{t}}}{\Delta t}, \tag{15}$$

where $U_\tau$ and $U_{\tau+\Delta\tau}$ refer to $U$ at pseudo-transient time $\tau$ and $\tau + \Delta\tau$ ($\Delta\tau$ is the increment of pseudo-transient time; $U_t$ and $U_{t+\Delta \mathrm{t}}$ refer to $U$ at (physical) time $t$ and $t + \Delta \mathrm{t}$ as in Eqn. (13)). We solve for $U_{\tau+\Delta\tau}$ and approximate $U_{t+\Delta \mathrm{t}}$ by $U_\tau$ to obtain an explicit update rule for each iteration:

$$U_{\tau+\Delta\tau} = U_\tau + \Delta\tau\, f(U_\tau) - \Delta\tau\, \frac{U_\tau - U_{\mathrm{t}}}{\Delta t}. \tag{16}$$

Converged iterations ($max|U_{\tau+\Delta\tau} - U_\tau| < \epsilon$) solve Equation (13), delivering the numerical solution for the differential equation (12). $U_{t+\Delta \mathrm{t}}$ is then given by the computed $U_\tau$.

Figure 2 shows the application of this method to the solution of the classical heat diffusion equation in 1D (the equation and the discrete implicit form are the same as Eqn. (12) and Eqn. (13), respectively; the unknown $U$ is the temperature; $f(U) = \frac{\lambda}{\rho c_p}\frac{\partial^2(U)}{\partial x^2}$, where $\frac{\partial^2}{\partial x^2}$ is the 2nd derivative in space; the initial heat profile is a Gaussian; temperature is kept constant in time at the boundaries; in the pseudo-transient iterations we denote $U_{\tau+\Delta\tau}$ by 'U' $U_\tau$ by 'U_tau', $U_{\mathrm{t}}$ by 'U_t', $f(U_\tau)$ by 'f', $\Delta\tau$ by 'dtau', $\Delta t$ by 'dt'; comments in the script explain the remaining employed notations).

The system of equations that is implemented in our solver is given in the following and its difference compared to the system of section 2.1 due to the usage of the pseudo-transient method are highlighted in red (the symbols are the same as in section 2.1 with exception of $\partial\tau_{\bar{P}}$, $\partial\tau_{P_f}$ and $\partial\tau_{V_i^s}$, which denote pseudo-transient time steps for $\bar{P}$, $P_f$ and $V_i^s$ respectively):

$$\nabla_k V_k^s + \frac{1}{K_d}\left(\frac{d^s\bar{P}}{dt} - \alpha\frac{d^f P_f}{dt}\right) + \frac{\bar{P}-P_f}{(1-\varphi)\eta_\varphi} = \frac{\partial\bar{P}}{\partial\tau_{\bar{P}}} \tag{17}$$

$$\nabla_k\left(\varphi(V_k^f - V_k^s)\right) - \frac{\alpha}{K_d}\left(\frac{d^s\bar{P}}{dt} - \frac{1}{B}\frac{d^f P_f}{dt}\right) - \frac{\bar{P}-P_f}{(1-\varphi)\eta_\varphi} = \frac{\partial P_f}{\partial\tau_{P_f}} \tag{18}$$

$$\nabla_j(\bar{\tau}_{ij} - \bar{P}\delta_{ij}) - g_i\bar{\rho} = \frac{\partial V_i^s}{\partial\tau_{V_i^s}} \tag{19}$$

$$\varphi(V_i^f - V_i^s) = -\frac{k}{\mu_f}(\nabla_i P_f + g_i\rho_f) \tag{20}$$

$$\nabla_i V_j^s + \nabla_j V_i^s - \frac{2}{3}(\nabla_k V_k^s)\,\delta_{ij} = \frac{1}{G}\frac{d^\nabla\bar{\tau}_{ij}}{dt} + \frac{\bar{\tau}_{ij}}{\mu_s} \tag{21}$$

$$\frac{1}{1-\varphi}\frac{d^s\varphi}{dt} = -\frac{1}{K_\varphi}\left(\frac{\partial\bar{P}}{\partial t} - \frac{\partial P_f}{\partial t}\right) - \frac{\bar{P}-P_f}{\eta_\varphi} \tag{22}$$

$$k = k_0\varphi^3 \tag{23}$$

$$\mu_s = \frac{\frac{\mu_0}{R}}{1+\left(\frac{\tau_{II}}{\tau_0}\right)^{(n-1)}} \tag{24}$$

$$R = \begin{cases} 1, & P_f \le \bar{P} \\ R_0, & P_f > \bar{P} \end{cases} \tag{25}$$

$$\eta_\varphi = \eta_0\frac{\mu_s}{\varphi} \tag{26}$$

$$\bar{\rho} = \varphi\,\rho_f + (1-\varphi)\rho_s \tag{27}$$

The solid velocities ($V_x^s$, $V_y^s$ and $V_z^s$), the total stresses ($\bar{\tau}_{xx}$, $\bar{\tau}_{yy}$, $\bar{\tau}_{zz}$, $\bar{\tau}_{xy}$, $\bar{\tau}_{xz}$ and $\bar{\tau}_{yz}$,), the total pressure ($\bar{P}$), the fluid pressure ($P_f$) and the porosity ($\varphi$) are the *unknown DOFs* (see glossary, section 5) of this system of equations, because they have a time (or pseudo time) derivative in the system. Unknown DOFs depend on their own evolution in time (physical or pseudo-transient) and cannot be computed uniquely via algebraic relations from other DOFs. The spatially heterogeneous viscosity at zero stress, $\mu_0$, is the only *known DOF* (see glossary). The remaining parameters are not DOFs, because they are either computable with algebraic relations from the DOFs or taken spatially homogeneous, i.e. as scalars. Table 1 lists the nature of each parameter of our solver.

| | |
|---|---|
| Unknown DOFs | $V_x^S, V_y^S, V_z^S, \bar{\tau}_{xx}, \bar{\tau}_{yy}, \bar{\tau}_{zz}, \bar{\tau}_{xy}, \bar{\tau}_{xz}, \bar{\tau}_{yz}, \bar{P}, P_f, \varphi$ |
| Known DOF | $\mu_0$ |
| Computable with algebraic relations | $k, \mu_s, R, \eta_\varphi, \bar{\rho}$ |
| Spatially homogenous | $G, K_\varphi, K_d, B, \alpha, k_0, \mu_f, \tau_{II}, \tau_0, n, \eta_0, R_0, \rho_s, \rho_f$ |

*Table 1. Nature of the parameters of our solver.*

```
clear
% Physics
lam = 1;                                          % Thermal conductivity
cp  = 1;                                          % Specific heat capacity
rho = 1;                                          % Density
lx  = 20;                                         % Length of the computational domain
tt  = 1;                                          % Total simulation time
% Numerics
nx  = 500;                                        % Number of grid points
nt  = 50;                                         % Number of time steps
eps = 1e-12;                                      % Convergence limit
% Preprocessing
dx  = lx/(nx-1);                                  % Grid spacing
x   = -lx/2:dx:lx/2;                              % Coordinates of grid points
dt  = tt/nt;                                      % Time step
% Initialisations
U   = exp(-x.^2)';                                % Initial temperature profile
% Action
for t = 0:dt:tt                                   % Time loop
    U_t=U;  U_tau=0;                              % Initializations for iterations
    while (max(abs(U-U_tau)) > eps)               % Pseudo-transient iterations
        U_tau = U;                                % U_tau is equal the previous U
        dtau  = 1/(2*lam/cp/rho/dx^2 + 1/dt);     % Pseudo-transient time step
        f     = lam/cp/rho*diff(U_tau,2,1)/dx^2;  % Computation of f(U_tau)
        f     = [0; f; 0];                        % Boundary conditions: no heat flux
        U     = U_tau + dtau*f - dtau*(U_tau-U_t)/dt;  % Computation of U following equation (16)
    end
end
```

*Figure 2. Application of our iterative method to the classical heat diffusion equation in 1D (the unknown U is the temperature). The initial heat profile is a Gaussian; no heat flux boundary conditions are applied. The while loop computes the implicit solution of U at time $t + \Delta t$ by applying every iteration the explicit update rule given in equation (16) (last line of loop) until $max|U_{\tau+\Delta\tau} - U_\tau| < \epsilon$. In the pseudo-transient iterations we denote $U_{\tau+\Delta\tau}$ by 'U' $U_\tau$ by 'U_tau', $U_t$ by 'U_t', $f(U_\tau)$ by 'f', $\Delta\tau$ by 'dtau', $\Delta t$ by 'dt'; comments in the script explain the remaining employed notations.*

It is important to note that our algorithm includes numerical damping of the solid velocities for faster convergence [30,44–46] and that the porosity is updated in the log space for better numerical stability.

Correct distributed memory parallelization, in our multi-GPU solver done with Cuda-Aware MPI [47,48], can be achieved by replacing in each local problem at each iteration the boundary values of the solid velocities ($V_x^s$, $V_y^s$ and $V_z^s$), of the porosity ($\varphi$), of the solid shear viscosity ($\mu_s$) and of the divergence of the Darcy flux ($\nabla_k\left(\varphi(V_k^f - V_k^s)\right)$) with the corresponding values computed in the neighbouring local problems. This is one of multiple possibilities to obtain a correct distributed memory parallelization. With *correct parallelization* we mean that the application's results do not depend on the number of employed processes and threads, i.e. any parallel run produces bitwise identical results as a fully serial run that solves the same problem with the same input parameters on the same hardware.

## 2.3 Performance measurement and parallel scaling

### 2.3.1 Flop-to-byte ratio

Today's hardware has a high *flop-to-byte ratio*. This ratio indicates how many floating point operations (FLOP) the hardware may execute per byte accessed from main memory. Current hardware has a flop-to-byte ratio of roughly 10 (the ratio is typically a bit higher for GPUs than for CPUs as well as for double precision than for single precision computations). A flop-to-byte ratio of 10 means that the hardware can perform up to 80 FLOPs while one floating point number is accessed from main memory, if we consider double precision computations (8 bytes per floating point number). The flop-to-byte ratio can also be estimated for an algorithm. It is computed as the ratio between the number of FLOPs the algorithm requires and the amount of bytes that must be transferred from main memory in order to perform these operations. If the flop-to-byte ratio of an algorithm is greater than the flop-to-byte ratio of the hardware it is executed on, then the algorithm may be considered bound by computation speed, else bound by memory access speed.

The algorithm of our 3D two-phase solver has a flop-to-byte ratio that is much below the one of current hardware as it computes spatial derivatives with finite difference stencils. The flop-to-byte ratio of finite difference stencil algorithms is in fact about 0.5 or lower [49], i.e. at least one order of magnitude below the flop-to-byte ratio of the available hardware. Our solvers performance is thus bound by the memory access speed. Memory throughput must therefore be studied to optimize our solver's performance. We do not consider performance optimization in this paper. We present instead in the following subsection (2.3.2) a simple memory throughput metric that permits to estimate how much a solver's performance is below the one of an ideal solver that implements the same system of equations. We will see moreover that the metric may be used to evaluate and compare any algorithms no matter if they are memory bound or compute bound. The metric is used in section 3 to present the results.

### 2.3.2 Effective memory throughput

Current processors' on-chip memory (registers and low-level cache) is not large enough to store all *degrees of freedom* (*DOFs*; see glossary, section 5, for the meaning in this context) of a meaningful 3D simulation, even if the resolution is low. We consider therefore only the case where the DOFs are stored off-chip, more precisely in main memory. In this case, all DOFs must be transferred at every iteration (for simplicity's sake, an iteration can refer to either an explicit physical time step or a numerical iteration in this article) between main memory and processor. More precisely, *unknown DOFs* (see glossary) must be updated, i.e. both read and written, and *known DOFs* (see glossary) must only be read. Ideally, each DOF must be read and if applicable written exactly once in a same iteration. The *minimally required main memory access per iteration* consists therefore in copying all the known DOFs from memory to the processor and copying all unknown DOFs forth and back. A non-ideal solver may however access main memory (and high-level cache that is off-chip) more than what is minimally required. It is thus useful to introduce the *effective main memory access per iteration*, measured in Gigabytes (GB), which is given by the minimally required main memory access per iteration:

$$A_{eff} \equiv 2 * D_{unknown} + D_{known}, \tag{28}$$

where $D_{unknown}$ and $D_{known}$ are the amount of Gigabytes that the unknown DOFs respectively the known DOFs occupy in main memory. The *effective memory throughput*, measured in Gigabytes per seconds (GB/s), is then given as

$$T_{eff} \equiv \frac{A_{eff}}{t_{it}}, \tag{29}$$

where $t_{it}$ is the execution time per iteration in seconds.

An ideal solver performs only the absolute minimally required memory access in the optimal manner for the given hardware and it may perfectly overlap all computations with the data transfer. Its effective memory throughput is therefore equal to the hardware's peak memory throughput, $T_{peak}$. For any solver we have therefore

$$T_{eff} \leq T_{peak}. \tag{30}$$

The effective memory throughput is therefore always to be compared with the peak memory throughput. It permits to quantify for a given solver and hardware how much the performance is below the one of an ideal solver *that implements the same system of equations*. The effective memory throughput metric may be used to this purpose no matter if the investigated solver's performance is bound by memory copy throughput or computation speed, because an ideal solver does not need to do more computations than it can perfectly overlap with the minimally required memory access, i.e. only the speed of memory access is relevant for its performance.

A great advantage of this metric over traditional flop and memory throughput metrics (see section 2.3.3) is its independency of the employed numerical algorithm and of implementation decisions; the metric is only based on the implemented system of equations. Any improvement or deterioration of execution time per iteration due to changes in the employed numerical algorithm or in the implementation is as a consequence always directly and fully reflected in $T_{eff}$: $T_{eff}$ is *directly proportional with the performance speedup that any algorithmic or implementation improvements may bring* (inverse proportional with execution time per iteration).

The metric is designed to quantify the performance of a solver independently of algorithmic and of implementation decisions; it is only relevant what system of equations is implemented. The metric is therefore helpful for the optimization of numerical algorithms and implementations. For the optimization of a given implementation, many additional hardware- and compiler-related aspects must nevertheless be taken into account (see for example the roofline model [49]) and the traditional flop and memory throughput metrics (see section 2.3.3) may be of great value to this purpose.

### 2.3.3 Hardware measured memory throughput

The term *memory throughput* refers commonly to the memory throughput measured by hardware counters (or by a profiler with a normally equivalent result). There can be an immense difference between effective memory throughput and memory throughput by hardware counters. We highlight it therefore here.

The hardware counters permit to count the *total amount of bytes* that travel through the memory bus in a given amount of time. The *total memory throughput* is thus measured. To measure the effective memory throughput with hardware counters would mean to count only the bytes of data transfer that are part of the minimally required memory access. This is however not possible.

Hardware measured throughput shows therefore simply how much *occupation of the memory bus* is achieved. A far from ideal solver that accesses main memory much more than what is minimally required and that achieves therefore a low effective memory throughput may nevertheless achieve a hardware measured throughput that is equal or close to the hardware's peak memory throughput. Memory throughput measurement with hardware is useful in the scope of performance optimization, but it is not sufficient to estimate how ideal the memory access of a solver is. This is why we suggest for such estimations the measurement of effective memory throughput.

### 2.3.4 Parallel scalability

We use the two common methods to measure parallel scalability: *strong scaling* and *weak scaling*. Strong scaling measures how the execution time depends on the number of processing elements (CPU/CPU-core/GPU/…) for a *fixed total problem size*. Weak scaling measures how the execution time depends on the number of processing elements *for a fixed problem size per processing element*. Amdahl's law shows that strong scaling is always bounded by the sequential part of the program [50]. Gustafson's law shows that weak scaling can be approximatively linear if the sequential part of the program takes a small fraction of the total execution time and if the sequential part does not take more time with a growing number of processing elements [51].

We use in addition the following common definition of *parallel efficiency* [52,53]: for a problem of size *N* on *P* processing elements, parallel efficiency is given as

$$E(N,P) \equiv \frac{1}{P}\frac{T_{seq}(N)}{T(N,P)}, \tag{31}$$

where $T(N,P)$ is the execution time of the parallel algorithm, and $T_{seq}(N)$ the execution time of the best sequential algorithm.

# 3 Results

## 3.1 Modelling

We present in this section results of a high resolution simulation of 3D deformation of fluid-filled viscoelastic porous media under a shear stress regime. We performed the simulation on 64 GPUs on the *Octopus* cluster (its technical specifications are found in section 3.2.1). Within two days 4800 physical time steps were computed, reaching a dimensionless physical time of 2.4e-2. The resolution was 511 x 511 x 994 grid points in respectively x, y and z (depth) dimensions. We visualized the results with ParaView [54] in parallel on 64 GPUs.

The simulation setup is a 3D cube of viscoelastic deformable porous media, saturated with buoyant pore fluid and containing an ellipsoidal high porosity anomaly.

The boundary conditions are

$$V_x = \frac{\tau_{xy}^{bg}}{\mu_0} y, \tag{32}$$

$$Vy = Vz = 0 \tag{33}$$

and zero normal component of the velocity difference:

$$\varphi\left(V_n^f - V_n^s\right) = 0, \tag{34}$$

where $\tau_{xy}^{bg}$ is the horizontal background shear stress. Figure 3 shows the simulation setup and the applied background shear stress boundary condition. Table 2 lists the employed non-dimensional parameters. In this table the “characteristic stress” is the stress at the transition between linear and power law viscous behaviour [36].

| Description | Symbol | Value |
|---|---|---|
| Domain size | | 30 x 30 x 60 |
| Elipsoid x-y-z-dimension axis lengths | | 14 x 14 x 3.5 |
| Horizontal background shear stress | $\tau_{xy}^{bg}$ | 1 |
| Gravity acceleration | $\vec{g}$ | [0, 0, 1] |
| Reference permeability | $k_0$ | 1 |
| Fluid shear viscosity | $\mu_f$ | 1 |
| Reference solid shear viscosity | $\mu_0$ | 1 + random noise (optional) |
| Reference bulk viscosity | $\eta_0$ | 1 |
| Fluid density | $\rho_f$ | 1 |
| Solid density | $\rho_s$ | 2 |
| Nonlinear shear viscosity exponent | $n$ | 3 |
| Characteristic stress | $\tau_0$ | 0.04 |
| Decompaction weakening factor | $R_0$ | 2000 |
| Effective bulk modulus | $K_\varphi$ | 333.33 |
| Skempton coefficient | $B$ | 0.89 |
| Biot-Willis coefficient | $\alpha$ | 0.9 |
| Elastic shear modulus | $G$ | 166.66 |

*Table 2. Non-dimensional simulation parameters.*

The initial high porosity ellipsoid contains more fluid than its surroundings and the fluid tends to move upwards as it is less dense than the solid [55]. The fluid overpressure peaks at the top borders of the ellipsoid and exceeds locally the confining pressure (total bulk pressure). In these areas pores widen, i.e. porosity increases (Eqn. (22) predicts here decompaction due to fluid overpressure). These local porosity increases results in the formation of multiple high porosity channels. The fluid underpressure develops along the major area of the channels (see blue areas in Figure 4, where the channels are shown by means of porosity isosurfaces that are coloured in function of the deviation from the fluid background pressure). The fluid underpressure relative to the near channel region causes fluid to be constantly drawn into the channels (see Eqn. (20)). The fluid that continuously enters the channels moves upwards due to buoyancy and maintains a fluid overpressure peak (red areas in Figure 4) at the channels' top, which widens pores and keeps the channels growing upwards. Underpressured areas (blue areas in Figure 4) are compacting. The result is a self-sustainable solitary-wave-like propagation of the porosity.

The channels tend to close at their bottoms, but asymmetry between the ease of decompaction and compaction hinders it (decompaction weakening [35]). Instead, the media near the bottoms, where fluid is drawn from, compacts as a result of mass balance and forms persistent walls.

Low shear viscosity regions appear as a result of elevated stresses (see Eqn. (24)), where shear deformation is localised (see red areas in Figure 5, where the channels are shown by means of porosity isosurfaces that are coloured in function of the nonlinear shear viscosity). These regions are preferentially located at the channels' top, where the decompaction process is important, and towards their tails, where viscous forces tend to compact the media with previously elevated porosity. Elevated shear and bulk viscosities (blue areas in Figure 5) are found at the tails of the waves where the media has undergone stress relaxation.

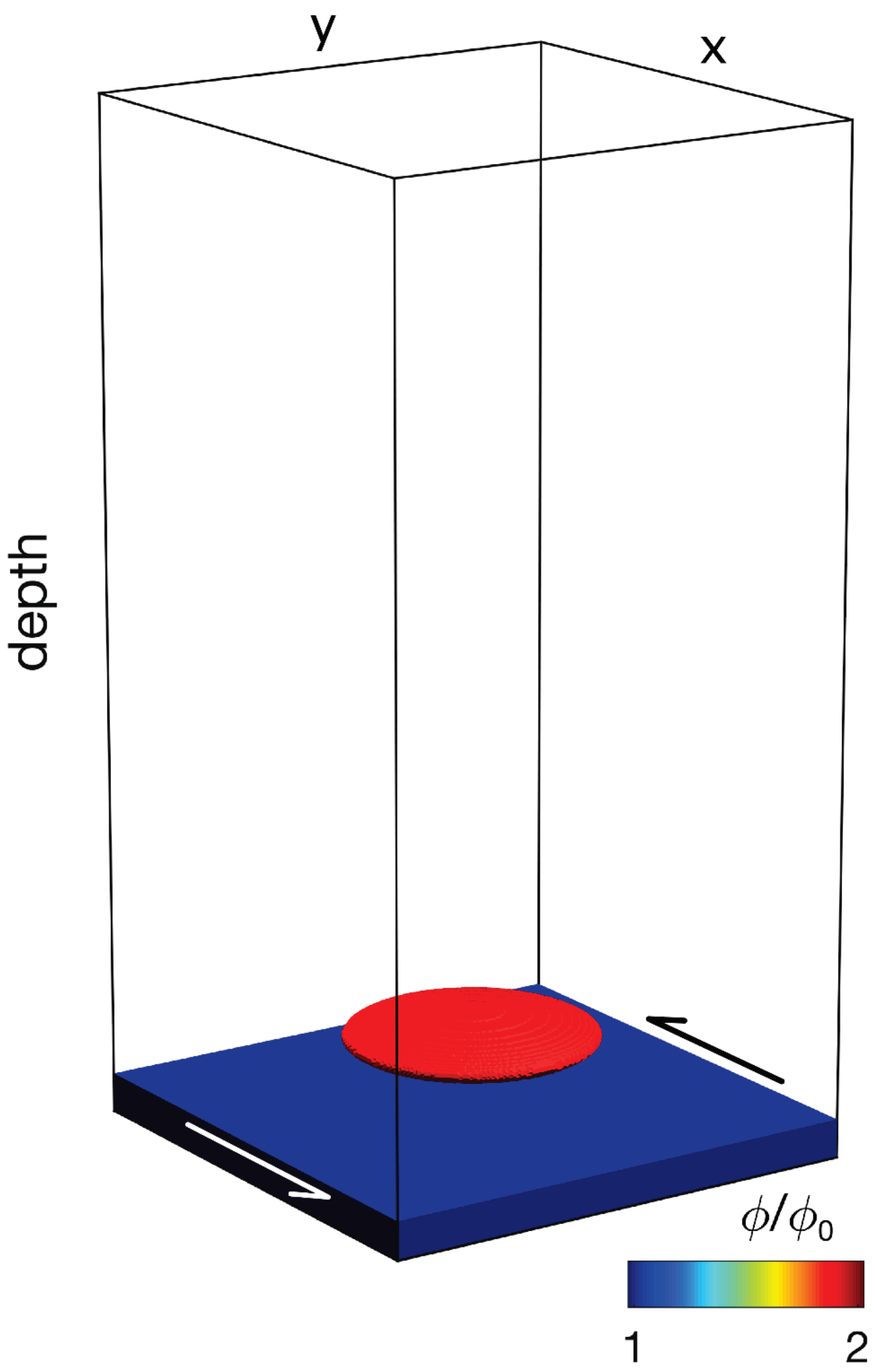

*Figure 3. The simulation setup and boundary conditions: a 3D cube of viscoelastic deformable porous media, saturated with buoyant pore fluid and containing an ellipsoidal high porosity anomaly at one fifth of the model depth; a background shear stress is applied. The porosity is in the anomaly twice as important as the background porosity, which is found in the rest of the cube. The figure shows the porosity distribution. The porosity values are normalized with the background porosity.*

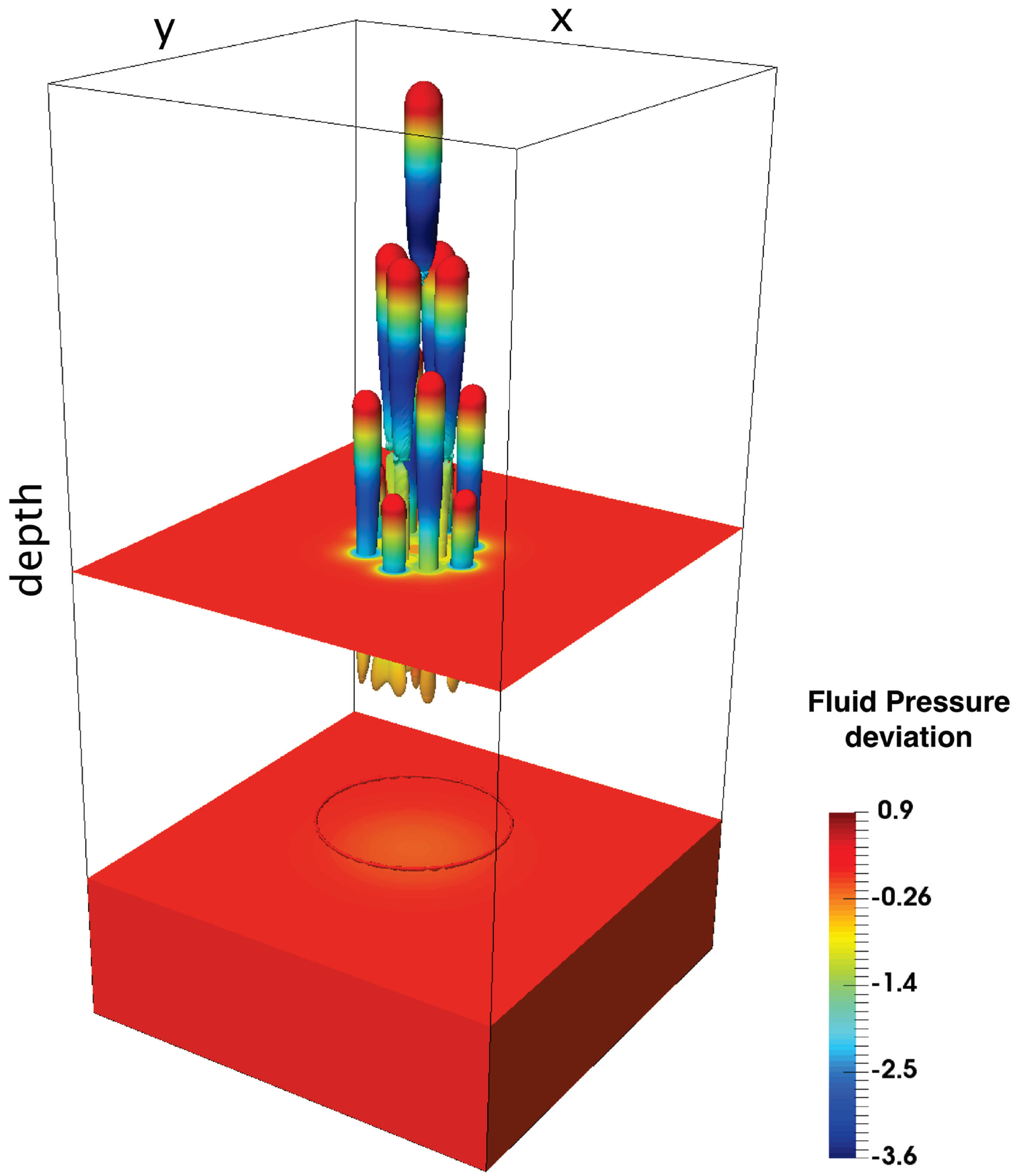

*Figure 4. The porosity channels at dimensionless physical time 2.4e-2: isosurface showing where the porosity is 4 times higher than the background. The isosurface is coloured in function of the deviation from the fluid background pressure.*

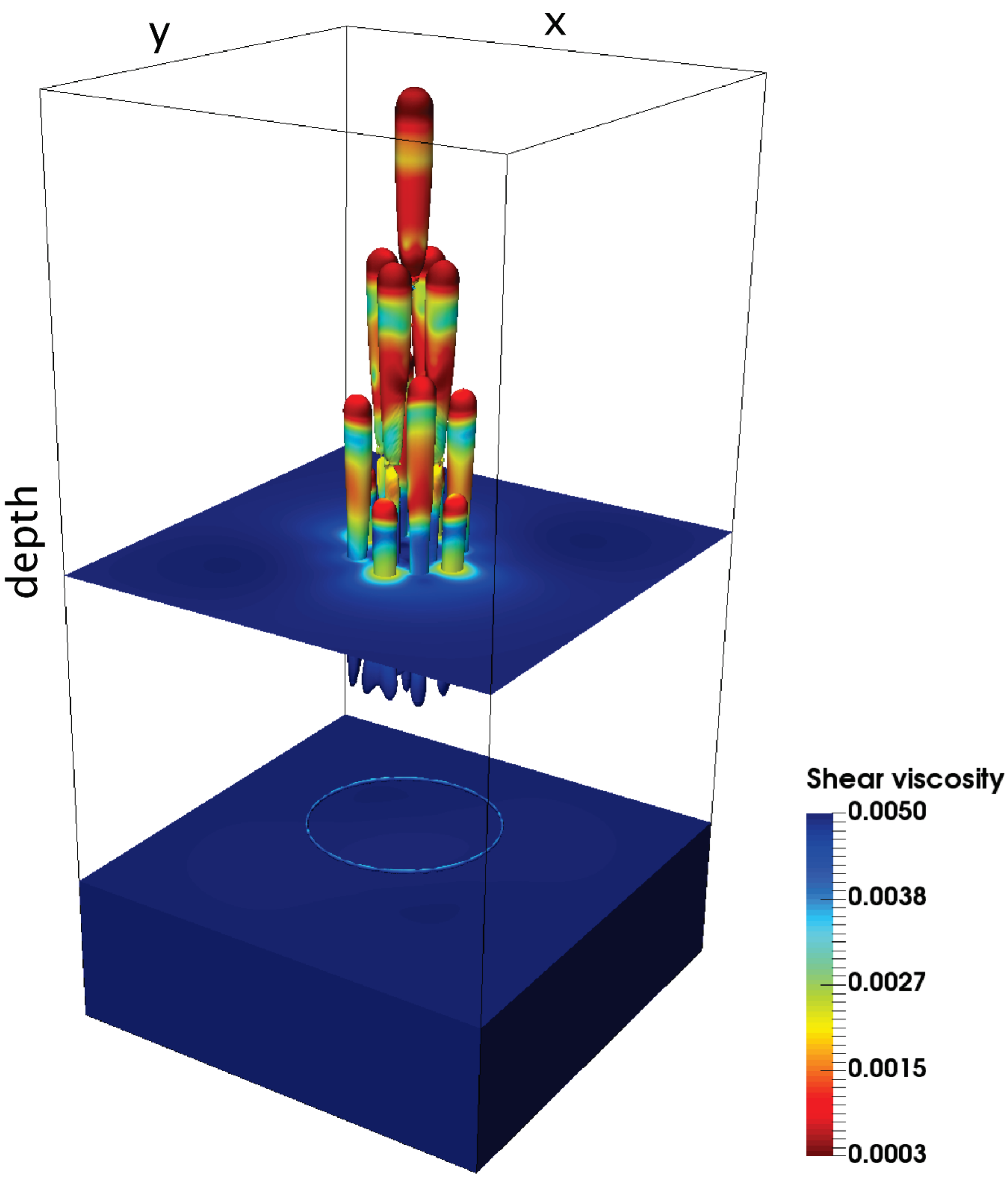

*Figure 5. The porosity channels at dimensionless physical time 2.4e-2: isosurface showing where the porosity is 4 times higher than the background. The isosurface is coloured in function of the shear viscosity.*

## 3.2 Performance

### 3.2.1 Our test systems and reported performance results

Our test systems are the Octopus cluster hosted by the Institute of Earth Science at the University of Lausanne (Lausanne, Switzerland) and the supercomputer *Piz Daint* at the Swiss National Center for Supercomputing (CSCS, Lugano, Switzerland)[2]. Octopus consists of 20 compute nodes, each containing 4 NVIDIA GeForce GTX TITAN X GPUs (Maxwell GM200 architecture; 3072 CUDA cores), 2 Intel Xeon E5-2620v3 (6 cores) and 64 GB system memory. The nodes are interconnected with dual rail FDR InfiniBand. Piz Daint is a Cray XC30 and consists of 5272 compute nodes, each containing 1 Nvidia Tesla K20X (Kepler GK110 architecture; 2688 CUDA cores), 1 Intel Xeon E5-2670 (8 cores) and 32 GB system memory. The nodes are interconnected with Cray's proprietary interconnect Aries. Piz Daint is currently listed as the number 8 on the TOP500 list of the world's top supercomputers [56]. To sum up, Octopus cumulates 80 GPUs (245 760 CUDA cores) and Piz Daint 5272 GPUs (14 171 136 CUDA cores).

All the reported performance results are the median of the obtained performance of 20 experiments, what represents in each case the typically obtained performance.

### 3.2.2 Our solver

The system of equations that we have implemented in our multi-GPU 3D two-phase solver contains 12 unknown DOFs and 1 known DOFs, listed in Table 1, page 25. Each DOF contains approximately $N = nx * ny * nz$ floating point numbers, where $nx$, $ny$ and $nz$ denote the dimensions of the problem (we assume the dimensions to be sufficiently large for the approximation to be valid). $N$ is the total number of nodes of the problem. All the unknown DOFs together contain therefore a total amount 12 N floating point numbers; all the known DOFs together contain a total amount of 1 N floating point numbers.

The amount of memory that a floating point number occupies in memory depends on the floating point number precision. There are two common standards: *single precision* occupying 4 bytes and *double precision* occupying 8 bytes for a floating point number. The amount of

[2] Both systems will be upgraded in close future and the specifications will change.

Gigabytes that all the unknown DOFs respectively known DOFs together occupy in main memory is therefore given by

$$D_{unknown} = 12\,N\,p * 10^{-9} \tag{35}$$

and

$$D_{known} = 1\,N\,p * 10^{-9}, \tag{36}$$

where $p$ is the precision in bytes. The effective main memory access per iteration (GB) is therefore given by

$$A_{eff} \equiv\; 2 * D_{unknown} + D_{known} = (2 * 12 + 1)\;N\,p * 10^{-9} \tag{37}$$

and the effective memory throughput (GB/s) by

$$T_{eff} \equiv \frac{A_{eff}}{t_{it}} = \frac{(2 * 12 + 1)\;N\,p * 10^{-9}}{t_{it}}. \tag{38}$$

The effective memory throughput for a problem of size 382³ on one GPU on Octopus (single precision) is

$$T_{eff} \equiv \frac{A_{eff}}{t_{it}} = \frac{(2 * 12 + 1)\;382^3\;4\,bytes * 10^{-9}}{0.106\;s} = 52.5\;GB/s. \tag{39}$$

This is 20 percent of the peak memory throughput $T_{peak} =\; 268$ GB/s (i.e. the same order of magnitude, see Figure 6).

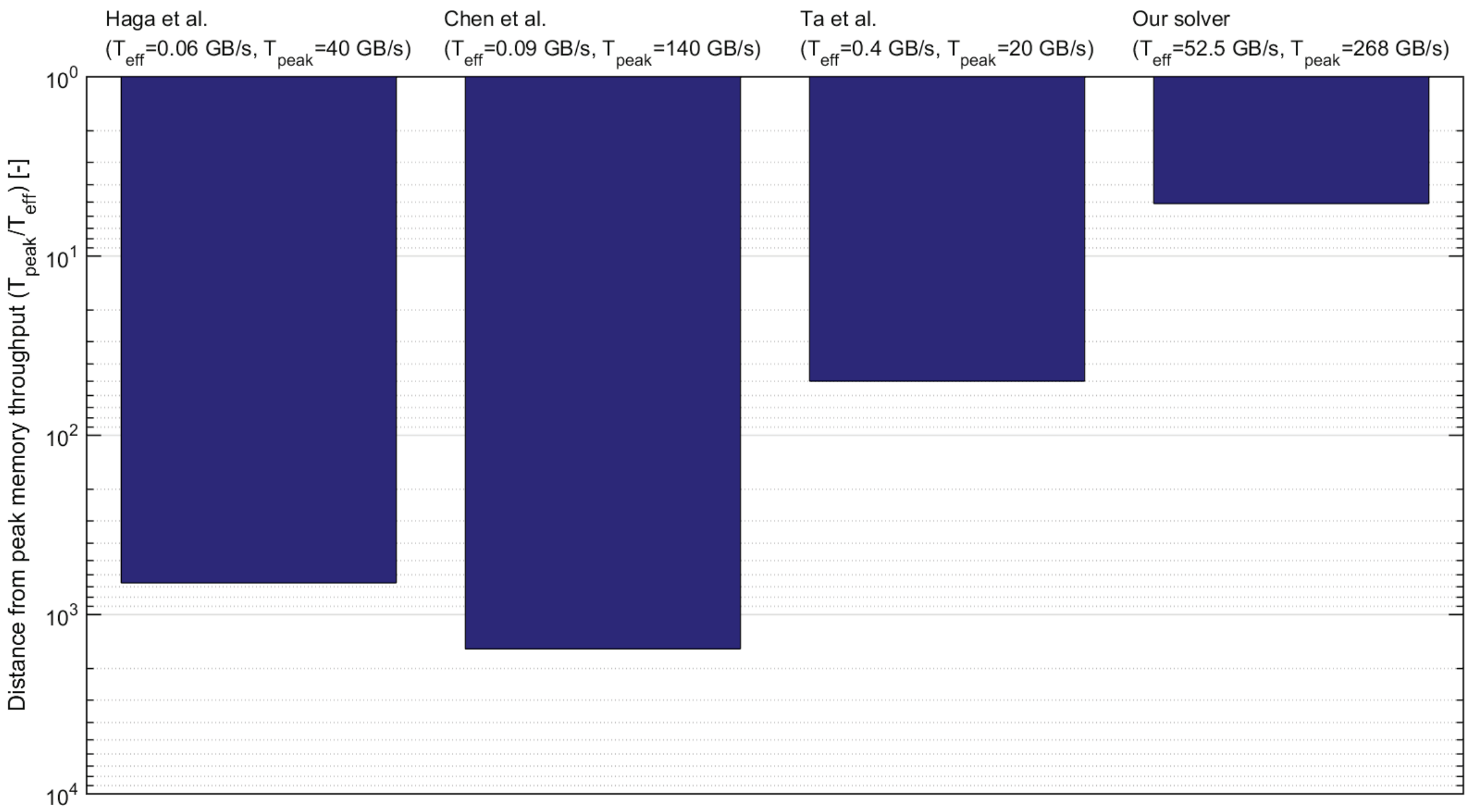

*Figure 6. Distance of each solver's effective memory throughput from the corresponding peak memory throughput (smaller bar is better: less distance from peak memory throughput).*

### 3.2.3 Comparison with other solvers

We compare in this section the performance of our solver with the one achieved by Haga et al. [25], by Chen et al. [26] and by Ta et al. [29]. Based on their respective publications it is difficult to exactly determine how many known and unknown DOFs the implemented systems of equations contain. The mathematical model that each of them use is similar to ours, with the exception that Haga et al. [25] and Chen et al. [26] do not consider viscous effects (the model Chen et al. [26] implemented on GPU does not contain the plasticity that they present in [27,28]) and Ta et al. [29] do not consider fluid pressure evolution, but include plasticity and solve additionally for temperature evolution. We assume in consequence that the system of equations that Haga et al. [25], by Chen et al. [26] and by Ta et al. [29] implemented, respectively, contains about the same amount of known and unknown DOFs as the system that we have implemented (12 unknown DOFs and 1 known DOFs, listed in Table 1, page 25).

The effective memory throughput (GB/s) is therefore for all solvers computed as for our solver:

$$T_{eff} \equiv \frac{A_{eff}}{t_{it}} \cong \frac{(2 * 12 + 1)\, N\, p * 10^{-9}}{t_{it}}. \quad (40)$$

Some of the estimations of effective memory throughput that we present in the following require a few more approximations. The estimations are nevertheless of high value, because they show the orders of magnitude of distance of the effective memory throughput from the corresponding peak memory throughput.

Haga et al. [25] implemented a CPU solver. For no test case in [25] is the execution time per iteration specified, nor can it be precisely deduced from the available information. The most accurate information with respect to this figures can be found for the 2D "test case III", a sedimentary basin simulation with 16 distinct layers of sediments, 8.4 × $10^6$ tetrahedral elements ($N$) and 1.7 × $10^6$ nodes. It was run on the *hexagon* Cray XT4 cluster located in Bergen, Norway [57]. The wall time one processor requires for one time step is 1.3 × $10^4$ seconds and a time step requires "500+ iterations" to converge [25]. The execution time per iteration is therefore approximatively 26 seconds (1.3 × $10^4$ seconds /500). The floating point number precision is not indicated. We assume it to be double precision (8 bytes), as it is the most commonly used standard in numerical calculations. The approximate effective memory throughput that Haga et al. [25] achieve in this simulation is therefore

$$T_{eff} \cong \frac{(2*12+1)\,N\,p*10^{-9}}{t_{it}} \cong \frac{(2*12+1)\,8.4*10^{6}\,8\,bytes*10^{-9}}{26s} = 0.06\,GB/s. \quad (41)$$

The peak memory throughput of the processor being roughly 40 GB/s, the effective memory throughput is three orders of magnitude below it (see Figure 6).

Chen et al. [26] present a GPU implementation (CPU offloading to GPU). They performed several consolidation analysis simulations with different preconditioners and varying material stiffness contrasts using a $24^3$ hexahedral mesh. The test system contained an Intel i7−920 CPU and a GeForce GTX 580 graphics card. The overall lowest time to solution was achieved with a partitioned block diagonal preconditioner, even though more sophisticated preconditioners converged in up to half the amount of iterations. The solver that uses the former preconditioner takes about 60 seconds for 2000 iterations. The execution time per iteration is therefore 0.03 seconds (60 seconds /2000). The floating point number precision is indicated as double precision (8 bytes). The approximate effective memory throughput that Chen et al. [26] achieve in this simulation is therefore

$$T_{eff} \cong \frac{(2*12+1)\,N\,p*10^{-9}}{t_{it}} = \frac{(2*12+1)\,24^3\,8\,bytes*10^{-9}}{0.03s} = 0.09\,GB/s. \quad (42)$$

The peak memory throughput of the GeForce GTX 580 graphics card being roughly 140 GB/s, the effective memory throughput is three orders of magnitude below it (see Figure 6).

Ta et al. [29] present an implementation that may be run on tightly coupled heterogeneous processors. They performed several performance tests varying the number of mesh elements from 7000 to 1.5 million. The most performant test system contained a tightly coupled AMD APU A10-7850 K heterogeneous processor, consisting of a quad-core CPU and a Radeon R7 GPU. The best performance was achieved with 1.5 million mesh elements: 1000 explicit time steps (without nested numerical iterations) took about 670 seconds. The execution time per iteration is therefore 0.67 seconds (670 seconds /1000). The floating point number precision is not indicated, so we assume it to be double precision (8 bytes) as before. The approximate effective memory throughput that Ta et al. (2015) achieve in this performance test is therefore

$$T_{eff} \cong \frac{(2*12+1)\,N\,p*10^{-9}}{t_{it}} \cong \frac{(2*12+1)\,1.5*10^6\,8\,bytes*10^{-9}}{0.67\ \mathrm{s}} = 0.4\,GB/s. \quad (43)$$

The peak memory throughput of the processor being roughly 20 GB/s, the effective memory throughput is two orders of magnitude below it (see Figure 6).

## 3.3 Parallel scalability

Haga et al. [25] present a CPU weak scaling test with a 3D simulation where each processor is responsible for a mesh of $16^3$ elements and a fixed amount of iterations is performed. The results are given in Figure 7. For comparison we show in the same figure weak scaling results of our 3D multi-GPU solver on Octopus (resolution per GPU: 382 x 382 x 382) and on Piz Daint (resolution per GPU: 291 x 293 x 293).

The *effective throughput in function of the number of processors $P$* can be computed from its trivial relation to the parallel efficiency and the effective throughput of a single processor run:

$$T_{eff(P)} = T_{eff(1)} * E(N,P). \quad (44)$$

Figure 8 shows $T_{eff(P)}$ corresponding to the parallel efficiencies given in Figure 7. For Haga et al. [25] we have computed $T_{eff(P)}$ based on Eqn. (44). We approximate to this purpose $T_{eff(1)}$ with the effective throughput computed in Eqn. (41), section 3.2.3 (page 38), i.e. $T_{eff(1)} = 0.07\ GBytes/s$.

Haga et al. [25] performed CPU strong scaling tests with a 3D simulation using a mesh with 8.4 × $10^6$ elements. The results are given in Figure 9 (we computed the effective throughput with the same method as in Figure 8). For comparison we show in the same figure strong scaling results of our 3D multi-GPU solver on Octopus. We chose as in Haga et al. [25] the mesh size after memory considerations: the mesh is of the maximum size that can fit on one GPU.

Chen et al. [26] and Ta et al. [29] do not present any parallel scaling results. Ta et al. [29] give nonetheless implicitly the scaling of their OpenMP implementation on the quad-core CPU of the AMD APU A10-7850K: while the serial code took 1700 seconds for 1000 explicit time steps with a 1.5 million elements mesh, the OpenMP implementation took 1000 seconds. The achieved parallel efficiency is therefore

$$E(N,P) \equiv \frac{1}{P}\frac{T_{seq}(N)}{T(N,P)} = \frac{1}{4}\frac{1700s}{1000s} = 0.425. \quad (45)$$

Ta et al. [29] performed the same test also with smaller meshes. The parallel efficiency is though worse the smaller the mesh, as one can expect it. The above computed parallel efficiency of 42.5% is therefore the best that Ta et al. [29] have achieved with their OpenMP implementation for four CPU cores.

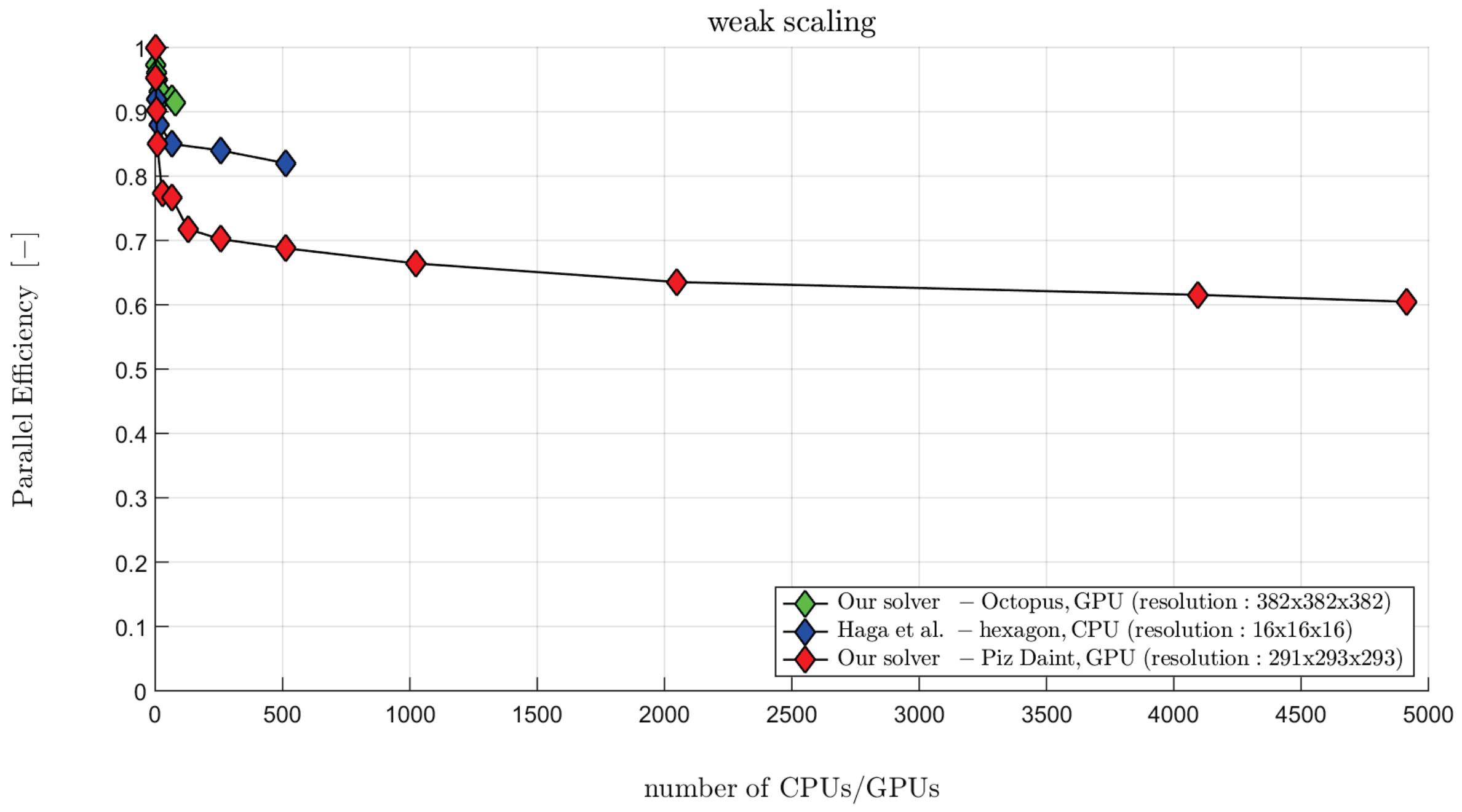

*Figure 7. Weak scaling of our 3D two-phase GPU solver and Haga et al's CPU solver (parallel efficiency).*

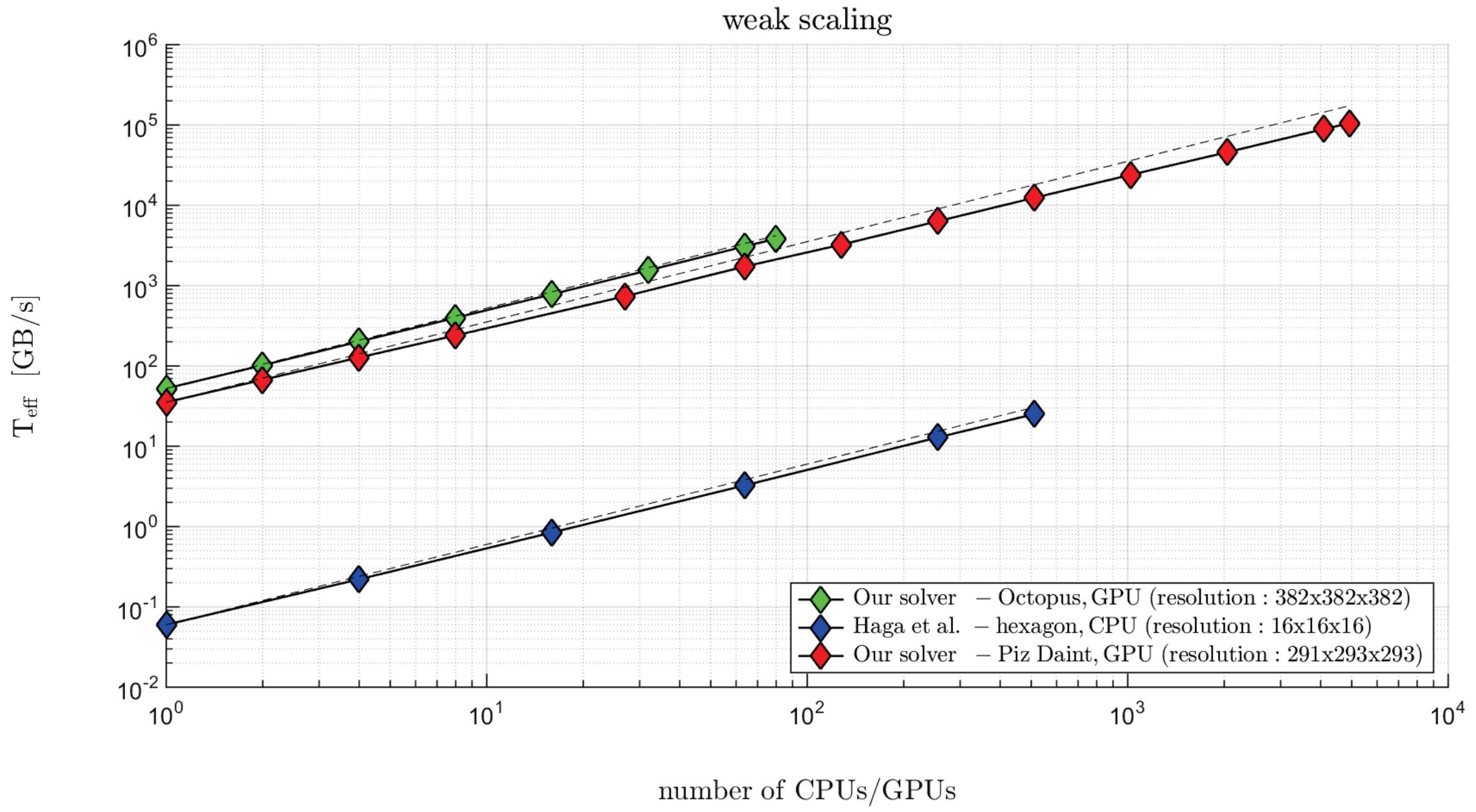

*Figure 8. Weak scaling of our 3D two-phase GPU solver and Haga et al's CPU solver (effective memory throuhput). The dashed lines shows the ideal scaling for the two solvers.*

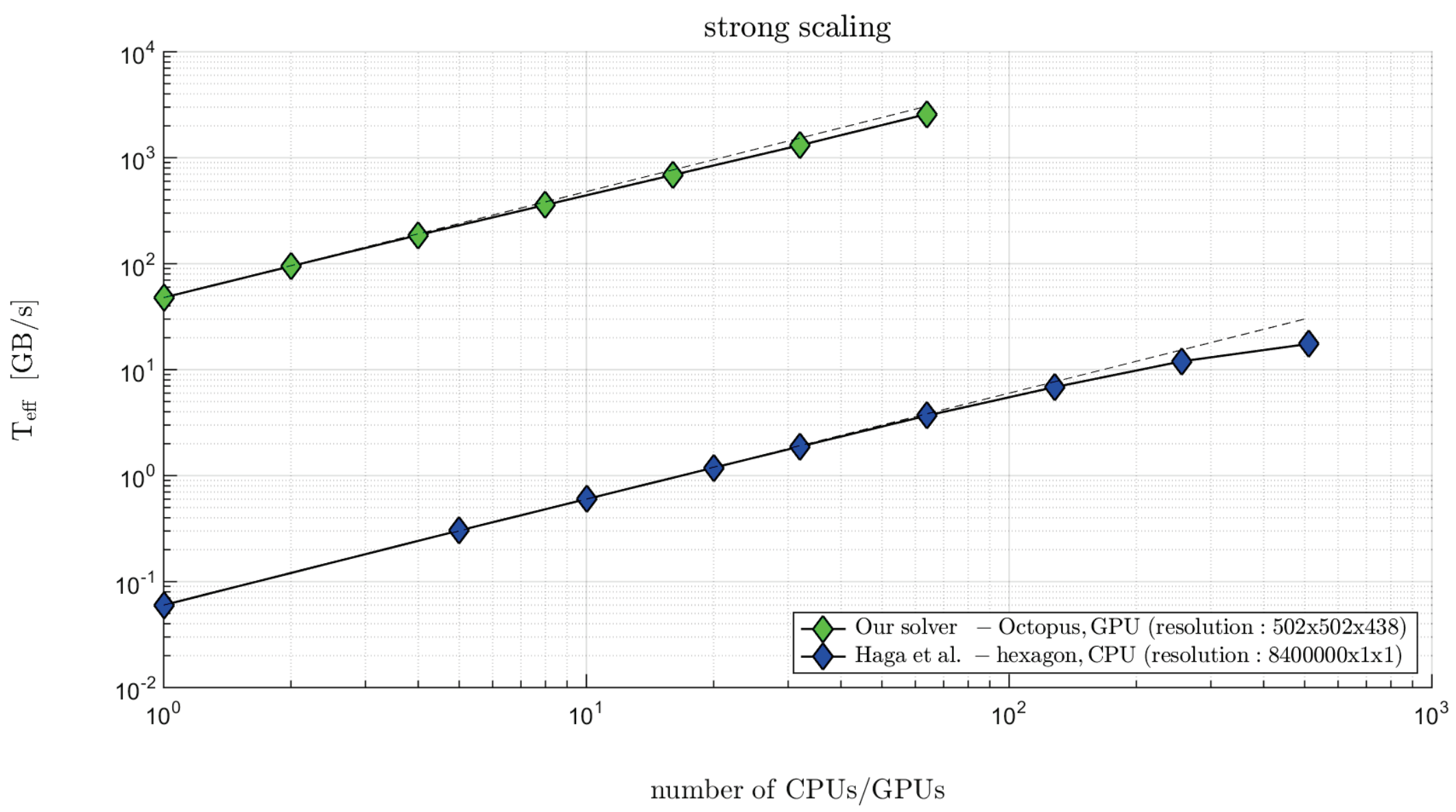

*Figure 9. Strong scaling of our 3D two-phase GPU solver and Haga et al's CPU solver (effective memory throuhput). The dashed lines shows the ideal scaling for the two solvers.*

## 4 Conclusions

We have conducted a high-resolution simulation of 3D deformation of fluid-filled viscoelastic porous media. The results show how multiple high porosity channels can form from a single high porosity ellipsoid. These channels propagate upwards as solitary waves; decompaction weakening [35] hinders the porosity restoration to its background value at the tail of the waves.

The obtained modelling results give new insights into the dynamics of channel formation in fluid-filled viscoelastic porous media, making an important next step after the computation of 3D spherically shaped porosity waves by Wiggins and Spiegelman [58]. The two key ingredients for high porosity channel generation are (1) an asymmetry in decompaction versus compaction bulk rheology [35] and (2) an appropriate observation length scale in order to capture the compaction length of the model. The modelled physical processes might be an answer to nonlinear behaviours observed in many sedimentary basins all over the world, and explain the formation of seismic chimneys in loosely consolidated reservoirs. The model could therefore find many applications in the domains of energy and risk assessment [59].

High-resolution in both space and time is needed to accurately resolve the porosity channel formation, because it is a priori unknown where and when the underlying physical processes occur and they are very localised in space and time. This demands scalable high performance solvers as the multi-GPU solver that we presented in this paper.

The effective memory throughput of our multi-GPU solver is close to the corresponding peak memory throughput, i.e. our solver runs close to hardware limit. The solver achieves a linear weak scaling on up to 5000 GPUs on Piz Daint at the Swiss National Supercomputing Center (CSCS, Lugano, Switzerland) and on the Octopus cluster, hosted by the Institute of Earth Science at the University of Lausanne (Lausanne, Switzerland), having 80 GPUs. Our algorithm is thus well suited as a basis for the building of high performance massively parallel 3D solvers for deformation of fluid-filled (visco-)elastic porous media.

Our solver solvers achieves a greatly superior ratio between effective and peak memory throughput than the other solvers that were considered in this paper: we achieve a more than two orders of magnitude better ratio than Haga et al. [25] and Chen et al. [26] and a one order of magnitude better ratio than Ta et al. [29].

Our solver is linearly scalable on thousands of nodes whereas of the other solvers only Haga et al.'s [25] has a scalable implementation. Haga et al.'s [25] solver scales linearly on up to 512 CPUs; the cumulated effective throughput of the 512 CPUs is though lower than the one of our solver on a single GPU. Moreover, Haga et al. [25] parallelization overhead makes up 18% on 512 CPUs, while the effective memory throughput is three orders of magnitude below peak memory throughput. If they improved the per CPU performance by two orders of magnitude to reach a performance that is comparable to ours, i.e. if they reduced the computation time by a factor 100, then the parallelization overhead would make up about 95 % on 512 CPUs[3]. This shows that our algorithm is better suited for massive parallelization.

We estimate that the data access pattern of our algorithm is much simpler and more regular and local than those of the other algorithms considered in this paper. We attribute the superior performance and parallelizability of our algorithm to this difference.

Following the reported results, we propose our algorithm as a performance benchmark to measure present and future algorithms against. It may help to better quantify the performance trade-offs that sophisticated algorithms normally have to face.

## 5 Glossary

- *degree of freedom (DOF)*: parameter that qualifies either as *unknown* or *known DOF* following the below definitions
- *unknown DOF*: independent *spatially heterogeneous* parameter for which a time derivative exists in the system of equations that the solver implements
- *known DOF*: independent *spatially heterogeneous* parameter from the system of equations that the solver implements, that is known previous to a simulation and thus for the entire simulation

[3] Calculation assuming that the computation time is reduced by a factor 100: 0.18/(0.18 + 0.82/100) = 0.956.

## 6 Acknowledgments

This work was supported by grants and computational resources from the Center for Advanced Modelling Science (CADMOS) and from the Swiss National Supercomputing Centre (CSCS) under project ID #s518. The financial support for CADMOS and the Blue Gene/Q system is provided by the Canton of Geneva, Canton of Vaud, Hans Wilsdorf Foundation, Louis-Jeantet Foundation, University of Geneva, University of Lausanne, and Ecole Polytechnique Fédérale de Lausanne.

## 7 References

[1] M.A. Biot, General solutions of the equations of elasticity and consolidation for a porous material, J. Appl. Mech. 23 (1956) 91–96.

[2] M.A. Biot, Theory of elasticity and consolidation for a porous anisotropic solid, J. Appl. Phys. 26 (1955) 182–185.

[3] M.A. Biot, Mechanics of deformation and acoustic propagation in porous media, J. Appl. Phys. 33 (1962) 1482–1498.

[4] M. a. Biot, General theory of three-dimensional consolidation, J. Appl. Phys. 12 (1941) 155–164. doi:10.1063/1.1712886.

[5] M.A. Biot, Thermoelasticity and irreversible thermodynamics, J. Appl. Phys. 27 (1956) 240–253.

[6] J.R. Rice, M.P. Cleary, Some basic stress diffusion solutions for fluid-saturated elastic porous media with compressible constituents, Rev. Geophys. Sp. Phys. 14 (1976) 227–241.

[7] K. von Terzaghi, Die berechnung der durchlassigkeitsziffer des tones aus dem verlauf der hydrodynamischen spannungserscheinungen, Sitzungsberichte Der Akad. Der Wissenschaften Wien, Math. Klasse, Abteilung IIa. 132 (1923) 125–138.

[8] A. Verruijt, Elastic Storage of Aquifers, in: R.J.M. De Wiest (Ed.), Flow through Porous Media, Academic Press, New York, 1969: pp. 331–376.

[9] F. Gassmann, Elastic waves through a packing of spheres, GEOPHYSICS. 16 (1951) 673–

685. doi:10.1190/1.1437718.

[10] J. Geertsma, The Effect of Fluid Pressure Decline on Volumetric Changes of Porous Rocks, in: Pet. Trans. AIME, Society of Petroleum Engineers, 1957: pp. 331–340. https://www.onepetro.org/general/SPE-728-G (accessed November 30, 2016).

[11] A.W. Skempton, The Pore-Pressure Coefficients A and B, Géotechnique. 4 (1954) 143–147. doi:10.1680/geot.1954.4.4.143.

[12] A. Verruijt, Theory and problems of poroelasticity, Delft University of Technology, Delft, The Netherlands, 2013.

[13] H. Wang, Theory of linear poroelasticity with applications to geomechanics and hydrogeology, Princeton University Press, 2000.

[14] R. de Boer, Theory of Porous Media : Highlights in Historical Development and Current State, Springer Science & Business Media, 2012.

[15] O. Coussy, Mechanics of Porous Continua, Wiley, New York, 1995.

[16] E. Detournay, A. Cheng, Fundamentals of Poroelasticity, Compr. Rock Eng. Princ. Pract. Proj. II (1993) 113–171. doi:10.1016/0148-9062(94)90606-8.

[17] V.M. Yarushina, Y.Y. Podladchikov, (De)compaction of porous viscoelastoplastic media: Model formulation, J. Geophys. Res. Solid Earth. 120 (2015) 4146–4170. doi:10.1002/2014JB011258.

[18] J.G. Berryman, Comparison of Upscaling Methods in Poroelasticity and Its Generalizations, J. Eng. Mech. 131 (2005) 928–936. doi:10.1061/(ASCE)0733-9399(2005)131:9(928).

[19] M. Schanz, Poroelastodynamics: Linear Models, Analytical Solutions, and Numerical Methods, Appl. Mech. Rev. 62 (2009) 30803. doi:10.1115/1.3090831.

[20] S. Doi, T. Washio, Ordering strategies and related techniques to overcome the trade-off between parallelism and convergence in incomplete factorizations, Parallel Comput. 25 (1999) 1995–2014. doi:10.1016/S0167-8191(99)00064-2.

[21] A. George, E. Ng, On the complexity of sparse QR and LU factorization of finite-element

matrices, SIAM J. Sci. Stat. Comput. 9 (1988) 849–861. doi:10.1137/0909057.

[22] O. Schenk, K. Gärtner, Solving unsymmetric sparse systems of linear equations with PARDISO, Futur. Gener. Comput. Syst. 20 (2004) 475–487. doi:10.1016/j.future.2003.07.011.

[23] J.B. Haga, H. Osnes, H.P. Langtangen, On the causes of pressure oscillations in low-permeable and low-compressible porous media, Int. J. Numer. Anal. Methods Geomech. 36 (2012) 1507–1522. doi:10.1002/nag.1062.

[24] J.B. Haga, H. Osnes, H.P. Langtangen, Efficient block preconditioners for the coupled equations of pressure and deformation in highly discontinuous media, Int. J. Numer. Anal. Methods Geomech. 35 (2011) 1466–1482. doi:10.1002/nag.973.

[25] J.B. Haga, H. Osnes, H.P. Langtangen, A parallel block preconditioner for large-scale poroelasticity with highly heterogeneous material parameters, Comput. Geosci. 16 (2012) 723–734. doi:10.1007/s10596-012-9284-4.

[26] X. Chen, Y. Jie, Y. Yu, GPU-accelerated iterative solutions for finite element analysis of soil-structure interaction problems, Comput. Geosci. 17 (2013) 723–738. doi:10.1007/s10596-013-9352-4.

[27] X. Chen, K.K. Phoon, Applications of Symmetric and Nonsymmetric MSSOR Preconditioners to Large-Scale Biot's Consolidation Problems with Nonassociated Plasticity, J. Appl. Math. 2012 (2012) 1–15. doi:10.1155/2012/352081.

[28] X. Chen, Y. Jie, J. Liu, Robust partitioned block preconditioners for large-scale geotechnical applications with soil-structure interactions, Int. J. Numer. Anal. Methods Geomech. 38 (2014) 72–91. doi:10.1002/nag.2199.

[29] T. Ta, K. Choo, E. Tan, B. Jang, E. Choi, Accelerating DynEarthSol3D on tightly coupled CPU–GPU heterogeneous processors, Comput. Geosci. 79 (2015) 27–37. doi:10.1016/j.cageo.2015.03.003.

[30] E. Choi, E. Tan, L.L. Lavier, V.M. Calo, DynEarthSol2D: An efficient unstructured finite element method to study long-term tectonic deformation, J. Geophys. Res. Solid Earth. 118 (2013) 2429–2444. doi:10.1002/jgrb.50148.

[31] M. Krotkiewski, M. Dabrowski, Efficient 3D stencil computations using CUDA, Parallel Comput. 39 (2013) 533–548. doi:10.1016/j.parco.2013.08.002.

[32] P. Micikevicius, 3D Finite Difference Computation on GPUs using CUDA, in: GPGPU-2 Proc. 2nd Work. Gen. Purp. Process. Graph. Process. Units, ACM Press, New York, New York, USA, 2009: pp. 79–84. doi:10.1145/1513895.1513905.

[33] T. Kaldewey, A. Di Blas, J. Hagen, E. Sedlar, S. Brandt, Memory matters, 29th IEEE Real-Time Syst. Symp. (2008) 1–4. http://www.kaldewey.com/pubs/Memory_Matters__RTSS08.pdf.

[34] M.A. Biot, J.E. Romain, Mechanics of Incremental Deformations, Phys. Today. 18 (1965) 68. doi:10.1063/1.3047001.

[35] J.A.D. Connolly, Y.Y. Podladchikov, Decompaction weakening and channeling instability in ductile porous media: Implications for asthenospheric melt segregation, J. Geophys. Res. Solid Earth. 112 (2007). doi:10.1029/2005JB004213.

[36] S.M. Schmalholz, Y.Y. Podladchikov, Tectonic overpressure in weak crustal-scale shear zones and implications for the exhumation of high-pressure rocks, Geophys. Res. Lett. 40 (2013) 1984–1988. doi:10.1002/grl.50417.

[37] R. Madariaga, Dynamics of an expanding circular fault, Bull. Seismol. Soc. Am. 66 (1976) 639–666.

[38] J. Virieux, SH-wave propagation in heterogeneous media: velocity-stress finite-difference method, Geophysics. 49 (1984) 1933–1942.

[39] J. Virieux, P-SV wave propagation in heterogeneous media: Velocity-stress finite-difference method, Geophysics. 51 (1986) 889–901.

[40] A.R. Levander, Fourth-order finite-difference P-SV seismograms, Geophysics. 53 (1988) 1425–1436. doi:10.1190/1.1442422.

[41] C.J. Randall, Absorbing boundary condition for the elastic wave equation: Velocity-stress formulation, Geophysics. 54 (1989) 1141–1152.

[42] K. Yomogida, J.T. Etgen, 3-D wave propagation in the Los Angeles basin for the Whittier-

Narrows earthquake, Bull. Seismol. Soc. Am. 83 (1993) 1325–1344.

[43] R.W. Graves, Simulating seismic wave propagation in 3D elastic media using staggered-grid finite differences, Bull. Seismol. Soc. Am. 86 (1996) 1091–1106.

[44] P.A. Cundall, Adaptive density-scaling for time-explicit calculations, in: 4th Int. Cod. Numer. Methods Geomech. 1, Edmonton, Canada, 1982: pp. 23–26.

[45] T. Heinze, G. Jansen, B. Galvan, S.A. Miller, Systematic study of the effects of mass and time scaling techniques applied in numerical rock mechanics simulations, Tectonophysics. 684 (2016) 4–11. doi:10.1016/j.tecto.2015.10.013.

[46] A.N.B. Poliakov, P.A. Cundall, Y.Y. Podladchikov, V.A. Lyakhovsky, An Explicit Inertial Method for the Simulation of Viscoelastic Flow: An Evaluation of Elastic Effects on Diapiric Flow in Two- and Three- Layers Models, in: D.B. Stone, S.K. Runcorn (Eds.), Flow Creep Sol. Syst. Obs. Model. Theory, Springer Netherlands, Dordrecht, 1993: pp. 175–195. doi:10.1007/978-94-015-8206-3_12.

[47] J. Kraus, An Introduction to CUDA-Aware MPI, (2013). https://devblogs.nvidia.com/parallelforall/introduction-cuda-aware-mpi/ (accessed October 27, 2016).

[48] W. Gropp, E. Lusk, A. Skjellum, Using MPI : portable parallel programming with the message-passing interface, Volume 1, MIT Press, 1999.

[49] S. Williams, A. Waterman, D. Patterson, Roofline: An Insightful Visual Performance Model for Floating-Point Programs and Multicore Architectures, Commun. ACM. 52 (2009) 65. doi:10.1145/1498765.1498785.

[50] G. Amdahl, The validity of the single processor approach to achieving large-scale computing capabilities, in: Proc. AFIPS Spring Jt. Comput. Conf., AFIPS, 1967: pp. 483–485.

[51] J.L. Gustafson, Reevaluating amdahl's law, Commun. ACM. 31 (1988) 532–533. doi:10.1145/42411.42415.

[52] G.C. Fox, M.A. Johnson, G.A. Lyzenga, S.W. Otto, J.K. Salmon, D.W. Walker, Solving

Problems on Concurrent Processors. Vol. 1: General Techniques and Regular Problems, Prentice-Hall, Inc., Upper Saddle River, NJ, USA, 1988.

[53] V. Kumar, A. Grama, A. Gupta, G. Karypis, Introduction to parallel computing: design and analysis of algorithms, Benjamin/Cummings Publishing Company Redwood City, CA, 1994. doi:10.1109/MCC.1994.10011.

[54] U. Ayachit, A. Bauer, B. Geveci, P. O'Leary, K. Moreland, N. Fabian, J. Mauldin, ParaView Catalyst: Enabling In Situ Data Analysis and Visualization, in: Proc. First Work. Situ Infrastructures Enabling Extrem. Anal. Vis. - ISAV2015, ACM Press, New York, New York, USA, 2015: pp. 25–29. doi:10.1145/2828612.2828624.

[55] J.A.D. Connolly, Y.Y. Podladchikov, A Hydromechanical Model for Lower Crustal Fluid Flow, in: D.E. Harlov, H. Austrheim (Eds.), Metasomatism Chem. Transform. Rock, Springer Berlin Heidelberg, Berlin, Heidelberg, 2013: pp. 599–658. doi:10.1007/978-3-642-28394-9_14.

[56] E. Strohmaier, J. Dongarra, H. Simon, M. Meuer, Top500 List - June 2016, TOP500.org. (2016). www.top500.org/list/2016/06/ (accessed November 7, 2016).

[57] The NOTUR computer cluster hexagon, (n.d.) http://www.notur.no/hardware/hexagon.

[58] C. Wiggins, M. Spiegelman, Magma migration and magmatic solitary waves in 3-D, Geophys. Res. Lett. 22 (1995) 1289–1292. doi:10.1029/95GL00269.

[59] L. Räss, V.M. Yarushina, N.S.C. Simon, Y.Y. Podladchikov, Chimneys, channels, pathway flow or water conducting features - an explanation from numerical modelling and implications for CO2 storage, in: T. Dixon, H. Herzog, S. Twinning (Eds.), Energy Procedia, ELSEVIER SCIENCE BV, NETHERLANDS, Austin, TX, 2014: pp. 3761–3774. doi:10.1016/j.egypro.2014.11.405.

# HPC.M: FROM MATLAB TO HPC ON GPU, CPU AND MIC

Samuel Omlin[1] and Yury Y. Podladchikov[1]

[1]Institute of Earth Sciences, University of Lausanne, Lausanne, Switzerland

### *Abstract*

Source-to-source translation has become a popular approach for the generation of high performance applications because it allows for automatic optimisations and parallelization. We present here the source-to-source translation based MATLAB HPC compiler *HPC.m*. It transforms simple MATLAB scripts into massively parallel near peak performance applications for GPU-, CPU- and MIC-supercomputers, clusters or workstations. The MATLAB scripts must employ basic syntax and follow a few simple rules. We have designed HPC.m in particular to iteratively solve systems of partial differential equations, computing spatial derivatives with finite differences. We illustrate the versatile use and great performance of HPC.m by deploying it to generate ten 2D and 3D multi-GPU solvers for a variety of physics across multiple disciplines. The specific mathematical models relevant to the ten solvers are simplifications of a beforehand described Thermo-Hydro-Mechanico-Chemical (THMC) master model. We outline modelling results and evaluate the performance and parallel scaling of the generated solvers. All profiled solvers run close to hardware's peak performance and scale linearly on the 80 GPUs of the Octopus cluster, hosted by the Institute of Earth Science at the University of Lausanne (Lausanne, Switzerland). They achieve moreover a speedup over the fully vectorised MATLAB input script of about 250x to 500x on one GPU, of 1000x to 2000x on one workstation with 4 GPUs and of 17 000x to 35 000x on 80 GPUs. Additionally, we show that our nonlinear poroviscoelastic two-phase flow solver scales also linearly on the 5000 GPUs of Piz Daint, a Cray XC 30 supercomputer at the Swiss National Supercomputing Centre (CSCS, Lugano, Switzerland), achieving a speedup over the fully vectorised MATLAB input script of over 500 000x. We expect a similar scaling for all the ten solvers. The source-to-source translator contained in HPC.m is, to the authors' knowledge, the first that can automatically perform all tasks required for the generation of a near peak performance supercomputing application from a code developed in classical prototyping environment as MATLAB.

### *Keywords*

Source-to-source translation; HPC; Supercomputing; Massive parallelization; Multi-GPU; MATLAB; Thermo-Hydro-Mechanico-Chemical modelling

# 1 Introduction

It takes many steps to build a near peak performance supercomputing application from a code developed in a classical prototyping environment as MATLAB. The three major tasks to accomplish are (1) the translation of the prototype to a compilable hardware-close language as C, (2) performance optimisation and (3) parallelization. Each of this tasks can take months or even years if they are performed manually and errors happen easily. Besides, code verification is difficult and time intensive.

The increasing availability of clusters and supercomputers in academia and industry over the last decades alongside with their increasing performance have risen a strong demand for tools that increase the programmers' productivity [1]. *"Software scalability is the most significant limiting factor in achieving the next 10x improvements in performance, and it remains one of the most significant factors in reaching 1000x"* is one of the key findings of the recent HPC market study by the U.S. Council on Competitiveness [1]. This situation gave rise to many projects that aim to automatize some of the steps in the building of a near peak performance supercomputing application from a high-level language prototype.

Source-to-source translation has become a popular approach for the generation of high performance applications as it allows for automatic optimisations and parallelization. It is common in this approach to support a limited set of applications or methods in order to allow for aggressive domain specific performance optimizations. Many of these source-to-source translators are designed for stencil-computations [2–5], which are for example required for finite difference computations in the solution of partial differential equations. The input source language for these source-to-source translators is typically a domain specific language (DSL), which has been developed on top of a compilable general purpose language like C or C++ [3,4].

We present in this article HPC.m – the MATLAB HPC compiler. HPC.m transforms simple MATLAB scripts within a few seconds into massively parallel near peak performance applications for GPU-, CPU- and MIC-supercomputers, clusters or workstations. The MATLAB scripts must employ basic syntax and follow a few simple rules. The essential part of HPC.m is a source-to-source translator. It is the first source-to-source translator known to the authors that can perform automatically all tasks that are required for the generation of a near peak performance supercomputing application from a code developed in classical prototyping environment as MATLAB.

We have designed HPC.m in particular to solve systems of partial differential equations (PDEs) in very high spatial and temporal resolution and thus to resolve many scales in space and time. This is especially relevant for real-world applications for which the time- and length-scale is unknown in advance, where events happen suddenly and very localized (typically the case for problems with important nonlinearities).

We have designed HPC.m to translate applications that compute spatial derivatives with finite differences and that use a regular Cartesian one, two or three dimensional grid which may be staggered. Possible applications reach from plain explicit solvers to more sophisticated iterative solvers with preconditioners, dynamic time step computation and potentially multigrid. Traditional solvers that require to generate the full coefficient matrix are not within the target applications of HPC.m. We have intentionally given priority to these rather simple solving methods (i.e. the usage of a regular Cartesian grid and explicit or iterative solvers) because they are massively parallelizable by construction: only local memory operations and local point to point communication are required for these methods; global communication is only occasionally required for optional features like dynamic time step computation. In consequence, HPC.m allows us to generate near peak performance real-world applications that scale linearly on the fastest supercomputers in the world. This holds a priori also on tomorrow's supercomputers (exascale machines).

HPC.m uses MABLAB scripts as input, because MATLAB provides an exceptional environment to prototype scientific applications that has made it the tool of first choice in many scientific communities [6,7]. MATLAB has an exceptionally user-friendly and intuitive syntax and very powerful graphics and debugger.

The objective of this article is to present (1) how HPC.m works and (2) what performance and parallel scaling we may achieve with it. To illustrate the versatile use and great performance of HPC.m we deploy it to generate 2D and 3D solvers for a variety of physics across multiple disciplines: shallow water, glaciar flow, convection, scalar and reactive porosity waves, poroviscoelastic two-phase flow, shear heating, seismics, acoustic wave propagation and heat diffusion.

The paper is organized as follows. In section 2, we present (1) the mathematical models for which we generate solvers, (2) the working of HPC$^{.m}$ and (3) our methodology to assess performance. In section 3, we report on modelling, performance and parallel scaling results of the generated solvers. In section 4, we give concluding remarks. Section 5 is a glossary.

# 2 Materials and Methods

## 2.1 Mathematical models

We deploy HPC$^{.m}$ in this paper to generate a multitude of solvers. We present here the specific mathematical models relevant to those solvers in comparative tables. These tables permit to find differences and similarities between the mathematical models. All those mathematical models are application-specific simplifications of a Thermo-Hydro-Mechanico-Chemical (THMC) master model that we introduce first. We colour temperature-related terms in red, hydrology-related terms in blue, chemistry-related terms in green and all remaining terms in black. All the symbols that are used in this article are listed in Table 1 (in the application-specific models, subscripts and bars are omitted where the context makes them redundant).

The following equations form the THMC master model:

$$
\begin{cases}
\varphi\rho_f \dfrac{d^f V_i^f}{dt} + (1-\varphi)\rho_s \dfrac{d^s V_i^s}{dt} = \nabla_j\left(\bar{\tau}_{ij} - \bar{P}\delta_{ij}\right) - \bar{\rho} g_i & (1) \\
\varphi(V_i^f - V_i^s) = -\dfrac{k}{\mu_f}\left(\nabla_i P_f + \rho_f g_i\right) & (2) \\
\dfrac{1}{T}\nabla_k q_k = -\dfrac{\overline{\rho c_p}}{T}\dfrac{dT}{dt} + \alpha_T \dfrac{d^s\bar{P}}{dt} - \alpha_{Tf}\dfrac{d^s P_f}{dt} - \dfrac{1}{T}\dfrac{\bar{\tau}_{ij}\,\bar{\tau}_{ij}}{\eta_s} & (3) \\
\nabla_k V_k^s = \alpha_T \dfrac{d\bar{T}}{dt} - \dfrac{1}{K_d}\dfrac{d^s\bar{P}}{dt} + \dfrac{\alpha}{K_d}\dfrac{d^f P_f}{dt} - \dfrac{\bar{P}-P_f}{(1-\varphi)\eta_\varphi} & (4) \\
\nabla_k\left(\varphi(V_k^f - V_k^s)\right) = -\alpha_{Tf}\dfrac{d\bar{T}}{dt} + \dfrac{\alpha}{K_d}\dfrac{d^s\bar{P}}{dt} - \dfrac{\alpha}{B\,K_d}\dfrac{d^f P_f}{dt} + \dfrac{\bar{P}-P_f}{(1-\varphi)\eta_\varphi} + R_X\dfrac{d^s X_s}{dt} & (5) \\
q_i = -\lambda(\nabla_i T) & (6) \\
\dfrac{\bar{\tau}_{ij}}{\mu_s} + \dfrac{1}{G}\dfrac{d^\nabla \bar{\tau}_{ij}}{dt} = \nabla_i V_j^s + \nabla_j V_i^s - \dfrac{2}{3}(\nabla_k V_k^s)\,\delta_{ij} \quad . & (7) \\
\dfrac{1}{1-\varphi}\dfrac{d^s\varphi}{dt} = -\dfrac{1}{K_\varphi}\left(\dfrac{d^s\bar{P}}{dt} - \dfrac{d^f P_f}{dt}\right) - \dfrac{\bar{P}-P_f}{\eta_\varphi} - R_p\dfrac{d^s X_s}{dt} & (8) \\
\dfrac{d^s X_s}{dt} = r\varphi^{2/3}\left(X_{seq} - X_s\right) & (9) \\
\bar{\rho} = \varphi\,\rho_f + (1-\varphi)\rho_s & (10) \\
\rho_s = \rho_{s_0}(1-\alpha_T T) & (11) \\
\rho_f = \rho_{f_0}(1-\alpha_T T) & (12) \\
\mu_s = \dfrac{\mu_0}{1 + \left(\dfrac{\tau_{II}}{\tau_0}\right)^{(n-1)}} & (13) \\
\eta_\varphi = \eta_0 \dfrac{\mu_s}{\varphi} & (14) \\
k = k_0\varphi^3 & (15)
\end{cases}
$$

Equations (1) and (2) conserve the total momentum (fluid and solid) and the fluid momentum, respectively; the latter includes a viscous rheology (Darcy flux). Equations (3) to (5) conserve energy, solid mass and fluid mass and include a visco-elastic volumetric bulk rheology and changing temperature and reaction induced expansion; they are fully coupled and have symmetric coefficients; the temperature of the solid is assumed to be equal to the temperature of the fluid. Equation (6) to (8) are the heat flux, a visco-elastic deviatoric rheology (following the Maxwell model) and a closure relation that describes the evolution of porosity. Equation (9) to (15) are for the evolution of the mass fraction of volatile species in solid; the total, solid and fluid density; the shear and bulk viscosity; and the nonlinear Carnman-Kozeny permeability.

| Symbol | Meaning |
|---|---|
| $V_k^s$ or $V_i^s$ | Solid velocity |
| $V_k^f$ or $V_i^f$ | Fluid velocity |
| $\bar{P}$ | Total pressure |
| $P_f$ | Fluid pressure |
| $\varphi$ | Porosity |
| $K_d$ | Drained bulk modulus |
| $\alpha$ | Biot-Willis coefficient |
| $B$ | Skempton's coefficient |
| $\eta_\varphi$ | Effective bulk viscosity |
| $\bar{\tau}_{ij}$ | Total stress deviator |
| $g_i$ | Gravitational acceleration |
| $\bar{\rho}$ | Total density |
| $\rho_f$ | Fluid density |
| $k$ | Permeability |
| $\mu_f$ | Fluid shear viscosity |
| $\mu_s$ | Solid shear viscosity |
| $G$ | Elastic shear modulus of the solid |
| $\delta_{ij}$ | Kronecker-delta |
| $K_\varphi$ | Effective bulk modulus |
| $k_0$ | Reference permeability |
| $\mu_0$ | Reference solid shear viscosity |
| $\eta_0$ | Reference bulk viscosity |
| $\tau_0$ | Characteristic stress at the transition between linear and power law viscous behaviour [8] |
| $\tau_{II}$ | Second invariant of the deviatoric stress |
| $n$ | Power law viscosity exponent |
| $\rho_s$ | Solid density |
| $\frac{d^s}{dt}$ | Time derivative with respect to the solid |
| $\frac{d^f}{dt}$ | Time derivative with respect to the fluid |
| $\frac{d^{\nabla}}{dt}$ | Objective time derivative (e.g. Jaumann) |
| $T$ | Temperature |
| $q_k$ or $q_i$ | Heat flux |
| $\bar{c}_p$ | Specific heat capacity |
| $\alpha_T$ | Thermal expansion coefficient |
| $\alpha_{Tf}$ | Thermal expansion coefficient specific for fluid |
| $\lambda$ | Thermal conductivity coefficient |
| $\rho_{s_0}$ | Reference solid density |
| $\rho_{f_0}$ | Reference fluid density |
| $X_s$ | Mass fraction of volatile species in solid |
| $X_{seq}$ | Mass fraction of volatile species in solid at equilibrium |
| $R_X$ | Effect of reaction on density |
| $R_p$ | Effect of reaction on porosity |
| $r$ | Reaction rate |
| $\beta$ | Compressibility coefficient |
| $H$ | Height of Water / Glacier |
| $Q_{SH}$ | Heat source due to shear heating |

*Table 1 List of symbols used in this article.*

Table 2 gives the systems of equation for heat diffusion, acoustic wave propagation, glacier flow and shallow water. It compares linear and nonlinear diffusion and wave propagation. Table 3 compares the linearized compressible Navier-Stokes to the Cauchy-Navier elasticity. Table 4 compares TM coupling (e.g. convection) and HT coupling (e.g. porous convection). Table 5 compares convection and shear heating [9], which are two applications of TM coupling. Table 6 compares Hm coupling (e.g. scalar porosity waves [10]) and HM coupling (poroviscoelastic two-phase flow; e.g. mechanical porosity waves [11]), the lower case m standing for scalar mechanics. Table 7 compares Hm coupling (e.g. scalar porosity waves [10]) and HmC coupling (e.g. reactive porosity waves [12]), the lower case letter 'm' standing again for scalar mechanics.

| | Diffusion | Wave propagation |
|---|---|---|
| Linear | $\begin{cases} \rho\, c_p \dfrac{\partial T}{\partial t} = -\nabla_k q_k \\ q_i = -\lambda\, \nabla_i T \end{cases}$ | $\begin{cases} \beta \dfrac{\partial P}{\partial t} = -\nabla_k V_k \\ \rho \dfrac{\partial V_i}{\partial t} = -\nabla_i P \end{cases}$ |
| Nonlinear | $\begin{cases} \dfrac{\partial H}{\partial t} = -\nabla_i (H V_i) \\ V_i = -\dfrac{H^2 g}{3\mu_f}\, \nabla_i (H+B) \end{cases}$ | $\begin{cases} \dfrac{\partial H}{\partial t} = -\nabla_k (H V_k) \\ \dfrac{\partial H V_i}{\partial t} = -\nabla_j \left(H V_j V_i\right) - \; Hg\, \nabla_i (H+B) \end{cases}$ |

*Table 2. Comparison of linear and nonlinear diffusion and wave propagation: Heat diffusion (top left), acoustic wave propagation (top right), glacier flow (bottom left) and shallow water (bottom right).*

| | Linearized compressible Navier-Stokes | Cauchy-Navier elasticity |
|---|---|---|
| Momentum conservation | $\rho \dfrac{\partial V_i}{\partial t} = \nabla_j \left(\tau_{ij} - P\delta_{ij}\right) - \rho g_i$ | |
| Mass conservation | $\beta \dfrac{\partial P}{\partial t} = -\nabla_k V_k$ | |
| Rheology | $\dfrac{\tau_{ij}}{\mu_s} = \; \nabla_i V_j + \nabla_j V_i - \dfrac{2}{3} (\nabla_k V_k)\, \delta_{ij}$ | $\dfrac{1}{G} \dfrac{\partial \tau_{ij}}{\partial t} = \; \nabla_i V_j + \nabla_j V_i - \dfrac{2}{3} (\nabla_k V_k)\, \delta_{ij}$ |

*Table 3. Comparison of linearized compressible Navier-Stokes and Cauchy-Navier elasticity.*

| | TM coupling | HT coupling |
|---|---|---|
| Momentum conservation | $0 = \nabla_j(\tau_{ij} - P\delta_{ij}) - \rho g_i$ | $\varphi(V_i^f - V_i^s) = -\frac{k}{\mu_f}(\nabla_i P_f + \rho_f g_i)$ |
| Mass conservation | $0 = -\nabla_k V_k$ | $\nabla_k\left(\varphi(V_k^f - V_k^s)\right) = -\frac{\alpha}{B\,K_d}\frac{d^f P_f}{dt}$ |
| Rheology | $\frac{\tau_{ij}}{\mu_s} = \nabla_i V_j + \nabla_j V_i - \frac{2}{3}(\nabla_k V_k)\,\delta_{ij}$ | |
| Heat diffusion | $\rho\, c_p\left(\frac{\partial T}{\partial t} + V_k \nabla_k T\right) = -\nabla_k q_k$<br>$q_i = -\lambda\, \nabla_i T$ | |
| Temperature dependent density | $\rho = \rho_0(1 - \alpha_T T)$ | |

*Table 4. Comparison of TM coupling (e.g. convection) and HT coupling (e.g. porous convection).*

| | Convection | Shear Heating |
|---|---|---|
| Momentum conservation | $0 = \nabla_j(\tau_{ij} - P\delta_{ij}) - \rho g_i$ | |
| Mass conservation | $0 = -\nabla_k V_k$ | |
| Rheology | $\frac{\tau_{ij}}{\mu_s} = \nabla_i V_j + \nabla_j V_i - \frac{2}{3}(\nabla_k V_k)\,\delta_{ij}$ | |
| Heat diffusion | $\rho\, c_p\left(\frac{\partial T}{\partial t} + V_k \nabla_k T\right) = -\nabla_k q_k + Q_{SH}$<br>$q_i = -\lambda\, \nabla_i T$ | |
| Temperature dependent density | $\rho = \rho_0(1 - \alpha_T T)$ | $\rho = \rho_0$ |
| Heat source | $Q_{SH} = 0$ | $Q_{SH} = \frac{\bar{\tau}_{ij}\,\bar{\tau}_{ij}}{\eta_s}$ |
| Gravity | $g_i = g_0$ | $g_i = 0$ |

*Table 5. TM coupling: comparison of convection and shear heating.*

| | Hm coupling (assuming $V^s = 0$) | HM coupling |
|---|---|---|
| Fluid momentum conservation | $\varphi(V_i^f - V_i^s) = -\frac{k}{\mu_f}(\nabla_i P_f + \rho_f g_i)$ | |
| Fluid mass conservation | $\nabla_k\left(\varphi(V_k^f - V_k^s)\right) = \frac{\alpha}{K_d}\frac{\partial \bar{P}}{\partial t} - \frac{\alpha}{B\,K_d}\frac{\partial P_f}{\partial t} + \frac{\bar{P} - P_f}{(1-\varphi)\eta_\varphi}$ | |
| Porosity closure relation | $\frac{1}{1-\varphi}\frac{d^s\varphi}{dt} = -\frac{1}{K_\varphi}\left(\frac{\partial \bar{P}}{\partial t} - \frac{\partial P_f}{\partial t}\right) - \frac{\bar{P} - P_f}{\eta_\varphi}$ | |
| Nonlinear Carnman-Kozeny permeability | $k = k_0\varphi^3$ | |
| Total density | $\bar{\rho} = \varphi\,\rho_f + (1-\varphi)\rho_s$ | |
| Total pressure approximation | $\bar{P} = \int \bar{\rho} g dz$ | |
| Solid momentum conservation | | $0 = \nabla_j\left(\bar{\tau}_{ij} - \bar{P}\delta_{ij}\right) - \bar{\rho} g_i$ |
| Solid mass conservation | | $\nabla_k V_k^s = -\frac{1}{K_d}\frac{d^s\bar{P}}{dt} + \frac{\alpha}{K_d}\frac{d^f P_f}{dt} - \frac{\bar{P} - P_f}{(1-\varphi)\eta_\varphi}$ |
| Visco-elastic deviatoric rheology (Maxwell) | | $\frac{\bar{\tau}_{ij}}{\mu_s} + \frac{1}{G}\frac{d^\nabla\bar{\tau}_{ij}}{dt} = \nabla_i V_j^s + \nabla_j V_i^s - \frac{2}{3}(\nabla_k V_k^s)\,\delta_{ij}$ |
| Shear viscosity | | $\mu_s = \frac{\mu_0}{1 + \left(\frac{\tau_{II}}{\tau_0}\right)^{(n-1)}}$ |
| Bulk viscosity | | $\eta_\varphi = \eta_0\frac{\mu_s}{\varphi}$ |

*Table 6. Comparison of Hm coupling (e.g. scalar porosity waves) and HM coupling (poroviscoelastic two-phase flow; e.g. mechanical porosity waves), the lower case letter 'm' standing for scalar mechanics.*

| | Hm coupling | HmC coupling |
|---|---|---|
| Fluid momentum conservation | $\varphi(V_i^f - V_i^s) = -\frac{k}{\mu_f}(\nabla_i P_f + \rho_f g_i)$ | |
| Fluid mass conservation | $\nabla_k\left(\varphi(V_k^f - V_k^s)\right) = \frac{\alpha}{K_d}\frac{\partial \bar{P}}{\partial t} - \frac{\alpha}{B\,K_d}\frac{\partial P_f}{\partial t} + \frac{\bar{P} - P_f}{(1-\varphi)\eta_\varphi} + R_X\frac{\partial X}{\partial t}$ | |
| Porosity closure relation | $\frac{1}{1-\varphi}\frac{\partial \varphi}{\partial t} = -\frac{1}{K_\varphi}\left(\frac{\partial \bar{P}}{\partial t} - \frac{\partial P_f}{\partial t}\right) - \frac{\bar{P} - P_f}{\eta_\varphi} - R_p\frac{\partial X_s}{\partial t}$ | |
| Nonlinear Carnman-Kozeny permeability | $k = k_0\varphi^3$ | |
| Total pressure approximation | $\bar{P} = \int \bar{\rho} g dz$ | |
| Total density | $\bar{\rho} = \varphi\,\rho_f + (1-\varphi)\rho_s$ | |
| Solid density | | $\rho_s = \rho_{s_0}(1 - \beta_s\bar{P} + \beta_X X_s)$ |
| Fluid density | | $\rho_f = \rho_{f_0}\left(1 - \beta_f P_f\right)$ |
| Mass fraction of volatile species in solid | | $\frac{\partial X_s}{\partial t} = r\varphi^{2/3}\left(X_{seq} - X_s\right)$ |
| Effect of reaction on density | $R_X = 0$ | $R_X = -(1-\varphi)\left(\frac{\rho_s - \rho_f}{\rho_f(1-X_s)} + \beta_X\right)$ |
| Effect of reaction on porosity | $R_p = 0$ | $R_p = (1-\varphi)\left(\beta_X - \frac{1}{1-X_s}\right)$ |
| Mass fraction of volatile species in solid at equilibrium | | $X_{seq} = X_{seq0} + \frac{\partial X_s}{\partial P_f}P_f$ |

*Table 7. Comparison of Hm coupling (e.g. scalar porosity waves) and HmC coupling (e.g. reactive porosity waves), the lower case letter 'm' standing for scalar mechanics.*

## 2.2 HPC$^{.m}$

### 2.2.1 Overview

HPC$^{.m}$ enables the following fully automatic steps: (1) translation to a compilable language (to Fortran 90 or C for CPU and MIC, to Cuda C for GPU), (2) shared memory parallelization (depending on step1: with OpenMP for Fortran 90 or C, with Cuda for Cuda C), (3) distributed memory parallelization (with MPI for any of the given options in step 1) and (4) deployment of a suitable compiler (depending on the language chosen in step 1). Figure 1 shows an overview of the workflow of HPC$^{.m}$.

HPC$^{.m}$ is called from within MATLAB. It is called like any normal MATLAB function. One must specify the name of the input script and the target language of the translation via the function arguments. HPC$^{.m}$ supports some compiler flags which are passed in form of optional arguments (explained in section 2.2.8). Most of the code generation of HPC$^{.m}$ is however steered directly from within the script. Within the script we define for example whether and how to use MPI and OpenMP or define parameters for the usage of a GPU.

HPC$^{.m}$ has been in particular designed to translate MATLAB prototypes for today's and tomorrow's supercomputing machines, which most, if not all, possess distributed memory. To allow for an automatic parallelization that is suitable for distributed memory, HPC$^{.m}$ requires that the problem to be parallelized can be spatially decomposed into a Cartesian grid of *local problems*, where each local problem defines for each adjacent local problem the *boundary conditions* of the common boundary (see 2D example in Figure 2). The size of a local problem is specified in the HPC$^{.m}$ input script. The size of the *global problem* is defined implicitly by the dimensions of the grid of local problems and the size of a local problem.

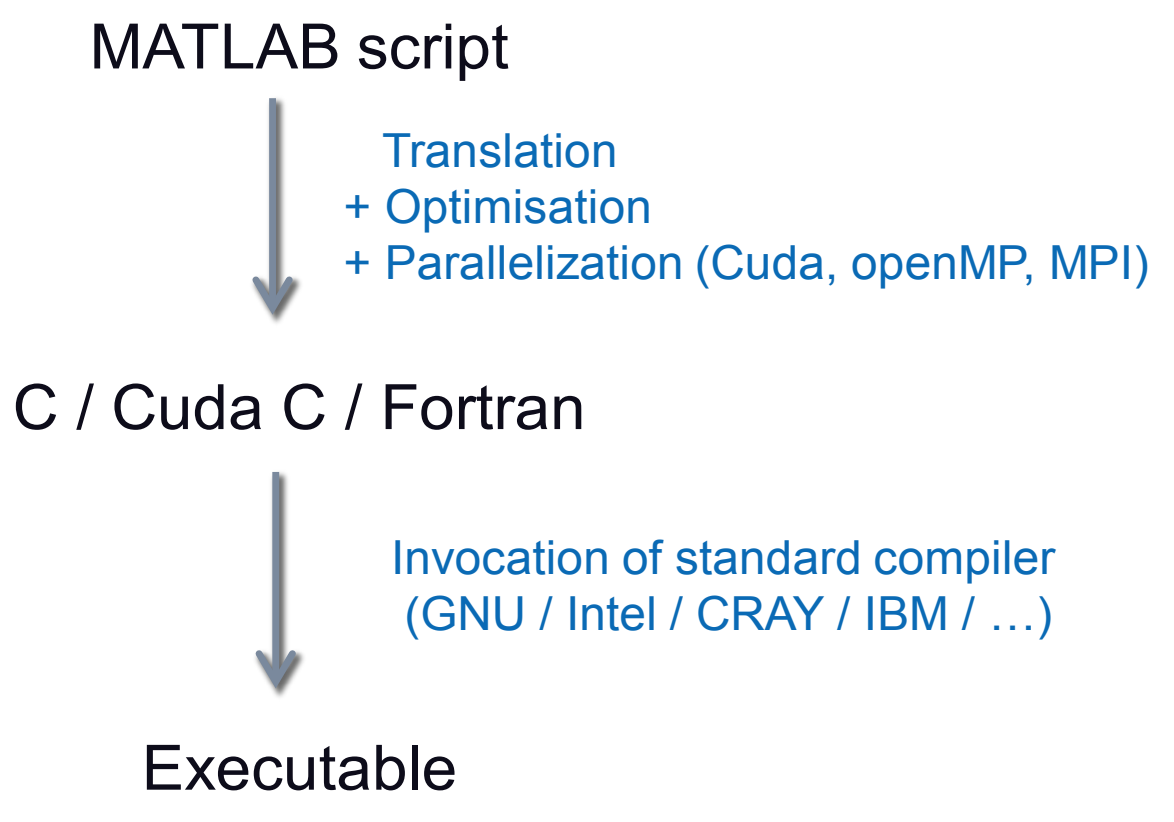

*Figure 1. Overview of the workflow of HPC$^{.m}$.*

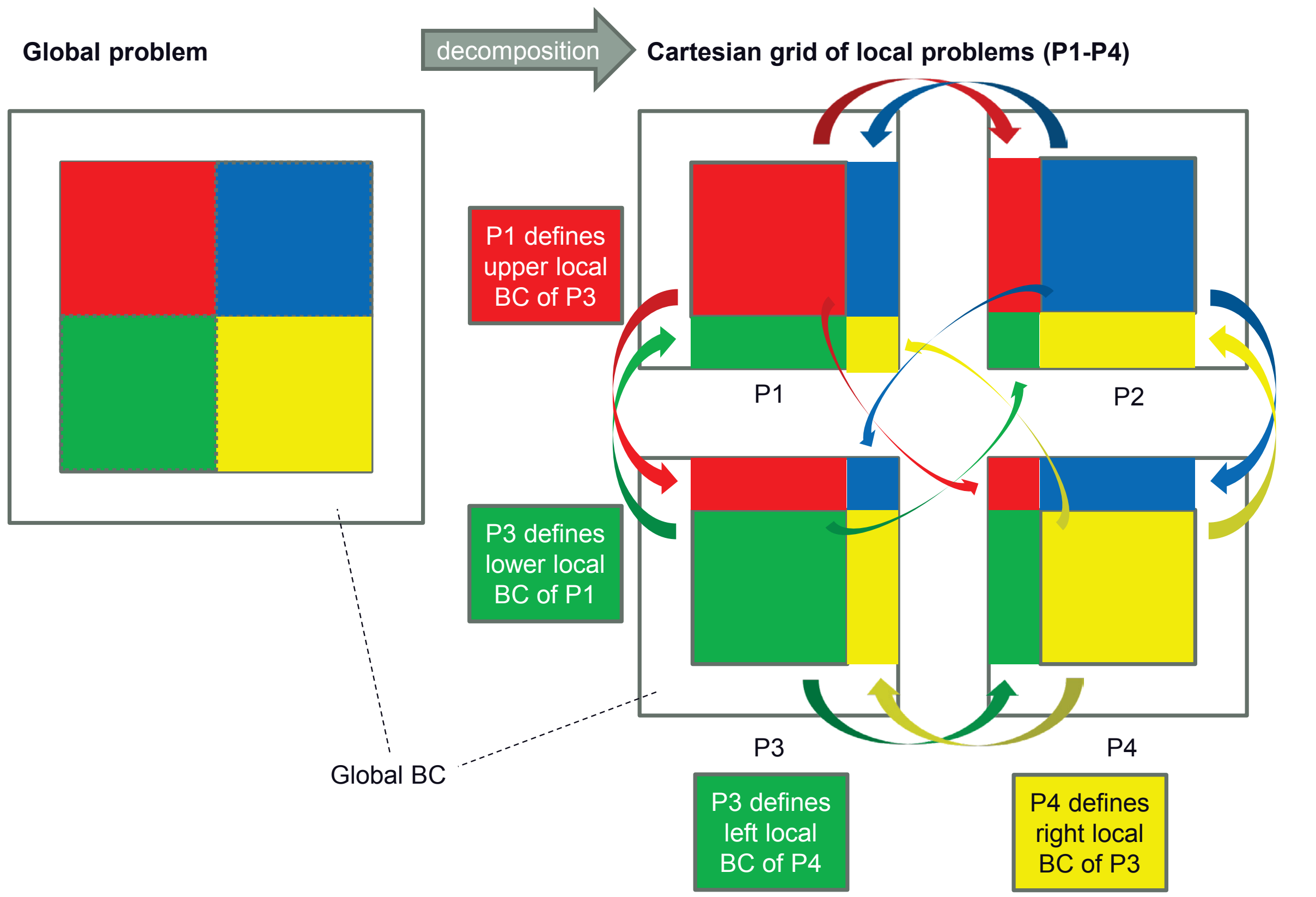

*Figure 2. Distributed memory parallelization: the global problem (2D) is spatially decomposed into a Cartesian grid of local problems, where each local problem defines for each adjacent local problem the boundary conditions (BC) of the common boundary.*

### 2.2.2 The input scripts for HPC·m

HPC·m translates scripts that are employing a small subset of the MATLAB language and some special MATLAB functions provided with the distribution of HPC·m. The scripts must in addition follow a few simple rules. The input scripts for HPC·m can be executed and debugged normally in MATLAB. The supported MATLAB language subset includes the following syntax, functions and language constructs:

- scalar and array assignments
- array ranges which may include the keyword *end* (e.g. *(:,2:end-1,:)* )
- scalar operators (e.g. *+*, *-*, ***, */*) with exception of the ^ operator (to be replaced with the function *power*).
- element-wise vector operators (e.g. *+*, *-*, *.**, *./*) with exception of the .^ operator (to be replaced with the function *power*).
- basic inbuilt compute functions (e.g. *max*, *min*, *mean*, *power*, *sqrt*, *exp*, *log*, etc.)

- the inbuilt functions *size*, *tic*, *toc* and *fprintf*
- control flow constructs (e.g. *for*, *while* and *if* constructs)

The special functions provided with the distribution of HPC.m include:

- functions for finite differences on a staggered, regular Cartesian 1D, 2D or 3D grid
- functions for file IO and for saving meta data of the application (its use for code verification is explained in section 2.2.10)
- functions for shared and distributed memory parallelization (explained in section 2.2.4 and 2.2.5)
- functions for global positioning (explained in section 2.2.5)
- functions for global reduction (e.g. to find the maximum of a field of the global problem; explained in section 2.2.5)

The rules that the scripts must respect are the following (compare with Figure 4):

1. Each script must define on the top a struct named *mandatory_params* that contains a few mandatory parameters. It must contain the fields (1) PRECIS (precision of the floating point numbers in bytes: 4 or 8), (2) *NDIMS* (number of Cartesian dimensions: 1, 2 or 3), (3-5) nx, ny and nz (number of grid cells of the local problem in the x, y and z dimension) and (6) OVERLAP (the amount of cells the local problems of size nx-ny-nz overlap; we are explaining in section 2.2.5, why and how the local problem must overlap). These parameters must then be defined via the function call "*define_params(mandatory_params);*".
2. If the code shall be translated for GPU (i.e. to Cuda C) then must next follow a struct named *gpu_params* that contains a few GPU-relevant parameters (details are given in section 2.2.4). These parameters must then be defined via the function call "*define_params(gpu_params);*".
3. Next must follow a struct named *sizes* that contains all array names and their corresponding sizes in form of tuples. The arrays must then be allocated via the function call "*allocate_sizes(sizes);*"
4. Optionally can be put calls to the functions *include_macros* and *include_functions* which are provided with HPC.m at the very beginning of the script (advanced feature). These functions take as argument the names of files containing user defined macros and functions which are written in the target language of the translation, respectively. HPC$^{.m}$ will copy them into the output code. This is an experimental feature of HPC$^{.m}$ that allows advanced users to make innovative use the software.

5. *nx*, *ny* and *nz* must never be used outside of the structs mandatory_params and sizes. To refer to the size of an array of the local problem one must use the MATLAB inbuilt function *size*. To refer to the size of an array of the global problem one can use the provided functions *nx_global*, *ny_global* and *nz_global*.
6. *ix*, *iy* and *iz* are special reserved words to which we are going to refer as *parallel indices* in the rest of the article. Parallel indices are to be used as indices for loops which can be fully parallelized, i.e. where each loop iteration is independent of the previous ones and the iterations can be executed in any order. The parallel indices must always appear in nested loops with no statement in between the beginning of each loop. More precisely, the beginning of a loop over iz must always stand immediately before the beginning of a loop over iy and a loop over iy must always stand immediately before the beginning of a loop over.

Figure 4 shows a 3D simulation of heat diffusion on a staggered grid in MATLAB that is a valid input script for HPC$^{.m}$ for the generation of a single process CPU application.  Figure 3 shows the initial setup and an advanced state of the simulation on a 64 x 64 x 64 grid.

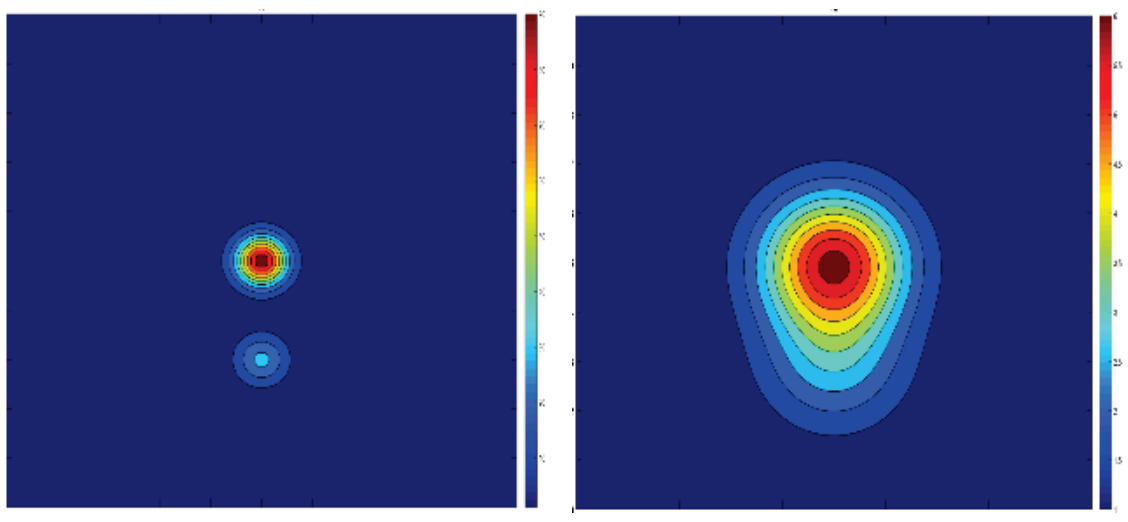

*Figure 3. Initial Setup (left) and advanced state of the 3D heat diffusion simulation (right).*

```
clear all
mandatory_params = struct('PRECIS' ,{'4'   }...                 % Mandatory parameters
                         ,'NDIMS'  ,{'3'   }...
                         ,'nx'     ,{'384' }...
                         ,'ny'     ,{'384' }...
                         ,'nz'     ,{'384' }...
                         ,'OVERLAP',{'2'   }...
                         );
define_params(mandatory_params);                                % Definition of the mandatory parameters
sizes = struct('Te'   ,{'nx  ','ny  ','nz  '}...               % Declaration of array sizes
              ,'Ci'   ,{'nx  ','ny  ','nz  '}...
              ,'dTedt',{'nx-2','ny-2','nz-2'}...
              ,'qx'   ,{'nx-1','ny-2','nz-2'}...
              ,'qy'   ,{'nx-2','ny-1','nz-2'}...
              ,'qz'   ,{'nx-2','ny-2','nz-1'}...
              );
allocate_sizes(sizes);                                          % Allocation of the arrays with the given sizes
% Physics
lam    = 1;                                                     % Thermal conductivity
c0     = 2;                                                     % Heat capacity
rho    = 1;                                                     % Density
lx     = 1;                                                     % Length of computational domain in x-dimension
ly     = 1;                                                     % Length of computational domain in y-dimension
lz     = 1;                                                     % Length of computational domain in z-dimension
te0    = 1;                                                     % Background temperature
zc1    = 0.5;                                                   % Z-location of 1st temperature anomaly
zc2    = 0.3;                                                   % Z-location of 2nd temperature anomaly
teA1   = 100;                                                   % Amplitude of 1st temperature anomaly
teA2   = teA1/3;                                                % Amplitude of 2nd temperature anomaly
rv     = 0.05*lz;                                               % Vertical size of the temperature anomaly
rh     = 0.05*lx;                                               % Horizontal size of the temperature anomaly
% Numerics
dx     = lx/(nx_global()-1);                                    % Space step in x-dimension
dy     = ly/(ny_global()-1);                                    % Space step in y-dimension
dz     = lz/(nz_global()-1);                                    % Space step in z-dimension
nt     = 1000;                                                  % Number of iterations
% Initial conditions
Ci(:) = 1/c0/rho;                                               % 1/heat capacity/density
for iz=1:size(Te,3)
  for iy=1:size(Te,2)
    for ix=1:size(Te,1)
      x_anom  = (ix-1)*dx - 0.5*lx;                             % X-distance from the Gaussian temperature anomalies
      y_anom  = (iy-1)*dy - 0.5*ly;                             % Y-distance from the Gaussian temperature anomalies
      z_anom1 = (iz-1)*dz - zc1*lz;                             % Z-distance from the 1st Gaussian temperature anomaly
      z_anom2 = (iz-1)*dz - zc2*lz;                             % Z-distance from the 2nd Gaussian temperature anomaly
      Te(ix,iy,iz) = te0          + teA1*exp(-power(x_anom/rh,2)-power(y_anom/rh,2)-power(z_anom1/rv,2));  % 1st anomaly
      Te(ix,iy,iz) = Te(ix,iy,iz) + teA2*exp(-power(x_anom/rh,2)-power(y_anom/rh,2)-power(z_anom2/rv,2));  % 2nd anomaly
    end
  end
end
time_phys=0;                                                    % Initialization of physical time
% Action
for it = 1:nt                                                   % Time loop
  start_of_parallel_iteration();                                % Declare start of parallel iteration
  dt = min(min(dx*dx,dy*dy),dz*dz)/lam/max_global(Ci)/8.1;      % Time step for 3D Heat diffusion
  qx    = -lam*d_xi(Te)/dx;                                     % Fourier's law of heat conduction
  qy    = -lam*d_yi(Te)/dy;                                     % ...
  qz    = -lam*d_zi(Te)/dz;                                     % ...
  dTedt = inn(Ci).*(-d_xa(qx)/dx - d_ya(qy)/dy - d_za(qz)/dz);  % Conservation of energy
  Te(2:end-1,2:end-1,2:end-1) = inn(Te) + dt*dTedt;             % Update of temperature
  time_phys = time_phys + dt;                                   % Update of physical time
  end_of_parallel_iteration();                                  % Declare end of parallel iteration
end
```

*Figure 4. Example of basic input script for HPC.m (for single process CPU application only): 3D simulation of heat diffusion on a staggered grid.*

### 2.2.3 Translation to a compilable language

HPC.m translates the scripts described in the section 2.2.2 to a compilable language, that is to Fortran 90 or C for CPU and MIC, and to Cuda C for GPU. This step includes syntax translation and automatic declaration and allocation of variables. The translation of basic inbuilt compute functions (e.g. max, min, mean, power, sqrt, exp, log etc.) and of the inbuilt function fprintf includes automatic type casting where required. The provided functions to compute finite differences on a staggered, regular Cartesian grid are expanded at the moment of translation much like the C pre-processor expands macros. The inbuilt function size is also expanded at the moment of translation.

The translation to Fortran 90 is overall the most straightforward task as this language allows to write computations with arrays in a vectorised form which is nearly identical as in MATLAB. The translation of vectorised computations to C is more elaborate as it requires to transform them into loop expressions. The translator must therefore analyse the ranges of the arrays which are involved in the vectorised computations. The translation of vectorised computations to Cuda C is very similar to the translation to C. If we start from a code that is already translated to C, we essentially have to do the following: the loops indices have to be replaced with Cuda thread ids and the corresponding loops must be removed.

#### *Data transfer between host and device – a particularity for the translation to Cuda C*

The translation to Cuda C shows several additional difficulties compared to the translation to C and Fortran 90. One difficulty is that it requires to handle the management of data transfer between host (CPU) and device (GPU). If we translate to Cuda 6.0 or later, this management can be delegated mostly or entirely to the Cuda runtime environment. The *unified memory* feature allows to do this: one simply has to declare all variables that must be accessible on both host and device as *managed*. Managed variables are stored on the device and are copied when required to the host. We measured however performances that are orders of magnitudes below the expected when we activated this feature (see section 3). To avoid this drastic loss in performance, the management of data transfer between host and device must be handled explicitly, i.e. the translator must generate the corresponding code. HPC.m employs the following strategy to handle this complexity in generic manner (note: it handles data transfer related to MPI communication differently):

- scalars and arrays are allocated both on the host and device

- scalars are only modified on the host (i.e. scalar assignments are always executed on the host)
- arrays are only modified on the device (i.e. array assignments are always executed on the device)
- whenever a scalar is modified, its new value is copied directly after to the corresponding variable on the device
- arrays are not copied to the corresponding variable on the host directly after they are modified, but only right before they are used in a host context

The reason why we chose to execute scalar assignments always on the host is that a scalar assignment is by nature always a job for one single thread. To execute it on the device being a totally multithreaded environment would hardly show any benefits, but would raise some important complexity (e.g. coherence of the data on all SMX processors). To copy a scalar to the device takes a negligible amount of time for a typical scientific real-world application. The advantages of the chosen strategy are therefore coherency and simplicity.

The reason why we chose to execute array assignments always on the device is that in the context of high performance programming an array assignment should always be parallelizable (at least to some parts) and therefore a job for multiple threads as they are available on the device. This choice may cause a fully serial access to an array to be executed slower than it would be on the host, yet it avoids very costly back and forth transfer of entire arrays between host and device; our strategy minimizes therefore the data transfer between host and device. Besides, applications with fully serial accesses to arrays are not our target applications.

The outlined strategy for the placement of assignment execution and the management of data transfer has in conclusion the great benefit that arrays are primarily stored on the device and are copied only in a few cases to the host (for example when doing a memory dump to the hard disk). The amount of data transfer between host and device is therefore minimal.

### 2.2.4 Automatic shared memory parallelization

HPC.m can automatically do *shared memory parallelization* when a MATLAB script is translated to a compilable language. Shared memory parallelization means to launch within a process multiple *threads* which may access the same memory while doing work concurrently. HPC$^{.m}$ does this parallelization with OpenMP [13] for Fortran 90 or C and with Cuda for Cuda C [14].

We assume that the major part of MATLAB input scripts for HPC$^{.m}$ is written in vectorised form using element-wise vector operators or with loops using the parallel indices ix, iy and iz (introduced in section 2.2.2). HPC$^{.m}$ does in consequence consider only statements written in one of these two forms for the shared memory parallelization. Statements in both forms consist exclusively of mutually independent work units that can be executed in any order and are therefore embarrassingly parallelizable (i.e. no communication is needed between the threads working on a same statement). HPC$^{.m}$ parallelizes in consequence these statements by distributing the contained work units simply in equal portions to the available threads. This holds for both Cuda and OpenMP.

To activate the parallelization with OpenMP in the generation of a C or Fortran 90 code, the HPC$^{.m}$ input script must simply call the provided function *setup_omp* in the beginning (after the definition of the mandatory parameters and sizes). The number of OpenMP threads to be launched must be given as argument.

The generation of a Cuda C code always requires that the HPC$^{.m}$ input script calls the provided function *setup_gpu* in the beginning and defines a struct named *gpu_params*. We are explaining the fields that gpu_params must contain and some related basics of Cuda programming in the following. We are listing after that all the fields that gpu_params must contain.

Like OpenMP, Cuda permits to launch threads to distribute the work among computational units of the available hardware. The Cuda threads are organized in a particular way to fit well to the target hardware i.e. to NVIDIA GPUs. The threads are organized in *blocks* and the blocks in turn form a *grid*. Both thread blocks and the grid can be 1D, 2D or 3D (Figure 5 shows 2D thread blocks in a 2D grid). Cuda provides automatically Cartesian coordinates for every thread in order to locate it within the resulting thread topology. These coordinates are then typically used to attribute to each thread a node in the computational domain. Each thread does then all the work related to its node.

One uses typically two triplets *block* and *grid*[1] to define the sizes of the blocks and of the grid. *block.x* defines then the number of threads each thread block has in x dimension; *block.y* and *block.z* do the same for the y and the z dimension. *grid.x* defines then the number of thread blocks the grid has in x dimension; *grid.y* and *grid.z* do the same for the y and the z dimension. To create the thread hierarchy shown in Figure 5 the triplets would have to take the following values: block.x=4, block.y=3, block.z=1, grid.x=3, grid.y=2, grid.z=1. The sizes of the block must not exceed a maximally allowed value which depends on the GPU architecture. The same is true for the total number of threads in a block (block.x*block.y*block.z). The maximum number of threads in a block is one of the fields of gpu_params (*MAX_NB_THREADS_PER_BLOCK*). The number of threads and their hierarchy can be different for every Cuda kernel call. HPC.m generates though the same thread hierarchy for the majority of kernels in an application (note: kernels related to MPI communication are for example handled differently). The triplets block and grid must therefore be defined only once for each application. They make up some more fields of gpu_params (BLOCK_X, BLOCK_Y, BLOCK_Z, GRID_X, GRID_Y and GRID_Z). The reason why HPC.m generates the same thread hierarchy for most kernels within an application is the following: HPC.m has been designed to translate applications that do stencil operations on a regular Cartesian grid and this kind of applications does normally not require a different thread hierarchy for each kernel.

The Cuda runtime scheduler distributes the threads (with its belonging work) in an automatic manner to the computational units on the GPU. Important to know in this concern is that a thread block is an inseparable unit for the scheduler. Only entire thread blocks are launched on the processors of the GPU (note: current NVIDIA GPU include several processors). The best load balancing between the processors is achieved if the number blocks is equal to the number of processors or a multiple of it (given that each block contains the same amount of work as it is the case in our target applications). It is therefore important to know the number of processors a GPU contains (this makes another field in gpu_params: *NB_PROCESSORS*).

[1] of type dim3

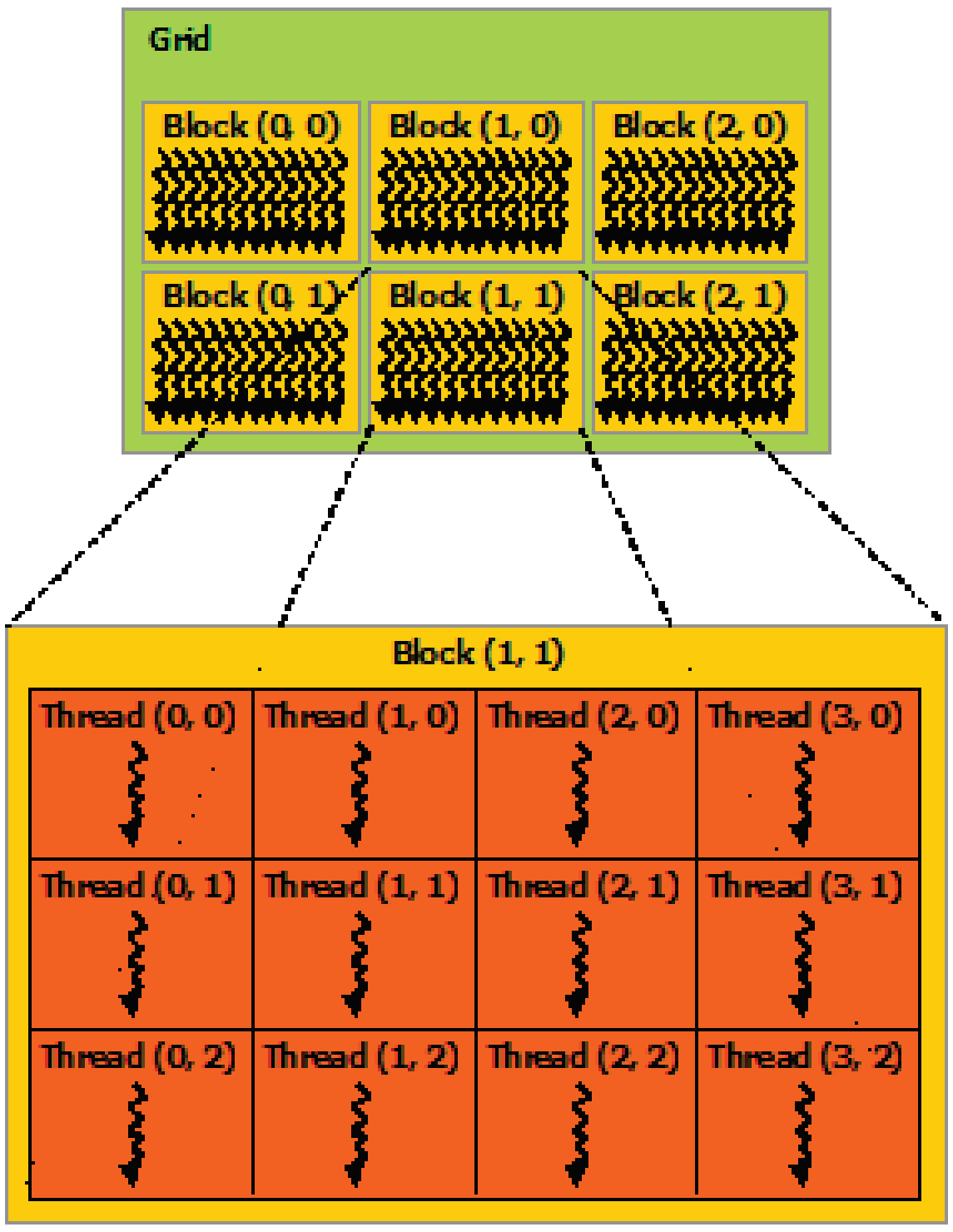

*Figure 5. 2D thread blocks in a 2D grid. Source: doc.nvidia.com – programming guide, 2.2. thread hierarchy.*

To summarize, the struct gpu_params must contain the following fields (the explanation of the last 3 fields has been omitted as it is not considered relevant for this article):

- *NB_PROCESSORS*: the number of *streaming processors* that are available on the GPU (this is a technical property of each NVIDIA GPU)
- *MAX_NB_THREADS_PER_BLOCK*: the maximum number of threads that can be executed as a single thread block (this is also a technical property of each NVIDIA GPU)
- *BLOCK_X*: the number of threads each thread block has in x dimension
- *BLOCK_Y*: the number of threads each thread block has in y dimension
- *BLOCK_Z*: the number of threads each thread block has in z dimension
- *GRID_X*: the number of thread blocks the grid has in x dimension
- *GRID_Y*: the number of thread blocks the grid has in y dimension
- *GRID_Z*: the number of thread blocks the grid has in z dimension
- *MAX_OVERLENGTH_X*: the difference between the largest array in x dimension and nx
- *MAX_OVERLENGTH_Y*: the difference between the largest array in y dimension and ny
- *MAX_OVERLENGTH_Z*: the difference between the largest array in z dimension and nz

These parameters allow HPC^.m to generate the Cuda code that launches the threads on the GPU and that defines their hierarchy. To choose the parameters that define the block and the grid dimensions in an optimal way is a complex task. It is though crucial for performance. HPC^.m helps in a quite interactive manner to choose these parameters: HPC^.m analyses the gpu parameters and the mandatory parameters and, if any of those parameters seem not to be chosen optimally, it gives some hints how to modify them to improve the hardware performance.

### 2.2.5 Automatic distributed memory parallelization

HPC^.m can automatically do *distributed memory parallelization* when a MATLAB script is translated to a compilable language. Distributed memory parallelization means to launch multiple *processes* which have each their private memory while doing work concurrently. The processes typically communicate via messages as they do not share the memory. HPC^.m does this parallelization with the Message Passing Interface (MPI) [15].

HPC^.m is able to do automatically distributed memory parallelization for problems that can be spatially decomposed into a *Cartesian grid of local problems*, where each *local problem* defines for each adjacent local problem the *boundary conditions* of the common boundary. The size of a local problem rather than the size of the *global problem* is specified in the HPC^.m input script (nx, ny and nz in mandatory_params) in order to make the domain decomposition of the global problem both the simplest possible and always optimal (see Figure 6). The size of the global problem is defined implicitly by the dimensions of the grid of local problems, the size of a local problem and the amount the local problems overlap. The amount local problems overlap must be specified in the HPC^.m input script (OVERLAP in mandatory_params). We are going to explain this parameter later in is section. The dimensions of the grid of local problems are specified via the arguments of the function *set_up_process_grid* (to be called directly after the definition of the mandatory parameters and sizes). This function sets up MPI and organizes the available processes in a *Cartesian grid topology*; a local problem is then attributed to each process (in function of the Cartesian coordinates) and the grid of local problems is therefore built. To define a grid of 2 x 2 x 1 processes and therefore a grid of 2 x 2 x 1 local problems, the function set_up_process_grid must be called with the arguments 2 (x dimension), 2 (y dimension) and 1 (z dimension), i.e. set_up_process_grid(2, 2, 1) (see Figure 6). The dimensions of the process grid are in general fixed for each executable that HPC^.m generates, i.e. if the function set_up_process_grid is called with the arguments 2, 2 and 1, then the executable will always have to be run with 4 MPI processes.

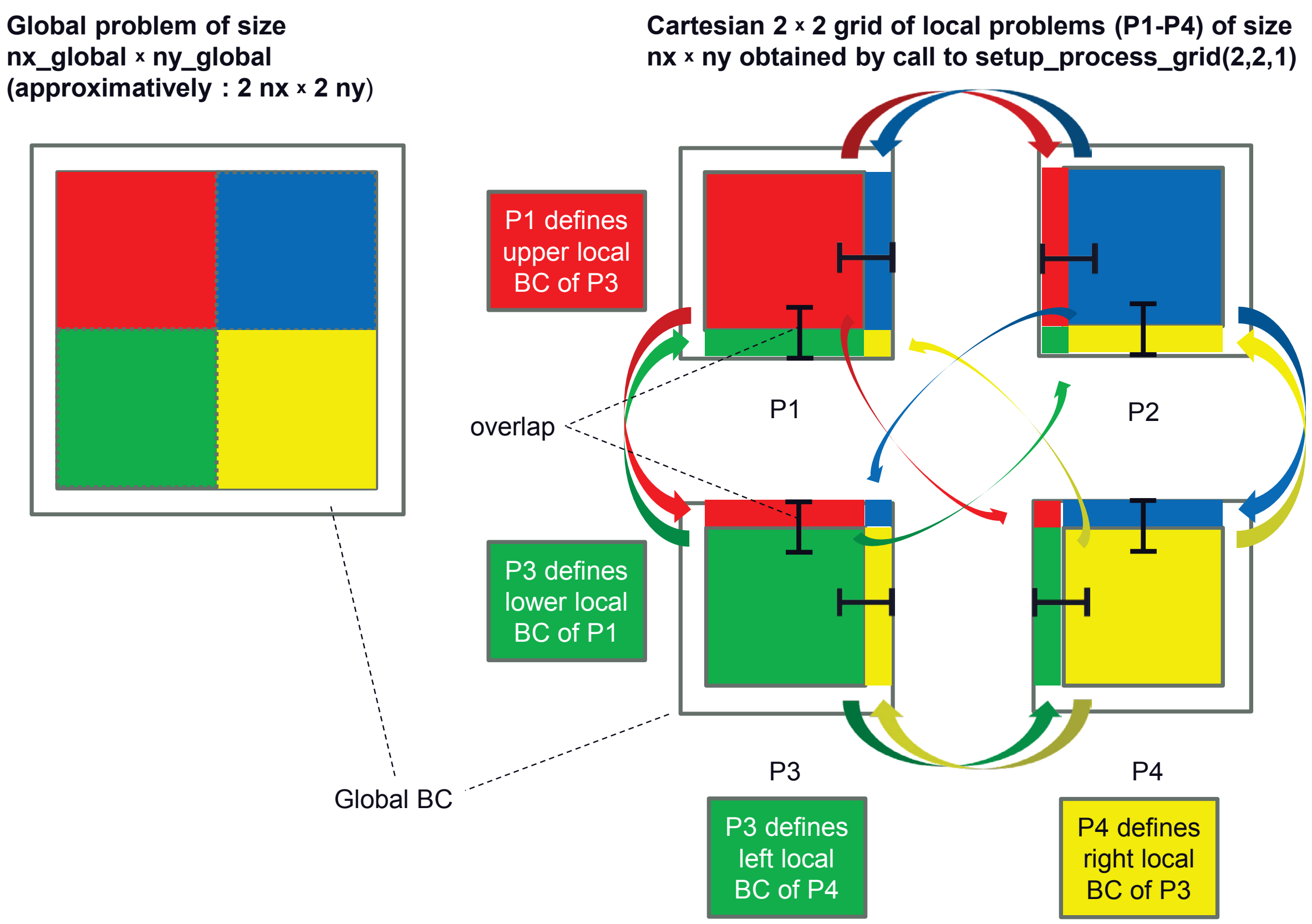

*Figure 6. Distributed memory parallelization: the global problem (2D) is spatially decomposed into a Cartesian grid of local problems, where each local problem defines for each adjacent local problem the boundary conditions (BC) of the common boundary.*

Boundary conditions of problems that are solved on a regular Cartesian grid can be defined by imposing the boundary values of some fields at each time step (or at each iteration for an iterative solver). For example for a heat diffusion problem, Dirichlet boundary conditions can be defined by imposing the boundary values of the temperature field. The boundary conditions of the global problem must be defined explicitly in the HPC.m MATLAB input script. HPC.m can generate the boundary conditions for each local problem. The idea is that each local problem defines for each adjacent local problem the boundary conditions of the common boundary. The local boundary conditions are imposed on the same physical fields as the global boundary conditions. It suffices therefore to replace every time step (or iteration) in each local problem the boundary values of these fields with the boundary values of the corresponding adjacent local problem. The user must indicate for which fields the local boundaries must be updated and when. The function *update_boundaries* serves to this purpose; the fields that have to be updated are given as arguments (see Figure 7). The fields for which the local boundaries have to be updated are the same as the fields on which the global boundaries are imposed. Whenever a global boundary condition is imposed on a field then its local boundaries must be updated before the field is again accessed. Else the grid of local problems does not anymore coherently represent the global problem. The user can though decide at what exact moment the local boundaries of a field shall be updated. This gives the possibility to the user to group most of the local boundary updates what typically leads to a more efficient MPI communication and an also in general more performant code.

HPC$^{.m}$ generates based on the users call(s) to the function update_boundaries the MPI code that allows to transfer the boundary data between the local problems. MPI permits to do the inter-process communication that is required as each local problem is solved by a different process. MPI allows to communicate between a group of processes no matter whether the processes are located on a same compute node and sharing the system memory or whether they are located on distinct compute nodes with physically distributed memory. MPI communicates between distinct nodes via the network that connects them.

The dimensions of the grid of local problems, the size of a local problem and the amount the local problems overlap define implicitly the size of the global problem. The amount the local problems need to overlap depends on the numerical methods that are employed. The required overlap is always 2 for finite differences using the smallest possible stencil.

HPC$^{.m}$ provides some functions to interrogate the size of the global problem. These functions are *nx_global*, *ny_global* and *nz_global,* which return the x, y and z dimensions of the global problem, respectively (see Figure 6). In the case of distributed memory parallelization the x dimension of the global problem is computed as *nx_global = dims(1)*(nx-OVERLAP) + OVERLAP*, where *dims(1)* is the x dimension of the grid of local problems. For the case that dims(1) is equal 1, the result is naturally *nx_global = nx*. The y and z dimension of the global problem are computed analogue to the x dimension.

HPC$^{.m}$ provides also some functions that return for an array cell in a local problem which is indexed as *(ix, iy, iz)* its Cartesian x, y or z coordinates in the global problem. These functions are *x_global*, *y_global* and *z_global*. In the case of distributed memory parallelization the x coordinate is computed as *x_global = (coords(1)*(nx-OVERLAP) + ix-1)*dx + x0*, where *x0 = (nx-nx_A)/2*dx*, *coords(1)* is the x coordinate of the MPI process which is solving the concerned local problem, *dx* is the distance between the grid points in x direction, *nx_A* is the x dimension of concerned array. The last term *x0* stands for the offset of the concerned array with respect to the grid nodes of the local problem. If nx_A is smaller than nx, then the offset is positive; if nx_A is greater than nx, then the offset is negative; and if nx_A is equal to nx, then the offset is zero. For the case that coords(1) is equal to 0, the formula for x_global simplifies to *x_global = (ix-1)*dx + x0*, which is same as for a problem that has not been decomposed into a grid of local problems. The y and z coordinates are computed analogue to the x coordinate.

Figure 7 shows a HPC.m input script to solve heat diffusion in 3D on multiple GPUs. It solves the same physics as the single process CPU application given in Figure 4 (page 71). The code parts that are different compared to single process CPU application are highlighted with a bold font.

### 2.2.6 Automatic optimizations

HPC.m can automatically do several kinds of optimizations when a MATLAB script is translated to a compilable language and parallelized. The automatic optimizations include:

- *Conversion of divisions* into multiplication of the nominator by the inverse of the denominator to avoid costly division operations;
- *Padding* of all arrays to make them equally sized and permit therefore an optimal data alignment and access;
- *On the fly computation* of derived variables where indicated by the user in order to avoid unnecessary access to the off-chip memory;
- *Fusion of loops* and of vectorised computation statements to improve the employment of the hardware's on-chip memory and reduce therefore redundant accesses to the off-chip memory;
- *Register queue* to further improve the improve the employment of the hardware's on-chip memory;
- *Minimization of required synchronization* points in shared memory parallelization to achieve more concurrency;
- *Pointer swaps* where indicated by the user in order to further permit reduction of required synchronization;
- *Loop tiling with ghost zones* to further reduce required synchronization and better fit into on-chip memory to again reduce redundant accesses to the off-chip memory.

```
clear all
mandatory_params = struct('PRECIS' ,{'4'   }...
                         ,'NDIMS'  ,{'3'   }...
                         ,'nx'     ,{'384' }...
                         ,'ny'     ,{'384' }...
                         ,'nz'     ,{'384' }...
                         ,'OVERLAP',{'2'   }...
                         );
gpu_params = struct('NB_PROCESSORS'           ,{'24'  }...  % GPU parameters
                   ,'MAX_NB_THREADS_PER_BLOCK',{'1024'}...
                   ,'BLOCK_X'                 ,{'32'  }...
                   ,'BLOCK_Y'                 ,{'8'   }...
                   ,'BLOCK_Z'                 ,{'1'   }...
                   ,'GRID_X'                  ,{'12'  }...
                   ,'GRID_Y'                  ,{'48'  }...
                   ,'GRID_Z'                  ,{'384' }...
                   ,'MAX_OVERLENGTH_X'        ,{'0'   }...
                   ,'MAX_OVERLENGTH_Y'        ,{'0'   }...
                   ,'MAX_OVERLENGTH_Z'        ,{'0'   }...
                   );
define_params(mandatory_params);
define_params(gpu_params);                                  % Definition of the GPU parameters
sizes = struct('Te'   ,{'nx  ','ny  ','nz  '}...
              ,'Ci'   ,{'nx  ','ny  ','nz  '}...
              ,'dTedt',{'nx-2','ny-2','nz-2'}...
              ,'qx'   ,{'nx-1','ny-2','nz-2'}...
              ,'qy'   ,{'nx-2','ny-1','nz-2'}...
              ,'qz'   ,{'nx-2','ny-2','nz-1'}...
              );
allocate_sizes(sizes);
set_up_gpu();                                               % Activate the use of GPU
set_up_process_grid(4,4,4);                                 % Activate distributed memory parallelization indicating...
% Physics                                                   % ...the dimensions of the grid of local problems
% (...)
% Numerics
% (...)
% Initial conditions
Ci(:) = 1/c0;
for iz=1:size(Te,3)
  for iy=1:size(Te,2)
    for ix=1:size(Te,1)
      x_anom  = x_global(ix,dx,size(Te,1)) - 0.5*lx;        % X-distance from the Gaussian temperature anomalies
      y_anom  = y_global(iy,dy,size(Te,2)) - 0.5*ly;        % Y-distance from the Gaussian temperature anomalies
      z_anom1 = z_global(iz,dz,size(Te,3)) - zc1*lz;        % Z-distance from the 1st Gaussian temperature anomaly
      z_anom2 = z_global(iz,dz,size(Te,3)) - zc2*lz;        % Z-distance from the 2nd Gaussian temperature anomaly
      Te(ix,iy,iz) = te0          + teA1*exp(-power(x_anom/rh,2)-power(y_anom/rh,2)-power(z_anom1/rv,2));  % 1st anomaly
      Te(ix,iy,iz) = Te(ix,iy,iz) + teA2*exp(-power(x_anom/rh,2)-power(y_anom/rh,2)-power(z_anom2/rv,2));  % 2nd anomaly
    end
  end
end
time_phys=0;
% Action
for it = 1:nt
    start_of_parallel_iteration();
    dt = min(min(dx*dx,dy*dy),dz*dz)/lam/max_global(Ci)/8.1;
    qx    = -lam*d_xi(Te)/dx;
    qy    = -lam*d_yi(Te)/dy;
    qz    = -lam*d_zi(Te)/dz;
    dTedt = inn(Ci).*(-d_xa(qx)/dx - d_ya(qy)/dy - d_za(qz)/dz);
    Te(2:end-1,2:end-1,2:end-1) = inn(Te) + dt*dTedt;
    update_boundaries(Te);                                   % Indicate that the local boundaries of Te must be updated
    time_phys = time_phys + dt;
    end_of_parallel_iteration();
end
```

*Figure 7. HPC.m input script to solve heat diffusion in 3D on multiple GPUs. The code lines that are different compared to the single process CPU application in Figure 4 are highlighted with a bold font. '%(…)' represents omitted code.*

Divisions of floating point numbers are very costly operations on many computer architectures. To compute the inverse of the denominator and multiply it by the nominator is normally much faster. HPC.m can do this conversion automatically if the denominator is a scalar.

The implementation of a staggered grid typically results in arrays of different sizes. Padding the arrays with zeros that are ignored for computations can make all arrays of an equal size chosen to guarantee an optimal data alignment (normally each dimension is best to be chosen as a multiple of 32). An optimal data alignment is the basis for an optimal access of the data. HPC.m can *pad* the arrays to have them all of equal size (nx+*MAX_OVERLENGTH_X, ny+MAX_OVERLENGTH_Y, nz+MAX_OVERLENGTH_Z*).

It is not necessary to ever store in off-chip memory variables that do not depend on their previous state but that are instead directly derived from other variables. This kind of variables can be (re-)computed every time that they are used, i.e. they can be computed *on-the-fly*. The on-the-fly computation of a variable is typically very advantageous if it requires only data which is available in the on-chip memory at this moment. Not a single off-chip memory access is necessary in this case. It is though more efficient to store a derived variable to off-chip memory and read it from there in later usage if its on-the-fly computation would generate more off-chip memory access. Whether the on-the-fly computation of a variable is advantageous has to be determined from case to case. HPC.m can generate code with on-the-fly computations of variables. The default code generation does not contain any on-the-fly computations of variables. If the user wants to have a variable computed on the fly then he can indicate that in the HPC.m MATLAB input script on behalf of the provided function *on_the_fly*. This function takes as argument the definition of the variable that is to be computed on the fly in form of an assignment (e.g. *on_the_fly('A = A + B')*, see Figure 8, page 85).

Loops and vectorised computation statements can be *fused* if they do not involve any dependencies. This fusion improves often the employment of the hardware's on-chip memory, i.e. more of the data that is required for the computations can be read from the on-chip memory rather than from the much slower off-chip memory. HPC.m goes linearly through the MATLAB input script and fuses subsequent loops and vectorised computation statements whenever there are no dependencies that inhibit it.

When a GPU thread reads in data sequentially from main memory it may store a small part of it in registers in order to perform some local operations as finite difference derivatives without redundant accesses to the off-chip memory. The content of the registers can be updated constantly as the sequential reading over the data advances. A very effective technique is to make the registers form a queue: register 1 in the queue receives the new data coming from main memory; register 2 receives the old value of the register 1; register 3 receives the old value of the register 2 and so on [16]. HPC.m is able to generate a register queue during the translation to Cuda C.

Shared memory parallelization with both Cuda and OpenMP requires normally *synchronization* whenever subsequent computations involve dependencies in order to assure correctness of the results. It is important to minimize the amount of synchronization in order to increase concurrency, i.e. the amount of operations that can be done in parallel. To set the required synchronization points is part of the automatic parallelization that HPC.m does. To set a synchronization point after every vectorised computation statement or loop over the parallel indices would be the simplest approach. It would however result in much unneeded synchronization. HPC.m goes linearly through the MATLAB input script and sets synchronization points only when it detects dependencies between subsequent computations. This way HPC.m minimizes the amount of synchronization.

Shared memory parallelization requires often synchronization between the reading and writing of a same variable. This synchronization requirement may be avoided by the following technique: (1) the variable A is allocated twice, for example as A and A2; (2) the variable is read from A and written to A2; (3) after writing to A2, the pointers A and A2 are swapped to have the new data again readable from A. HPC.m provides the function *swap_pointers* to this aim (For the given example the function call in the MATLAB input script would be *[A2, A] = swap_pointers(A2, A)* ).

*Loop tiling with halo* (also referred to as *loop tiling with ghost zones*) allows to reduce the amount of synchronization that is required in shared memory parallelization to the same amount as in a distributed memory parallelization. This can often be one single point of synchronization per time step (or per iteration for an iterative solver) for a real-world application. The cost of halo updates is negligible if the tile size is well chosen. This minimization of synchronization translates to a maximization of concurrency. Another equally important aspect of tiling with halo is that the tiles can be sized to fit as well as possible onto the on-chip memory of the available hardware. The aim is to read the maximum amount of data that is required for the computations from the on-chip memory rather than from the much slower off-chip memory.

HPC.m is able to do automatically loop tiling with halo for problems that can be spatially decomposed into a *Cartesian grid of local problems*, where each *local problem* defines for each adjacent local problem the *boundary conditions* of the common boundary. The obtained spatial decomposition splits the physical fields into *tiles* (also called blocks) and a halo is added where required. The tiles are stored in shared memory (tiling is a shared memory optimization). This grid of local problems can represent either the global problem or a bigger local problem which itself is part of a grid of local problems in a distributed memory parallelization. HPC.m can therefore apply loop tiling with halo and automatic distributed memory parallelization to the same kind of problems and both at a time. To activate tiling with halo the HPC.m MATLAB input script must contain a call to the provided function *set_up_tiling_with_halo* (to be called before the definition of the mandatory parameters and sizes). The dimensions of the corresponding grid of local problems are specified via the function arguments. The parameters nx, ny and nz specify the size of the local problems and therefore of the data tiles, and the parameter OVERLAP defines the amount local problems overlap (the parameters are part of mandatory_params). Boundary conditions are handled analogue to the case of distributed memory parallelization: the boundary conditions of the local problems are imposed on the same physical fields as the global boundary conditions; every time step (or iteration) and in each local problem, the boundary values of these fields are replaced with the boundary values of the corresponding adjacent local problem. The function *update_boundaries* serves like for distributed memory parallelization to indicate for which fields the boundaries of the local problems must be updated and when. HPC.m generates based on the users call(s) to this function the code that allows to transfer the boundary data between the local problems. The transfer can be achieved with simple memory copy, as the grid of local problems is located in shared memory. The provided functions *nx_global*, *ny_global*, *nz_global*, *x_global*, *y_global* and *z_global* are designed to return the correct value no whether distributed memory parallelization or loop tiling with halo or both are activated.

```
%(...)
sizes = struct('Te'  ,{'nx  ','ny  ','nz  '}...          % Declaration of array sizes
              ,'Te2' ,{'nx  ','ny  ','nz  '}...
              ,'Ci'  ,{'nx  ','ny  ','nz  '}...
              );
allocate_sizes(sizes);                                    % Allocation of the arrays with the given sizes
%(...)
% Action
for it = 1:nt
    start_of_parallel_iteration();
    if (mod(it,100)==1)                                   % Update the size of the time step every 100th iteration
        dt = min(min(dx*dx,dy*dy),dz*dz)/lam/max_global(Ci)/8.1;
    end
    on_the_fly('qx    = -lam*d_xi(Te)/dx;');              % Indicate calculation to be exectuted on-the-fly
    on_the_fly('qy    = -lam*d_yi(Te)/dy;');              % ...
    on_the_fly('qz    = -lam*d_zi(Te)/dz;');              % ...
    on_the_fly('dTedt = inn(Ci).*(- d_xa(qx)/dx - d_ya(qy)/dy - d_za(qz)/dz);'); % ...
    Te2(2:end-1,2:end-1,2:end-1) = inn(Te) + dt*dTedt;
    [Te2, Te] = swap_pointers(Te2,Te);                    % Swap pointers to permit the on-the-fly computation of dTedt
    update_boundaries(Te);
    time_phys = time_phys + dt;
    end_of_parallel_iteration();
end
```

*Figure 8. Performance optimisation of the code defining a time step in Figure 7 with calls to the functions on_the_fly and swap_pointers and with a less frequent update of the size of the time step 'dt'. The optimisation leads also to a modification of the sizes struct. The differences compared to the non-optimized code are highlighted with a bold font. '%(…)' represents omitted code.*

HPC.m permits to activate optimizations independently. This permits to determine with relatively little effort the best combination of optimizations for each particular application. Figure 8 shows how the code defining a time step in Figure 7 (page 81) can be optimized in performance with calls to the functions *on_the_fly* and *swap_pointers* and with a less frequent update of the size of the time step 'dt' (here every $100^{th}$ time step). The differences compared to the non-optimized code are highlighted with a bold font. We will give performance results for the above explained automatic optimizations in section 3.

### 2.2.7 Automatic statistics on generated code

HPC.m automatically does some statistics on the generated code. There are two types of statistics: global statistics, for the entire application, and local statistics, per loop for a C code and per kernel for a Cuda C code. Both global and local statistics include:

- Amount of reads and writes (from on- or off-chip memory)
- Amount of non-redundant reads (from off-chip memory)
- Amount of FLOPs
- FLOPs to bytes ratio (number of executed floating point operations per transferred byte)

### 2.2.8 Compiler flags

HPC$^{.m}$ is called like any normal MATLAB function. HPC$^{.m}$ takes two obligatory arguments: (1) the name of the input script and (2) the target language of the translation. The target language is specified with the arguments 'F90' for Fortran 90, 'c' for C and 'cu' for Cuda C (the abbreviations are the filename extension of the corresponding languages). HPC$^{.m}$ supports some compiler flags which are passed as optional arguments. The most important flags are (the flags that are marked with a star are only valid if Cuda C is the target language of translation):

- -divisionconvert
- -padding
- -loopfusion
- -registerqueue*
- -cudaMPI*
- -hideMPI
- -unifiedmemory*
- -groupstatistics
- -gpustyle

The first four flags activate the automatic optimizations (1) conversion of divisions, (2) padding of arrays, (3) loop fusion and (4) register queue as described in section 2.2.6. The flag -*cudaMPI* is to generate Cuda-Aware MPI code. The flag -hideMPI is to generate code that executes the MPI boundary updates simultaneously with the computations, i.e. to *hide* communication behind computation. It is though only possible in combination with loop tiling with halo (see section automatic optimizations). The flag -*unifiedmemory* is to generate GPU code that uses the unified memory feature which is available for Cuda 6.0 or newer (see in subsection "Data transfer between host and device – a particularity for the translation to Cuda C" in section 2.2.3). Code that makes use of the unified memory feature is much shorter and more readable. The overall performance of a code that HPC$^{.m}$ generates should theoretically not depend on the use of that feature, yet performance tests showed that the activation of this feature leads to very weak performance. The flag -*groupstatistics* is to print local statistics, per loop for a C code and per kernel for a Cuda C code (global statistics, for the entire application, is always printed). The flag -*gpustyle* is only valid if C is the target language of translation and makes

HPC.m translate to a C code for CPU that is very similar to a Cuda C code for GPU. This option helps for debugging and makes the understanding of the Cuda C codes for new users easier.

### 2.2.9 Output and transparency

HPC.m outputs more than just the final executable in order to provide maximum transparency. When it compiles an application, it outputs the result of the code generation, i.e. a Fortran 90, C or Cuda C code. HPC.m can in addition document the code generation: it can write the results of different code generation steps to files. These files are all executable MATLAB scripts which produce equivalent results as the HPC.m MATLAB input script. The following table indicates the steps after which HPC.m may write a file if demanded.

Step 0: HPC.m removes all comments of the input script (to simplify the processing of file), handles calls to the function *on_the_fly* and converts divisions (if the flag *-divisionconvert* is set).

Step 1: HPC.m expands the provided functions to compute finite differences on a staggered, regular Cartesian grid, much like the C pre-processor expands macros (see section 2.2.3, page 72). In addition, all arrays are padded and the array ranges adapted correspondingly if the flag *-padding* is set.

Step 2: HPC.m does loop tiling with halo if the user has activated it.

Step 3: HPC.m expands the inbuilt function size.

HPC.m does not generate multiple outputs in the target translation language in order not to overwhelm the user. The user can though easily verify what HPC.m does in each part of the code generation as he can activate independently the automatic shared and distributed memory parallelization and the automatic optimizations.

### 2.2.10 Code verification

HPC.m provides functions for file IO and saving meta data (see section 2.2.2, page 68). These functions' output can be easily read in to MATLAB and be compared visually and bitwise. It is therefore very simple to compare the results of any code that can be generated by HPC.m with the results of the HPC.m MATLAB input script or with any other code generated by HPC.m.

We have designed HPC.m to make the MATLAB input script and the corresponding C code produce *bitwise identical results.* A generated Cuda C code and a generated Fortran code may not produce bitwise identical results when comparing with the MATLAB input script. These differences have though been negligible in any conducted test (see section 3.1). Moreover, HPC.m has been programmed such that any parallelized code produces bitwise identical results with the corresponding code without parallelization[2].

HPC.m provides C-functions using the MATLAB MEX interface that can replace the built-in MATLAB functions *power, sqrt, log* and *exp*. In some cases the MATLAB input script and the corresponding C code produce only bitwise identical results if these MEX-functions are used. They can be used in MATLAB simply by calling *power_c, sqrt_c, log_c* and *exp_c* instead of power, sqrt, log and exp.

### 2.2.11 The programming approach

HPC.m is programmed in MATLAB. The main reason for this choice is that parts of the input scripts can be executed during code generation. This facilitates the programming of certain tasks and opens the possibility to generate codes that are particularly well adapted to the parameters in the input script. Another reason for the usage of MATLAB to program HPC.m is MATLAB's capacity for powerful and user-friendly text processing. MATLAB includes in particular a set of functions for regular expressions. HPC.m is as a lightweight program making use of regular expressions and some procedural text processing.

Many of the translation tasks that are part of HPC.m can be programmed with regular expressions. For some translation tasks the usage of regular expressions alone is though not sufficient. This is why we chose to do procedural text processing in addition. Regular expressions do for example not allow to find a function call's closing parenthesis if the function's arguments may contain themselves an unknown number of parenthesis. The function call's closing parenthesis can though easily be found by going character by character through the code following the function's opening parenthesis, keeping track of the number of parenthesis that remain to be closed. This is an example of the kind of procedural text processing that HPC.m contains.

---

[2] We assume obviously that HPC.m receives a correct MATLAB script as input for the code verifications.

The programming approach that we have chosen for HPC.m would probably not be the best choice for the development of a very general translator. The construction of a formal grammar for the supported subset of the MATLAB language and the usage of a parser that would use that grammar would certainly be more adequate for that purpose. The usefulness of our programming approach is therefore limited to the translation of rather small subsets of the MATLAB language.

## 2.3 Performance evaluation

Today's *hardware's' flop-to-byte ratio* is high: roughly ten to hundred floating point operations can be performed per floating point number accessed from main memory (the exact ratio depends on the number precision and the specific hardware). Today's *algorithms' flop-to-byte ratio* is typically low: roughly one to ten floating point operations are required per floating point number accessed from main memory. Nowadays applications are in consequence most often bound by the memory access speed rather than by the speed of floating point operations.

As a consequence, we use a simple memory throughput metric that is useful to quantify the difference in hardware performance between a given solver and an ideal solver that implements the same system of equations. It may be employed to evaluate and compare any applications no matter if they are memory bound or compute bound. An important property of the metric is its independence of the employed numerical algorithm and of implementation decisions; the metric takes only the implemented system of equations into account. As a result, it is *directly proportional with the performance speedup that any algorithmic or implementation improvements may bring*. We present the metric here briefly.

We define the *effective memory throughput*, measured in Gigabytes per seconds (GB/s), as

$$T_{eff} \equiv \frac{A_{eff}}{t_{it}}, \qquad (16)$$

where $t_{it}$ is the execution time per iteration in seconds and $A_{eff}$ is the *effective main memory access per iteration*, measured in Gigabytes *(GB)*. $A_{eff}$ is given by the minimally required main memory access per iteration: *unknown DOFs* (see glossary) must be updated, i.e. both read and written, and *known DOFs* (see glossary) must only be read. We have therefore

$$A_{eff} \equiv 2 * D_{unknown} + D_{known}, \qquad (17)$$

where $D_{unknown}$ and $D_{known}$ are the amount of Gigabytes that the unknown DOFs and the known DOFs occupy in main memory, respectively. The effective memory throughput is bound by the hardware's peak memory throughput ($T_{peak}$ ):

$$T_{eff} \leq T_{peak}. \tag{18}$$

The system of equations describing heat diffusion that we have implemented contains one unknown DOF ($T$) and one known DOF ($c_p$), given that we choose λ and $\rho$ to be spatially homogenous. Both DOFs contain a total amount of $N$ nodes, i.e. $N$ floating point numbers. The amount of Gigabytes that the unknown DOF and the known DOF occupy in main memory is therefore given by

$$D_{unknown} = 1\, N\, p * 10^{-9} \tag{19}$$

and

$$D_{known} = 1\, N\, p * 10^{-9}, \tag{20}$$

where $p$ is the precision in bytes. The effective main memory access per iteration (GB) is therefore given by

$$A_{eff} \equiv\ 2 * D_{unknown} + D_{known} = (2 * 1 + 1)\, N\, p * 10^{-9} \tag{21}$$

and the effective memory throughput (GB/s) by

$$T_{eff} \equiv \frac{A_{eff}}{t_{it}} = \frac{(2 * 1 + 1)\, N\, p * 10^{-9}}{t_{it}}. \tag{22}$$

Figure 9 shows how the effective memory throughput can be measured in a HPC.m compatible MATLAB script.

```
%(...)
% Action
for it = 1:nt
    if (it==3); tic(); end  % Start of the chronometer at iteration 3, because lazy library loading may occur before.
    start_of_parallel_iteration();
    %(...)
    end_of_parallel_iteration();
end
time_s = toc();
% Performance
A_eff = (2*1+1)*1/1e9*nx_global()*ny_global()*nz_global()*PRECIS;  % Effective main memory access per iteration [GB]
t_it  = time_s/(nt-2);                                             % Execution time per iteration [s]
T_eff = A_eff/t_it;                                                % Effective memory throughput [GB/s]
if (me==0); fprintf('\ntime_s=%.4f T_eff=%.4f\n',time_s,T_eff); end
```

*Figure 9. Measurement of the effective memory throughput in a HPC.m compatible MATLAB script ('%(…)' represents omitted code).*

## 2.4 Our test systems and reported performance results

Our test systems are the Octopus cluster hosted by the Institute of Earth Science at the University of Lausanne (Lausanne, Switzerland), the *Piz Daint* supercomputer at the Swiss National Center for Supercomputing (CSCS, Lugano, Switzerland)[3] and the *Lemanicus* supercomputer at the Swiss Federal Institute of Technology Lausanne (EFPL, Lausanne, Switzerland). Octopus consists of 20 compute nodes, each containing 4 NVIDIA GeForce GTX TITAN X GPUs (Maxwell GM200 architecture; 3072 CUDA cores), 2 Intel Xeon E5-2620v3 (6 cores) and 64 GB system memory. The nodes are interconnected with dual rail FDR InfiniBand. Piz Daint is a Cray XC30 and consists of 5272 compute nodes, each containing 1 Nvidia Tesla K20X (Kepler GK110 architecture; 2688 CUDA cores), 1 Intel Xeon E5-2670 (8 cores) and 32 GB system memory. The nodes are interconnected with Cray's proprietary interconnect Aries. Piz Daint is currently listed as the number 8 on the TOP500 list of the world's top supercomputers [17]. Lemanicus is a IBM Blue Gene/Q and consists of 1024 compute nodes, each containing 1 PowerA2 CPU (16 cores) and 16 GB system memory. The nodes are interconnected with a custom IBM 5D-torus interconnect. To sum up, Octopus cumulates 80 GPUs (245 760 CUDA cores), Piz Daint 5272 GPUs (14 171 136 CUDA cores) and Lemanicus 1024 CPUs (16 384 CPU cores).

All the reported performance results are the median of the obtained performance of 20 experiments, what represents in each case the typically obtained performance.

[3] Both systems, Octopus and Piz Daint, will be upgraded in close future and the specifications will therefore change.

# 3 Results

## 3.1 Modelling

We present here results of high-resolution simulations for several of the 2D and 3D multi-GPU solvers that we have generated with HPC.m. All simulations were run on 80 GPUs of Octopus. A detailed discussion of the results is beyond the scope of this article. We content ourselves with the demonstration of the feasibility of these high-resolution multi-GPU simulations thanks to HPC.m.

Figure 10 to Figure 15 show an advanced simulation state of 2D glaciar flow, 2D shallow water, 2D convection, 2D shear heating, 3D reactive porosity waves and 3D visco-elasto-plastic two-phase flow, respectively. The resolution is 20 462 x 10 354 for the 2D glaciar flow and the 2D shallow water simulations, 32 762 x 16 378 for the 2D convection simulation, 32 768 x 32 768 for the 2D shear heating simulation, 506 x 506 x 2042 for the 3D reactive porosity waves simulation and 511 x 511 x 994 for the 3D viscoelastoplastic two-phase flow simulation.

We have verified all the generated multi-GPU solvers against their MATLAB input script by comparing low resolution results after 100 time steps or iterations numerically and visually. The verifications were done in double precision as this is the native precision in MATLAB. In any case the modulus of the maximum difference was beyond 1e-10. This negligible differences show up because some mathematical operations are implemented differently in CUDA than in MATLAB[4].

[4] This does not mean that either of them is wrong.

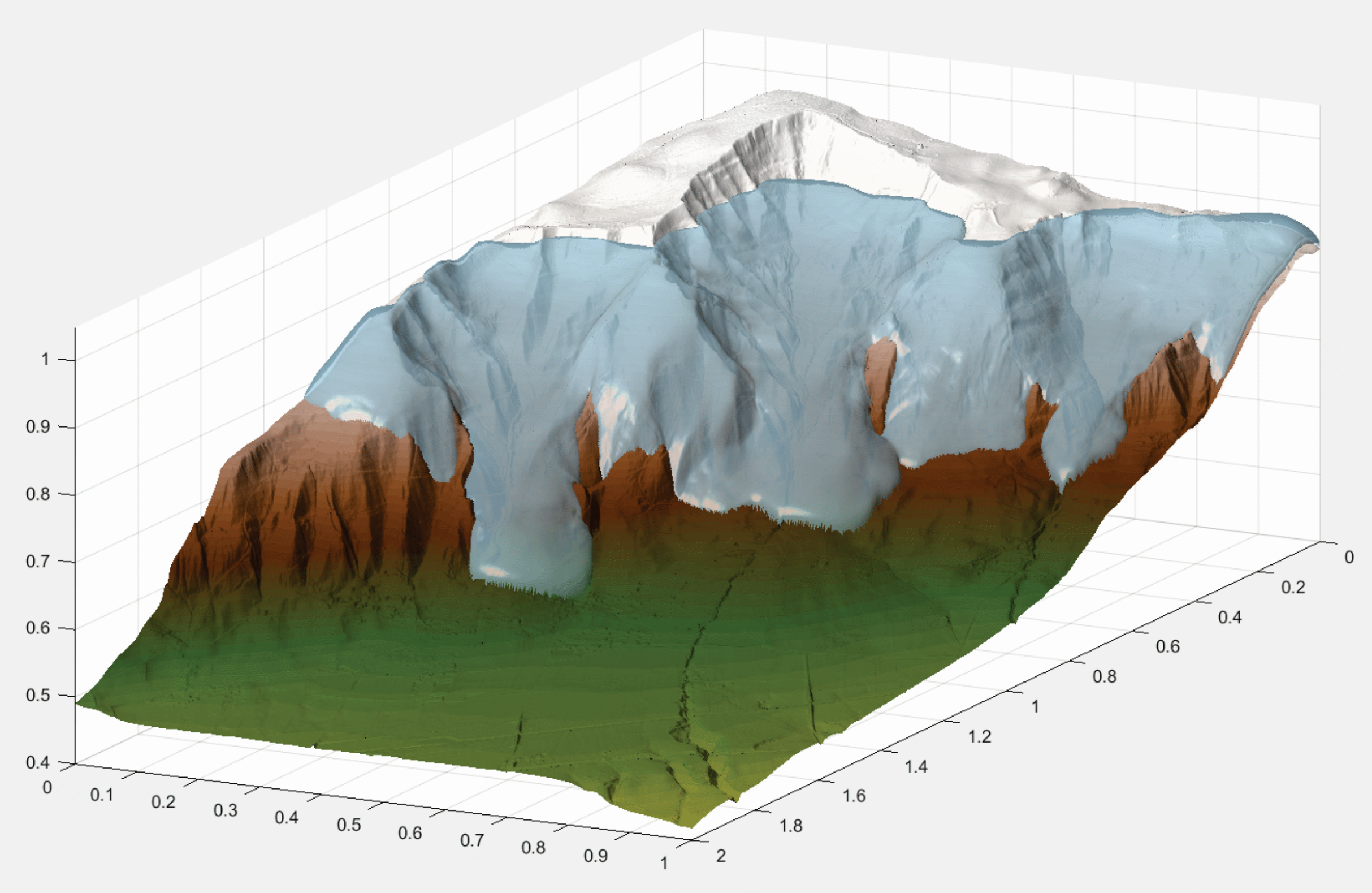

*Figure 10. Advanced state of 2D glaciar flow simulation on 80 GPUs (resolution 20 462 x 10 354). Glacier flow equations are given in Table 2. Zero glacier thicknesses were set at the model sides as the boundary conditions. A glacier with constant thickness, present only between two altitudes, was the initial condition. See the respective animation in the supplementary material.*

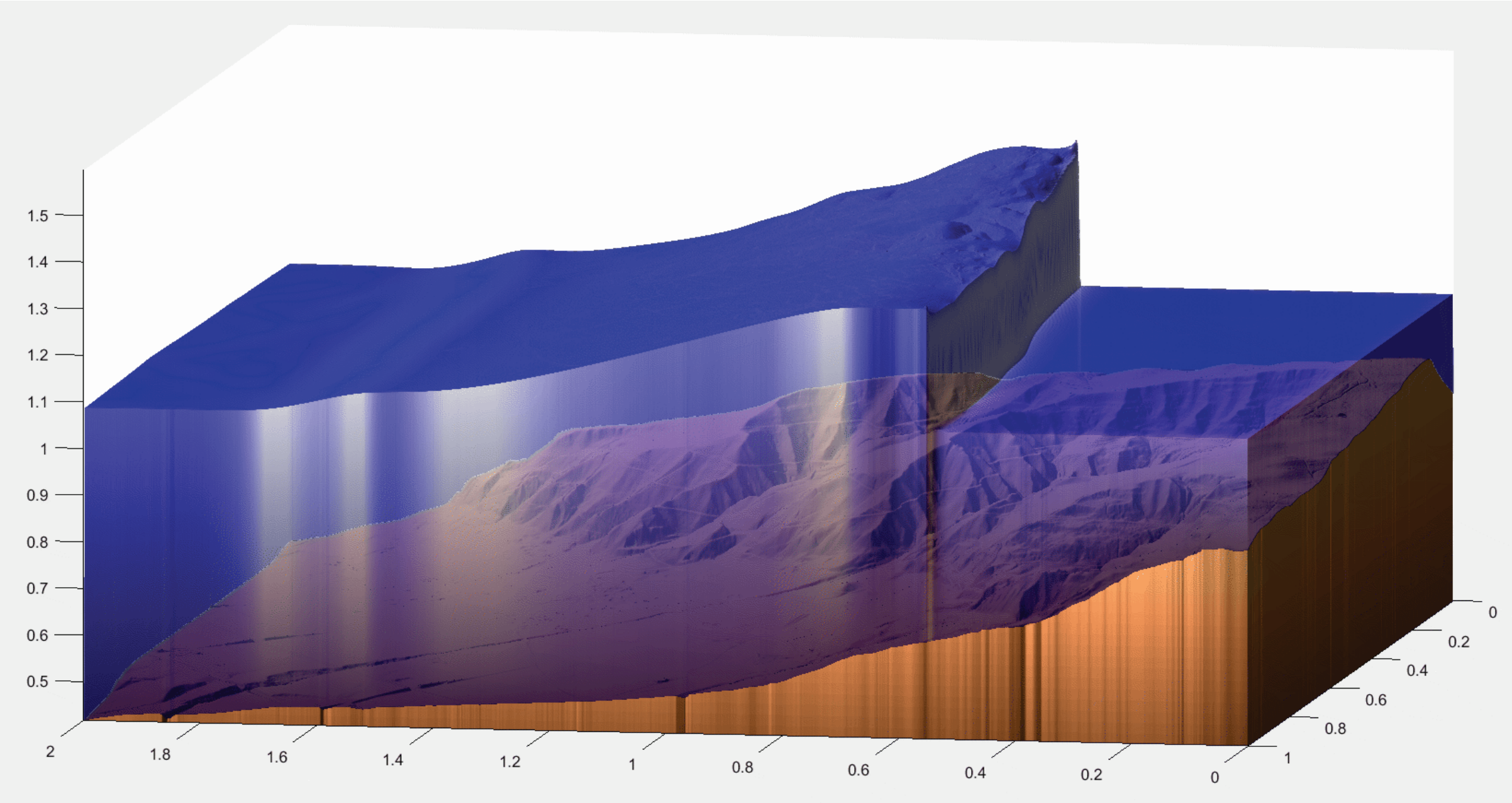

*Figure 11. Advanced state of 2D shallow water simulation on 80 GPUs (resolution 20 462 x 10 354). Shallow water equations are given in Table 2. No flux boundary conditions. Two initial Gaussian perturbations of the water surface (one large and one small perturbation) were set at the left part of the model. See the respective animation in the supplementary material.*

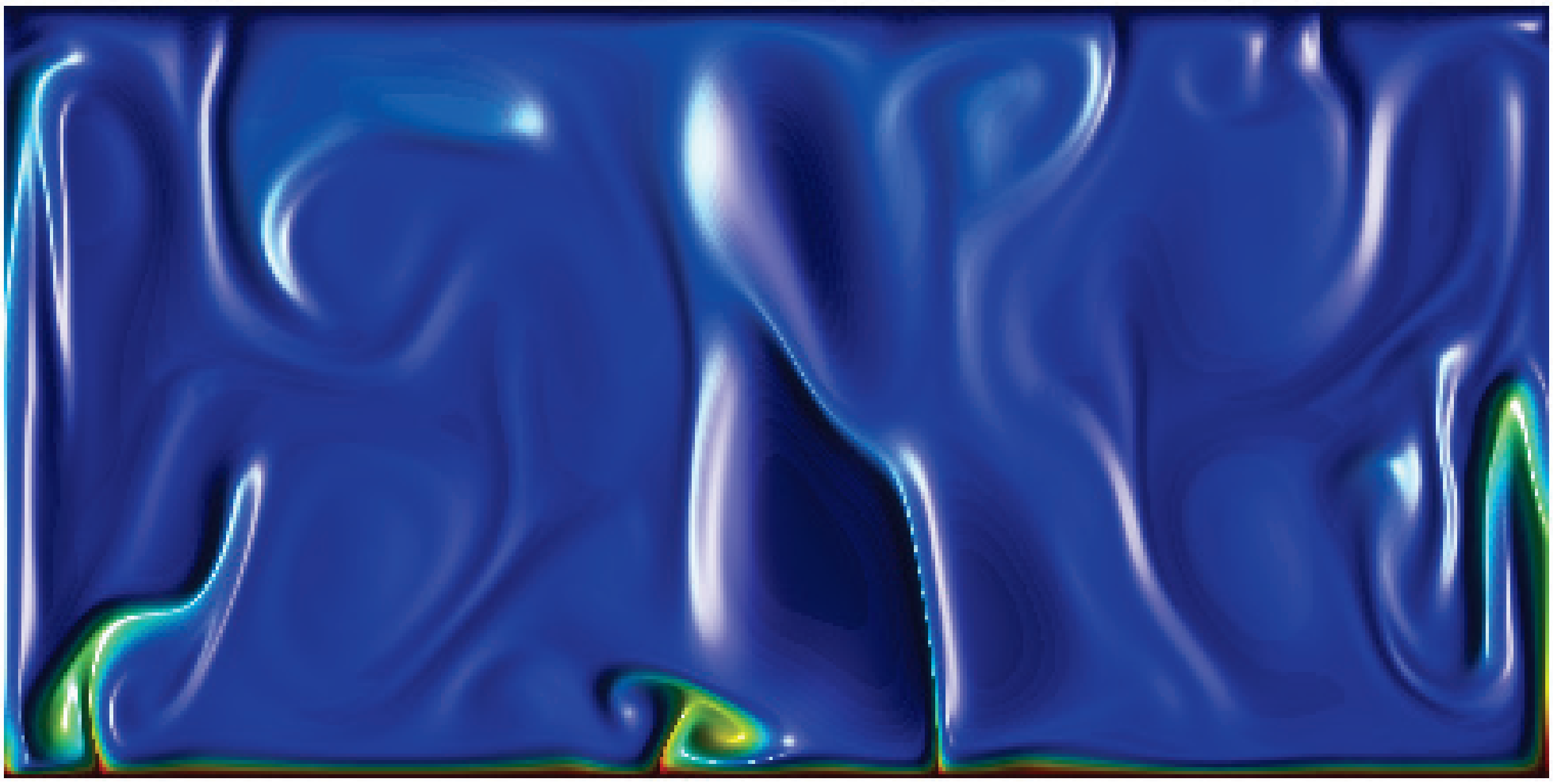

*Figure 12. Advanced state of 2D thermal convection simulation on 80 GPUs (resolution 32 762 x 16 378). TM coupling equations for convection are given in Table 4. Free slip was used as the mechanical boundary conditions, fixed temperatures were set at the top and bottom, no heat flux was implemented at the side boundaries. A Gaussian perturbation was used as initial condition to trigger convective motion (not visible anymore). See the respective animation in the supplementary material.*

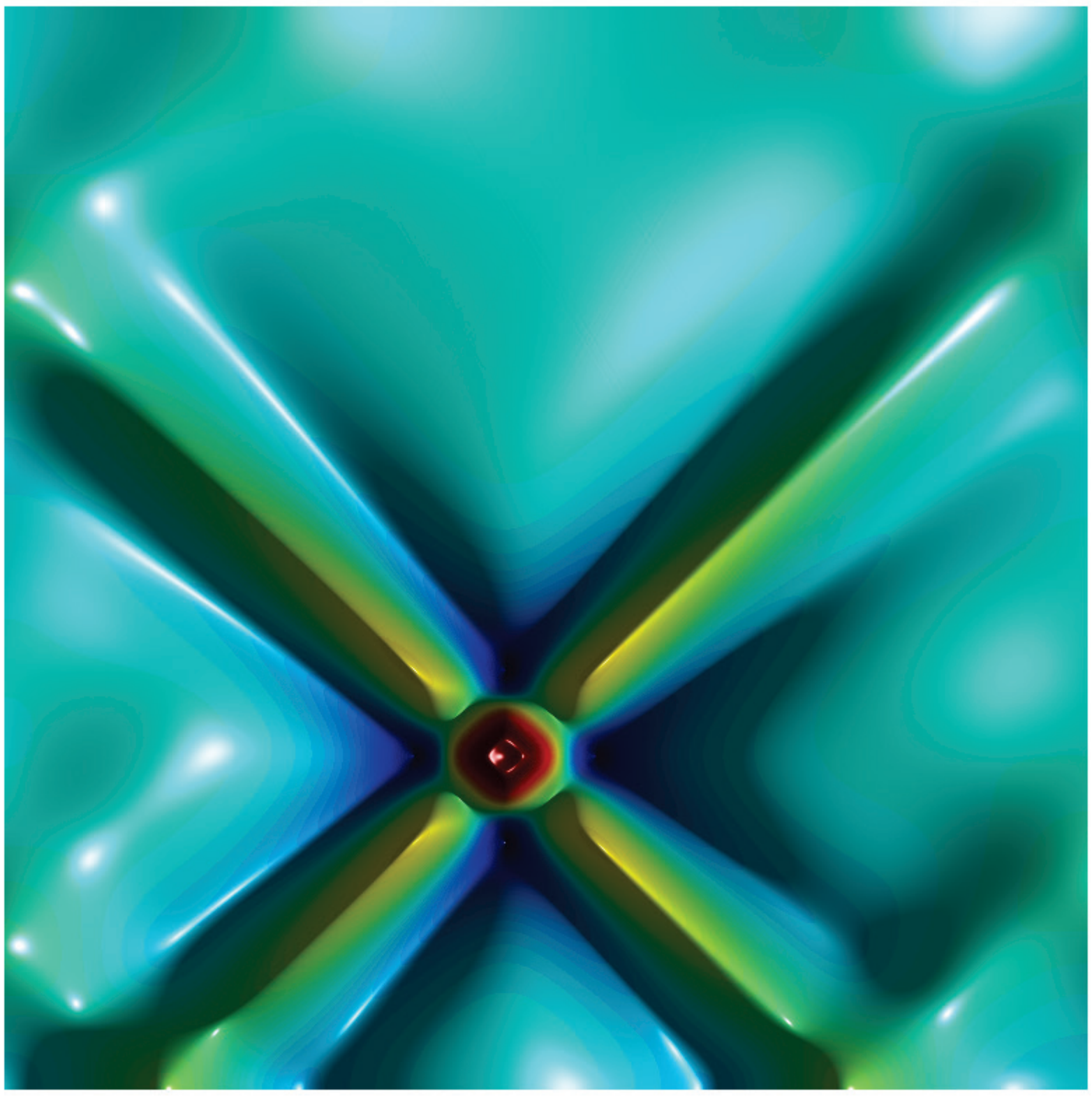

*Figure 13. Advanced state of 2D shear heating simulation on 80 GPUs (resolution 32 768 x 32 768): thermal shear bands under pure shear deformation. Shear heating equations are given in Table 5. Free slip was used as the mechanical boundary conditions and no heat flux was implemented at the side boundaries. A Gaussian perturbation was used as initial condition to trigger strain localization (visible as the small red spot slightly off the centre of the model). See the respective animation in the supplementary material.*

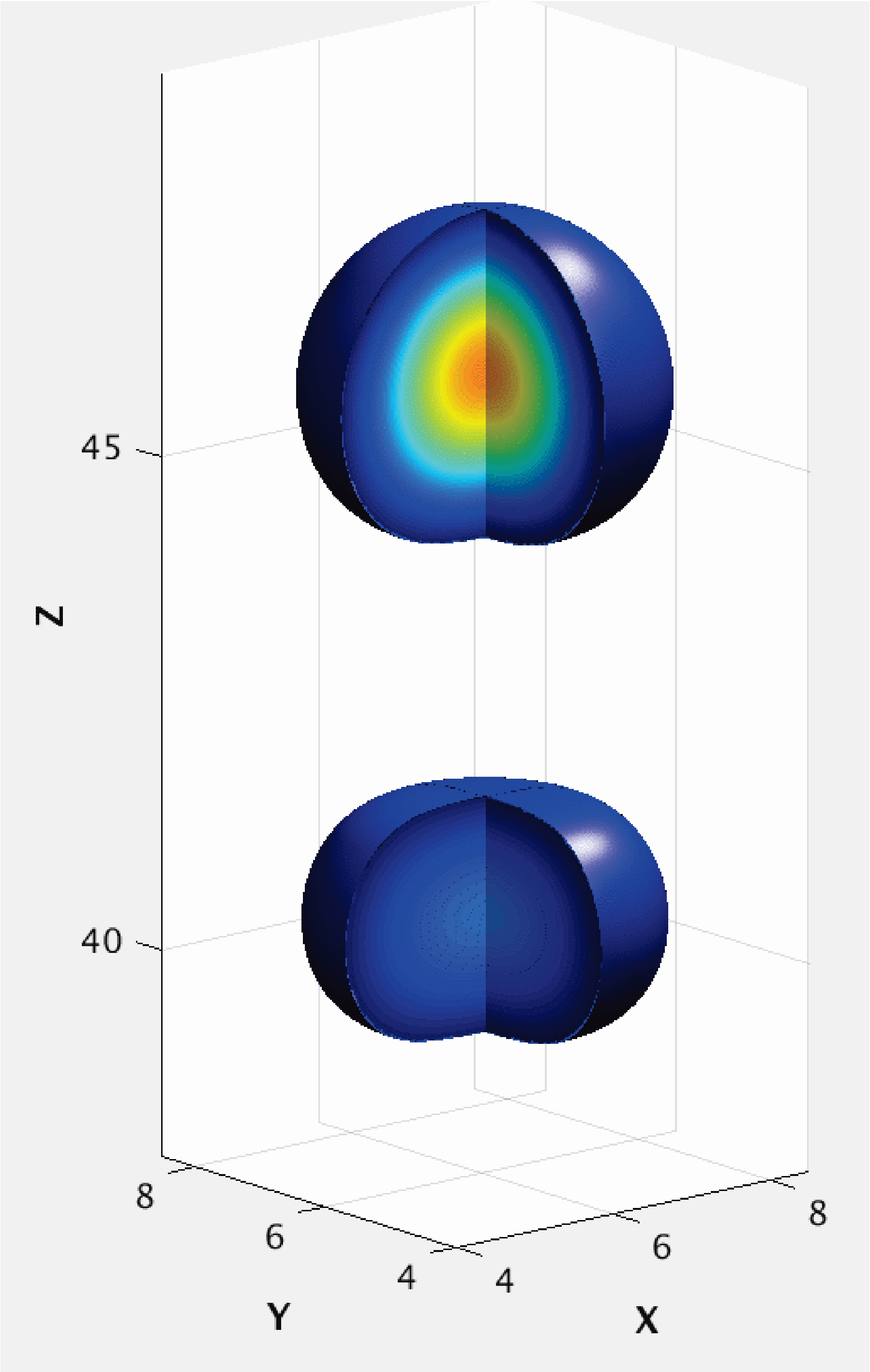

*Figure 14. Advanced state of 3D reactive porosity waves simulation on 80 GPUs (resolution 506 x 506 x 2042). HmC coupling equations are given in Table 7. Zero pressure anomaly is set at the boundaries. Equilibrium solid chemical composition, $X_S$, computed at zero fluid pressure anomaly, is set initially in the entire domain. Two Gaussian porosity anomalies were used as initial condition, arranged similar as the temperature anomalies shown in Figure 3. More details are given in the next chapter (article 3). See the respective animation in the supplementary material.*

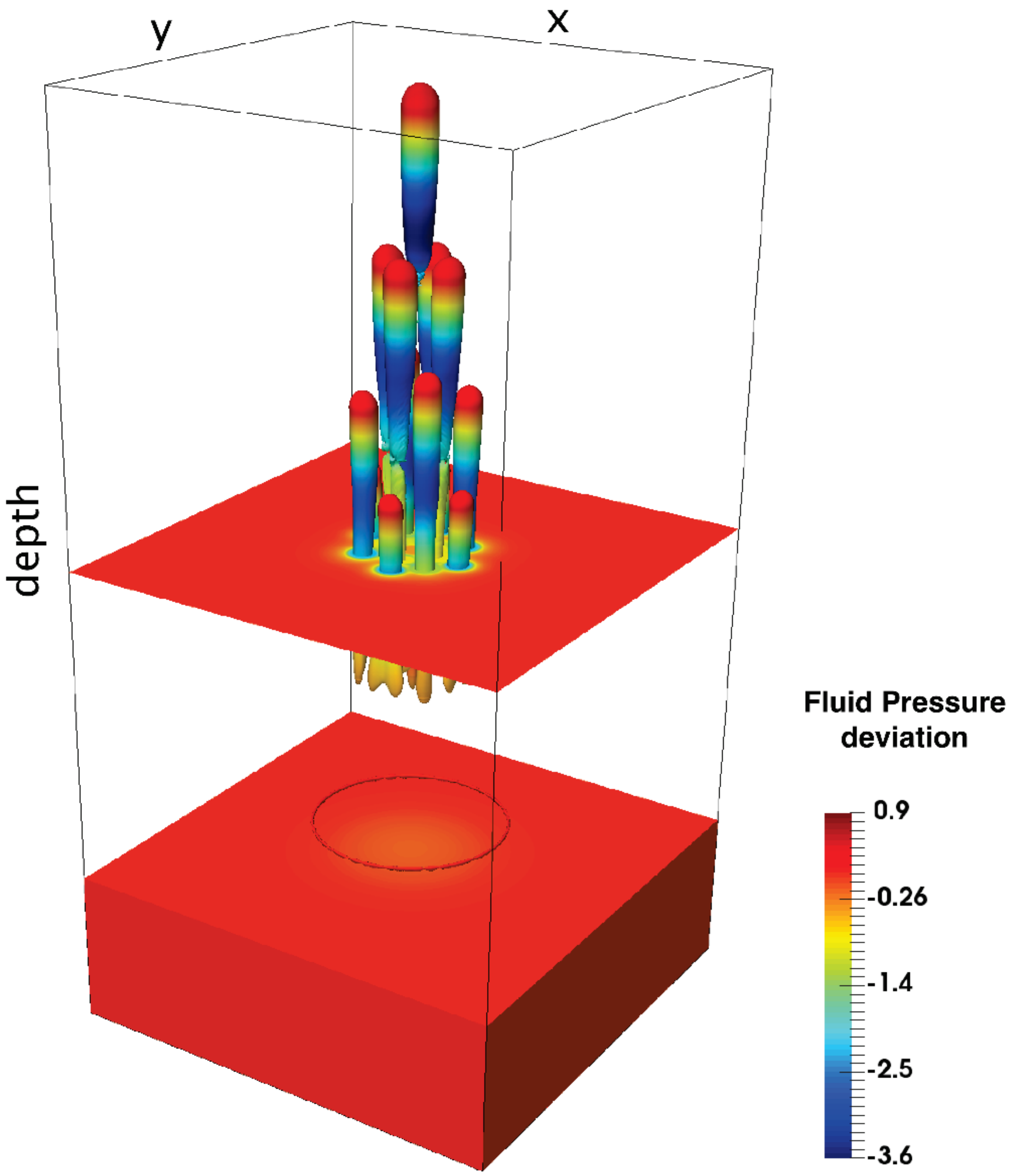

*Figure 15. Advanced state of 3D viscoelastoplastic two-phase flow simulation on 80 GPUs (resolution 511 x 511 x 994). HM coupling equations are given in Table 6. The initial setup, the boundary conditions are described in the previous chapter (article 1). See the respective animation in the supplementary material.*

## 3.2 Performance and parallel scaling

We compute for each solver the effective memory throughput $T_{eff}$ (see section 2.3, page 89). Table 8 lists the known and unknown DOFs and the specific formula of $T_{eff}$ for each solver (the symbols' meanings are listed in Table 1). Figure 16 gives the performance of the 3D Diffusion solver in MATLAB, C and Cuda C with different optimisations on one CPU/GPU of Octopus. Both $T_{eff}$ and the speedup over the fully vectorised MATLAB solver are shown. $T_{eff}$ of the GPU solver with all optimisations is very close to $T_{peak}$, i.e. very close to the fastest possible memory copy the GPU can do. The speedup over MATLAB is 463x.

Figure 17 gives $T_{eff}$ and its scaling on Octopus for all multi-GPU solvers[5] generated with HPC.m. $T_{eff}$ of all solvers is close to $T_{peak}$, i.e. all solvers run close to peak performance. All solvers scale linearly and close to ideally. Figure 18 shows the speedup of all solvers over the corresponding fully vectorised MATLAB solver on Octopus; the speedup is given in function of the number of GPUs. The measured speedup of each solver over MATLAB is approximatively 250x to 500x on one GPU and 17 000x to 35 000x on 80 GPUs.

Figure 19 shows the scaling of the nonlinear poroviscoelastic two-phase flow solver – our most complex solver – on up to 5000 Nvidia Tesla K20X GPUs on the Piz Daint supercomputer and on 16 384 CPU cores on the Lemanicus supercomputer. The scaling on Octopus is repeated in the same figure as a reference. The scaling is also on the supercomputers linear and close to the ideal scaling. On the 4913 GPUs of the Piz Daint supercomputer that we could access, the total speedup over the corresponding fully vectorised MATLAB solver is about 552 000x.

[5] We consider only single precision solvers.

| | Unknown DOFs | Known DOFs | $T_{eff}$ |
|---|---|---|---|
| Memcopy | $U$ | $K$ | $\frac{(2*1+1)\,N\,p*10^{-9}}{t_{it}}$ |
| 3D Heat diffusion | $T$ | $c_p$ | $\frac{(2*1+1)\,N\,p*10^{-9}}{t_{it}}$ |
| 2D Glaciar flow | $H$ | $B$ | $\frac{(2*1+1)\,N\,p*10^{-9}}{t_{it}}$ |
| 3D Acoustic[6] | $P$ | $\beta$ | $\frac{(2*1+1)\,N\,p*10^{-9}}{t_{it}}$ |
| 2D Shallow water | $H, HV_x, HV_y$ | $B$ | $\frac{(2*3+1)\,N\,p*10^{-9}}{t_{it}}$ |
| 3D Scalar porosity waves | $P_f, \varphi$ | | $\frac{(2*2+0)\,N\,p*10^{-9}}{t_{it}}$ |
| 3D Reactive porosity waves | $P_f, \varphi, X_s$ | $\beta_x$ | $\frac{(2*3+1)\,N\,p*10^{-9}}{t_{it}}$ |
| 2D Convection | $V_x, V_y, P, T$ | | $\frac{(2*4+0)\,N\,p*10^{-9}}{t_{it}}$ |
| 2D Shear heating | $V_x, V_y, P, T$ | | $\frac{(2*4+0)\,N\,p*10^{-9}}{t_{it}}$ |
| 3D Incompressible Cauchy-Navier elasticity | $V_x, V_y, V_z, \bar{\tau}_{xx}, \bar{\tau}_{yy}, \bar{\tau}_{zz}, \bar{\tau}_{xy}, \bar{\tau}_{xz}, \bar{\tau}_{yz}$ | $\rho, \eta, \lambda$ | $\frac{(2*9+3)\,N\,p*10^{-9}}{t_{it}}$ |
| 3D Poroviscoelastic two-phase flow | $V_x^S, V_y^S, V_z^S, \bar{\tau}_{xx}, \bar{\tau}_{yy}, \bar{\tau}_{zz}, \bar{\tau}_{xy}, \bar{\tau}_{xz}, \bar{\tau}_{yz}, \bar{P}, P_f, \varphi$ | $\mu_s$ | $\frac{(2*12+1)\,N\,p*10^{-9}}{t_{it}}$ |

*Table 8. Known and unknown DOFs and the specific formula of $T_{eff}$ for each solver generated with HPC.m.*

[6] The implemented equations use 2nd derivative in time.

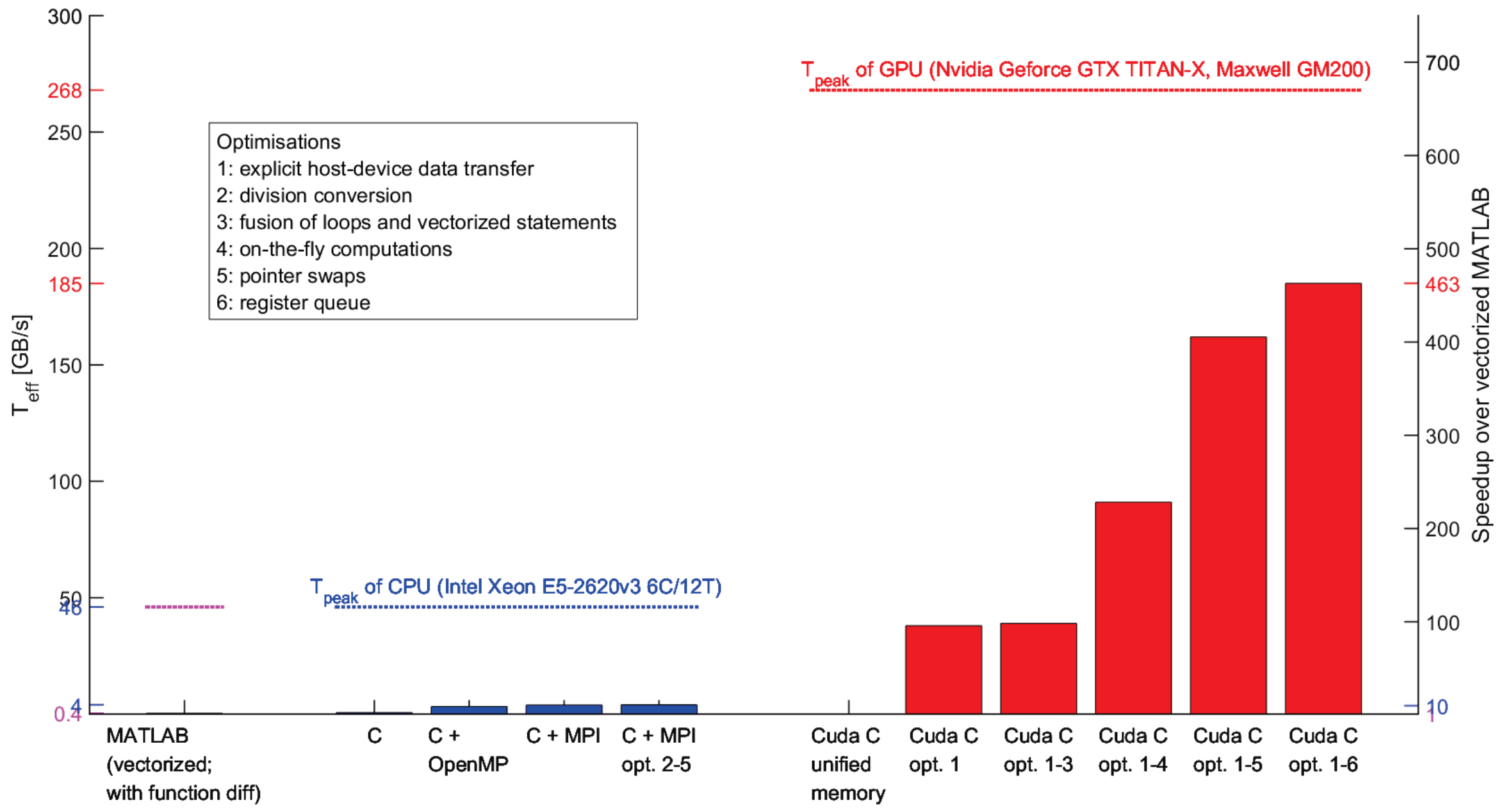

*Figure 16. Performance of the 3D Diffusion solver in MATLAB, C and Cuda C with different optimisations. The left vertical axis shows the effective memory throughput, the right vertical axis the speedup over the fully vectorised MATLAB solver.*

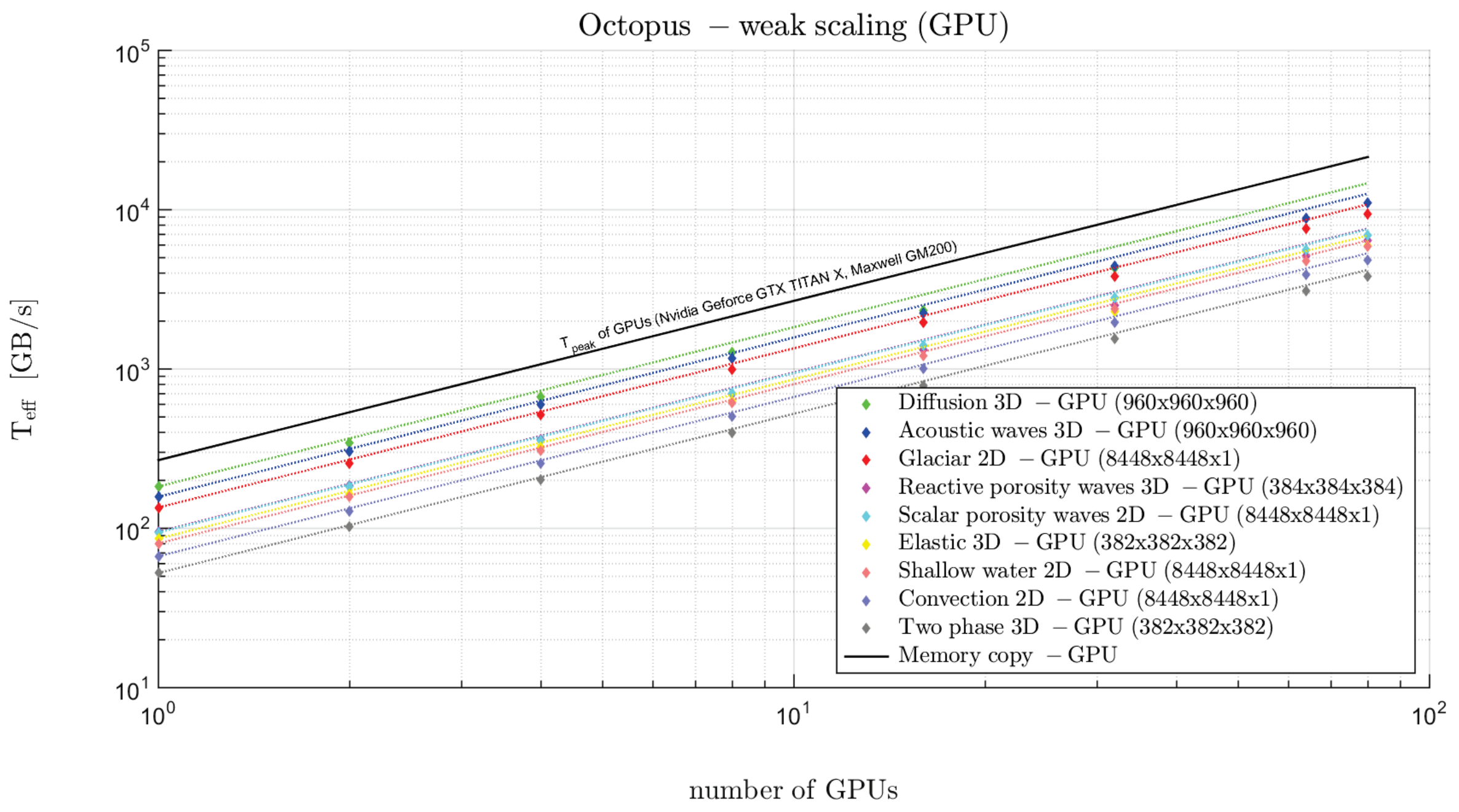

*Figure 17. Performance and parallel scaling of all multi-GPU solvers generated with HPC.m. The vertical axis shows the effective memory throughput. The pointed coloured lines give the ideal scaling of each solver. The employed resolution is given in parenthesis in the legend.*

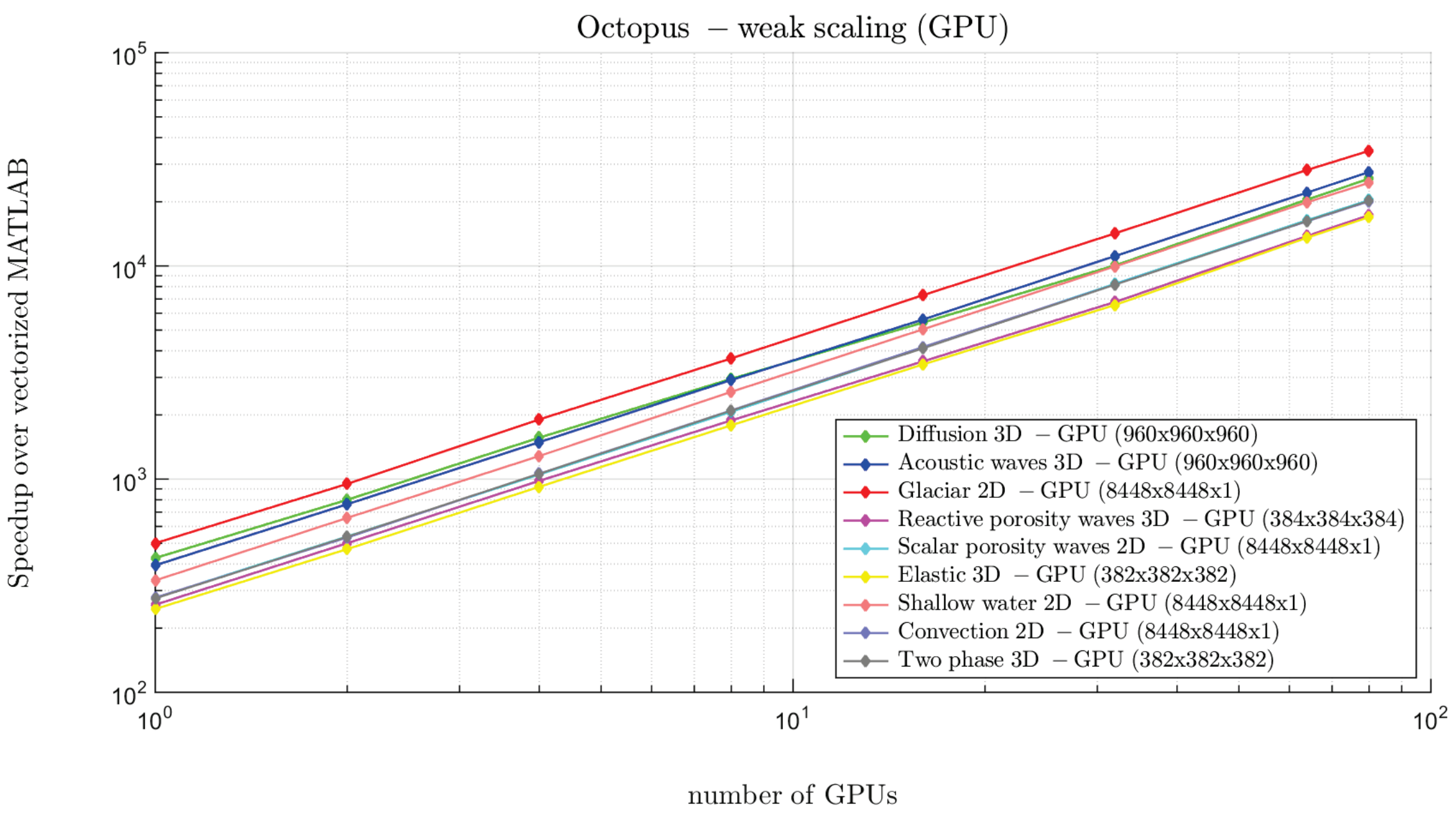

*Figure 18. Speedup of all multi-GPU solvers generated with HPC.m over the corresponding fully vectorised MATLAB solver. The speedup is given in function of the number of GPUs. The employed resolution is given in parenthesis in the legend.*

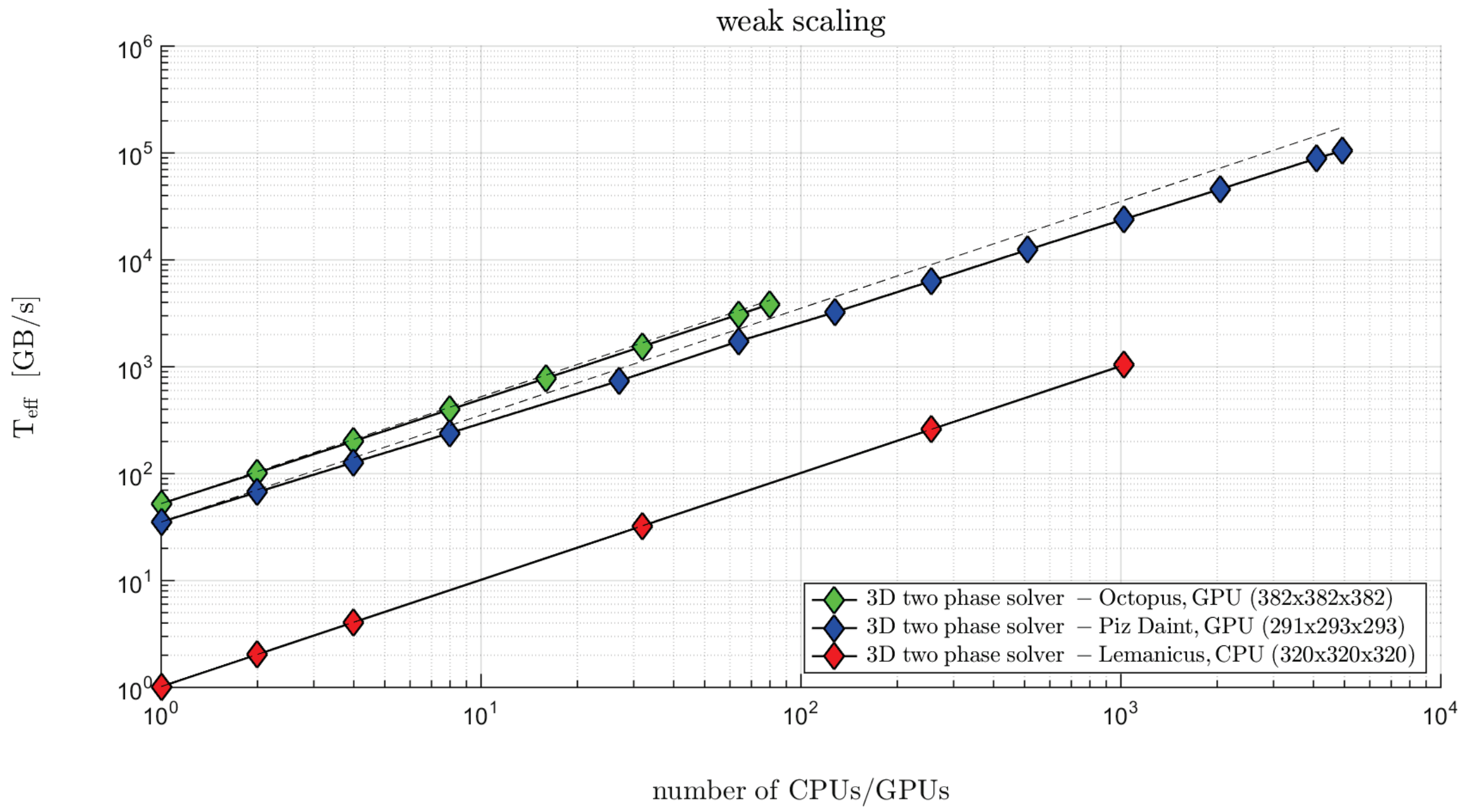

*Figure 19. Parallel scaling of the nonlinear poroviscoelastic two-phase flow solver on Piz Daint, Lemanicus and Octopus. The dashed lines give the ideal scaling of each application. The employed resolution is given in parenthesis in the legend.*

# 4 Conclusions and outlook

HPC.m reduces the time required for transforming a prototype to a near peak performance supercomputing application from typically months or years to a few seconds. It is the first source-to-source translator, to the authors knowledge, that can perform in an automatic fashion all task that are needed for the generation of a near peak performance supercomputing application from a code developed in classical prototyping environment as MATLAB. We have shown the versatile use and great performance of HPC.m by deploying it to generate 2D and 3D multi-GPU solvers for a variety of physics across multiple disciplines. We verified all generated solvers against the MATLAB input script.

All profiled applications achieve on a single GPU a 250x to 500x better performance than the fully vectorised MATLAB input script; on one workstation with 4 GPUs[7], the speedup is about 1000x to 2000x. The generated applications' performance is close to hardware's peak. All applications scale linearly on 80 GPUs of the Octopus cluster and achieve on 80 GPUs a 17 000x to 35 000x better performance than the fully vectorised MATLAB input script. Moreover, we have shown linear weak scaling for our nonlinear poroviscoelastic two-phase flow solver on the full Piz Daint supercomputer[8] at the Swiss National Supercomputing Centre (CSCS, Lugano, Switzerland), i.e. on nearly 5000 GPUs and on the full Lemanicus supercomputer at the Swiss Federal Institute of Technology Lausanne (EFPL, Lausanne, Switzerland), i.e. on up to 1024 CPUs (16 384 CPU cores). On the 4913 GPUs of the Piz Daint supercomputer that we could access, the total speedup over the corresponding vectorised MATLAB solver (executed on Octopus) is over 500 000x. We expect a similar scaling on Piz Daint for all the generated applications.

Future work may also include the automatic instrumentation of the HPC.m output codes for in-situ visualization with a software like *Visit* [18] or *ParaView* [19]. In-situ visualisation software enables the user to connect directly to a running simulation and permits to visualize intermediate or final simulation results continually, *accessing the data directly in memory*. Simulation data never needs to be written to hard disk in order to pass it to the visualization software, as it is traditionally being done.

[7] The compute nodes of Octopus are usable as workstations and contain each four GPUs.
[8] Piz Daint is currently listed as the number 8 on the TOP500 list of the world's top supercomputers [17].

## 5 Glossary

- *degree of freedom (DOF)*: parameter that qualifies either as *unknown* or *known DOF* following the below definitions
- *unknown DOF*: independent *spatially heterogeneous* parameter for which a time derivative exists in the system of equations that the solver implements
- *known DOF*: independent *spatially heterogeneous* parameter from the system of equations that the solver implements, that is known previous to a simulation and this for the entire simulation

## 6 Acknowledgments

This work was supported by grants and computational resources from the Center for Advanced Modelling Science (CADMOS) and from the Swiss National Supercomputing Centre (CSCS) under project ID #s518. The financial support for CADMOS and the Blue Gene/Q system is provided by the Canton of Geneva, Canton of Vaud, Hans Wilsdorf Foundation, Louis-Jeantet Foundation, University of Geneva, University of Lausanne, and Ecole Polytechnique Fédérale de Lausanne.

## 7 References

[1] R. Stefano, T.S. Hailu, Solve. The Exascale Effect: Benefits of Supercomputing Investment for U.S. Industry, Washington, D.C., 2014. http://www.compete.org/reports/all/2695-solve.

[2] D. Unat, X. Cai, S.B. Baden, Mint: Realizing CUDA performance in 3D Stencil Methods with Annotated C, in: Proc. Int. Conf. Supercomput. - ICS ’11, 2011: p. 214. doi:10.1145/1995896.1995932.

[3] N. Maruyama, Physis: An Implicitly Parallel Programming Model for Stencil Computations on Large-Scale GPU-Accelerated Supercomputers, in: Proc. 2011 Int. Conf. High Perform. Comput. Networking, Storage Anal. - SC ’11, ACM Press, New York, New York, USA, 2011. doi:10.1145/2063384.2063398.

[4] T. Gysi, C. Osuna, O. Fuhrer, M. Bianco, T.C. Schulthess, STELLA: A Domain-specific Tool for Structured Grid Methods in Weather and Climate Models, in: Proc. Int. Conf. High

Perform. Comput. Networking, Storage Anal. - SC '15, ACM Press, New York, New York, USA, 2015: pp. 1–12. doi:10.1145/2807591.2807627.

[5] D.A. Orchard, M. Bolingbroke, A. Mycroft, Ypnos: Declarative, Parallel Structured Grid Programming, in: Proc. 5th ACM SIGPLAN Work. Declar. Asp. Multicore Program. - DAMP '10, ACM Press, New York, New York, USA, 2010: pp. 15–24. doi:10.1145/1708046.1708053.

[6] A.E. Trefethen, V.S. Menon, C.-C. Chang, G. Czajkowski, C. Myers, L.N. Trefethen, MultiMATLAB: MATLAB on Multiple Processors, (1996). http://hdl.handle.net/1813/7242 (accessed October 24, 2016).

[7] C.B. Moler, Numerical computing with MATLAB: Revised Reprint, SIAM, 2008.

[8] S.M. Schmalholz, Y.Y. Podladchikov, Tectonic overpressure in weak crustal-scale shear zones and implications for the exhumation of high-pressure rocks, Geophys. Res. Lett. 40 (2013) 1984–1988. doi:10.1002/grl.50417.

[9] T. Duretz, S.M. Schmalholz, Y.Y. Podladchikov, Shear heating-induced strain localization across the scales, Philos. Mag. 95 (2015) 3192–3207. doi:10.1080/14786435.2015.1054327.

[10] J. Connolly, Y. Podladchikov, Compaction-driven fluid flow in viscoelastic rock, Geodin. Acta. 11 (1998) 55–84. doi:10.1016/S0985-3111(98)80006-5.

[11] L. Räss, V.M. Yarushina, N.S.C. Simon, Y.Y. Podladchikov, Chimneys, channels, pathway flow or water conducting features - an explanation from numerical modelling and implications for CO2 storage, in: T. Dixon, H. Herzog, S. Twinning (Eds.), Energy Procedia, ELSEVIER SCIENCE BV, NETHERLANDS, Austin, TX, 2014: pp. 3761–3774. doi:10.1016/j.egypro.2014.11.405.

[12] B. Malvoisin, Y.Y. Podladchikov, J.C. Vrijmoed, Coupling changes in densities and porosity to fluid pressure variations in reactive porous fluid flow: Local thermodynamic equilibrium, Geochemistry, Geophys. Geosystems. 16 (2015) 4362–4387. doi:10.1002/2015GC006019.

[13] L. Dagum, R. Menon, OpenMP: an industry standard API for shared-memory

programming, IEEE Comput. Sci. Eng. 5 (1998) 46–55. doi:10.1109/99.660313.

[14] J. Nickolls, I. Buck, M. Garland, K. Skadron, Scalable parallel programming with CUDA, Queue. 6 (2008) 40. doi:10.1145/1365490.1365500.

[15] W. Gropp, E. Lusk, A. Skjellum, Using MPI : portable parallel programming with the message-passing interface, Volume 1, MIT Press, 1999.

[16] P. Micikevicius, 3D Finite Difference Computation on GPUs using CUDA, in: GPGPU-2 Proc. 2nd Work. Gen. Purp. Process. Graph. Process. Units, ACM Press, New York, New York, USA, 2009: pp. 79–84. doi:10.1145/1513895.1513905.

[17] E. Strohmaier, J. Dongarra, H. Simon, M. Meuer, Top500 List - June 2016, TOP500.org. (2016). www.top500.org/list/2016/06/ (accessed November 7, 2016).

[18] VisItusers.org, (n.d.). www.visitusers.org (accessed November 14, 2016).

[19] U. Ayachit, A. Bauer, B. Geveci, P. O'Leary, K. Moreland, N. Fabian, J. Mauldin, ParaView Catalyst: Enabling In Situ Data Analysis and Visualization, in: Proc. First Work. Situ Infrastructures Enabling Extrem. Anal. Vis. - ISAV2015, ACM Press, New York, New York, USA, 2015: pp. 25–29. doi:10.1145/2828612.2828624.

# Porous fluids extraction by reactive solitary waves in 3-D

Samuel Omlin[1], Benjamin Malvoisin [1], and Yury Y. Podladchikov[1]

[1]Institute of Earth Sciences, University of Lausanne, Lausanne, Switzerland

## *Abstract*

Below the brittle/ductile transition, viscous compaction is known to produce solitary porosity and fluid pressure waves allowing for efficient fluid transport in the lower crust. Metamorphic reactions of (de-)volatilization can also induce porosity changes in response to the propagating fluid pressure anomalies. Here, we performed high resolution simulations with GPU parallel processing in three dimensions with a model including both viscous and reaction-induced (de-)compaction. The Damköhler number (Da) quantifies the respective roles of viscous deformation (low Da) and reaction (high Da) on wave propagation. 3-D waves were found to abandon their source region and to propagate at constant speed. During collision the waves were going through each other in the soliton-like fashion for all the investigated Da numbers. Solitary wave propagation controlled by metamorphic reactions at high Da number provides an additional mechanism for fluid extraction from the Earth's crust through the domains of undeformed rocks due to their high viscosity and/or strength above the brittle/ductile transition. It is estimated that this mechanism takes place at the meter-scale in the lower crust and at the kilometer-scale in the upper crust providing explanations for both metamorphic veins formation and fluid extraction in subduction zones.

# 1 Introduction

Equations describing porous fluid flow in a viscously deforming solid skeleton admit solutions in the form of travelling waves of fluid-filled porosity having stationary or slowly evolving porosity profiles. Higher amplitude waves travel with greater velocity leading to wave collisions. Viscous rheology results in soliton-like propagation of bigger waves through smaller waves and spectacular recovering of the waves pre-collisional shapes after highly nonlinear interaction [1–3]. In two and three dimensions, the waves have blob-like shapes and, due to their ability to detach from their source region, provide an efficient mechanism for fluid and melt extraction or sediments compaction [4–9]. Transition from viscous to elastic-plastic-brittle skeleton rheology leads to wave's coalescence after collision and makes them unable to detach from their source region. This results in decaying amplitude of the propagating waves due to geometric spreading [6]. Such loss of solitary solutions and efficient mechanism of fluid extraction can be associated with viscosity rise due to temperature decrease towards the Earth's surface [10].

Volume changes and fluid consumption/release during chemical reactions also induce porosity and fluid pressure changes [11–15]. Reaction modelling was first developed by assuming constant and small porosity to predict element transfer during metasomatism [16–20]. Then, the need for predictive tools led to the development of reactive models considering reaction kinetics and reaction-induced changes in porosity [21–23]. These models did not consider the interplays between reaction and fluid pathways generation by deformation whereas processes such as reaction-induced fracturing can strongly modify rock hydraulic properties [24–27]. The couplings between reaction and deformation play a first order role in magma ascent. Rock dissolution into magma was indeed found to favour channelization [28]. This infiltration instability mechanism led to numerous numerical and experimental studies aimed at deciphering the influence of rock composition, melting rate and geometry on the organization, stability and efficiency of fluid extraction [29–40]. Flow organization into channels is not as efficient as solitary waves for fluid extraction since waves do not detach from their source and thus need continuous fluid feeding for propagating. Fluid generation is abundant in magmatic systems where the whole solid can turn into fluid at elevated temperatures. In metamorphic systems, fluids are released in smaller amounts and in narrow P-T domains [41]. Infiltration instability mechanism may thus be insufficient for explaining the large fluid fluxes expected during slab devolatilization in subduction zones [42,43]. Solitary waves generated by viscous deformation have been proposed as an alternative process to explain metamorphic fluid extraction [44–47]. However, viscous solid compaction occurs below the brittle-ductile transition whereas metamorphic reactions are known to produce fluids above this transition [41,48]. The process of fluid extraction in the Earth's upper crust and the role of metamorphic reactions on deformation and fluid flow are thus still not well understood [49].

Here, we used GPU parallel processing to efficiently calculate the 3-D time evolution of porosity in a reactive system at high resolution. We draw a parallel between viscous deformation and kinetically controlled fluid release by reaction. Both processes are shown to generate 3-D solitary waves travelling through each other with stationary profiles in a soliton-like fashion. These simulations indicate that porosity waves is a mechanism relevant for fluid transport not only in ductile rocks but in the whole Earth's crust releasing metamorphic fluids.

## 2 Equations and Numerics

The system considered here is a porous and viscous rock which can undergo volatilization/devolatilization reactions. In this system, the volatile species (e.g. water or $CO_2$ molecules) are either chemically bound to the solid with a mass ratio of $X_s$ or constitute the fluid filling the porosity ($\phi$). The fluid moves relative to solid in a response to fluid pressure ($P_f$) gradients. In such a system, metamorphic reactions occur as a result of changes in fluid pressure and will lead to exchange of volatile species between the fluid and the solid and thus to modifications of the porosity and the solid density ($\rho_s$). Equations describing fluid flow and deformation in such a reactive system have been derived in details in [50]. Here, these equations are extended by introducing the simplest first order reaction kinetics. Only viscous solid rheology (no elasticity) is considered for simplicity reason.

The resulting equations are for effective pressure ($P_e = P_l - P_f$ with $P_l$ the lithostatic pressure defined as $\nabla P_l = (-\rho_s(1-\phi) + \rho_f\phi)g$)

$$\nabla\left(\frac{k_0}{\mu_f}\left(\frac{\phi}{\phi_0}\right)^3\left(\nabla P_e + (1-\phi)(\rho_s-\rho_f)g\vec{e}_z\right)\right) = \frac{P_e}{(1-\phi)\eta_\phi} + \frac{1-\phi}{1-X_s}\left(1+\Delta X-\frac{\rho_s}{\rho_f}\right)\frac{dX_s}{dt}, \quad (1)$$

porosity

$$\frac{1}{1-\phi}\frac{d\phi}{dt} = -\frac{P_e}{(1-\phi)\eta_\phi} - \frac{1+\Delta\mathrm{X}}{1-\mathrm{X_s}}\frac{dX_s}{dt}, \quad (2)$$

and $X_s$

$$\frac{dX_s}{dt} = \frac{X_{s_{eq}} - X_s}{\tau}, \quad (3)$$

where $\frac{d}{dt} = \frac{\partial}{\partial t} + \vec{v}_s \cdot \nabla$, $k_0$ is the permeability coefficient, $\phi_0$ is the background porosity, $\mu_f$ is the fluid viscosity, $\rho_f$ is the fluid density, $\eta_\phi$ is the bulk rock viscosity, $\Delta X$ is the change in solid density during reaction ($\Delta X = -\frac{1-X_s}{\rho_s}\frac{d\rho_s}{dX_s}$), $X_{s_{eq}}$ is the amount of fluid bound to the solid at the equilibrium and $\tau$ is the kinetic parameter defining characteristic reaction time.

Field data [51] and experiments [52] indicate that the pressure controlling kinetics of metamorphic reactions is the fluid pressure. Therefore, $X_{s_{eq}}$ was taken here as a function of the fluid pressure:

$$X_{s_{eq}} = \beta_r\left(P_f - P_{ref}\right) + X_{s_{ref}}, \quad (4)$$

with $\beta_r$ defining the change in $X_{s_{eq}}$ with fluid pressure and $P_{ref}$ and $X_{s_{ref}}$ reference pressure and $X_s$, respectively.

The reactive model presented here allows us to study the couplings between reaction, deformation and fluid flow and, in particular, the transition from viscous to reacting porosity waves. Non-reactive and equilibrium limits can be recovered from Equations (1) and (2) when $\tau$ tends to infinity and zero, respectively. This will lead, for the non-reactive case, to $\frac{dX_s}{dt} = 0$ and, for the equilibrium case, to $X_s = X_{s_{eq}}$ and thus to $\frac{dX_s}{dt} = \beta\frac{dP_f}{dt}$ which is equivalent to Equation (21) in [50].

Nondimensionalization of Equations (1), (2) and (3) is performed by choosing as scales the viscous compaction length

$$\delta = \sqrt{\frac{k_0 \eta_\phi}{\mu_f}}, \tag{5}$$

the compaction pressure

$$p^* = \delta\left(\rho_{s_0} - \rho_f\right)g, \tag{6}$$

with $\rho_{s_0}$ the reference solid density, and the characteristic time for viscous compaction

$$t^* = \frac{\eta_\phi}{p^*}. \tag{7}$$

Introducing these scales leads to the following nondimensional equations:

$$\nabla\left(\left(\frac{\phi}{\phi_0}\right)^3\left(\nabla p + (1-\phi)\frac{\rho_s - \rho_f}{\rho_{s_0} - \rho_f}\vec{e}_z\right)\right) = \frac{p}{1-\phi} + \frac{1-\phi}{1-X_s}\left(1 + \Delta X - \frac{\rho_s}{\rho_f}\right)\frac{dX_s}{dt_c}, \tag{8}$$

$$\frac{1}{1-\phi}\frac{d\phi}{dt_c} = -\frac{p}{1-\phi} - \frac{1+\Delta \mathrm{X}}{1-\mathrm{X_s}}\frac{dX_s}{dt_c}, \tag{9}$$

and

$$\frac{dX_s}{dt_c} = Da\left(X_{s_{eq}} - X_s\right), \tag{10}$$

where $p$ and $t_c$ are the nondimensional effective pressure and time, respectively, and $Da = \frac{t^*}{\tau}$ is the Damköhler number comparing the timescale of deformation with the timescale of reaction. For $Da \ll 1$, the system is controlled by viscous deformation whereas for $Da \gg 1$ it is controlled by reaction. In the Earth's crust, values for $t^*$ range from 10 000 yr in the low viscosity lower crust to over 100 Myr above the brittle/ductile transition. $\tau$ also spans over a wide range of values comprised between 0.1 yr, as shown by experiments performed in the laboratory (e.g. [26,53]), and over 100 Myr, as documented by the preservation of non-equilibrated metamorphic rocks up to the Earth's surface. Therefore, $Da$ spans in the Earth's crust over several orders of magnitude ($10^{-4}$ to $10^{16}$) and processes controlled by reaction and viscous deformation are both expected to contribute to fluid transport in the Earth's crust.

Here, we compared the impact of metamorphic reactions and viscous deformation on fluid flow by running simulations at both high ($10^2$) and low ($10^{-2}$) Damköhler numbers, respectively, while fixing the other parameters. One dimensional waves are unstable in three dimensions where they evolve towards spherical blobs [2,4,54]. Therefore, simulations were performed here in three dimensions. Porosity waves result from complex non-linear interactions generating sharp and non-static solution gradients. Simulations were in consequence performed at high resolution in time (time steps $10^{-5}t^*$ and $10^{-6}t^*$ for the low and high Damköhler numbers, respectively) and space (506x506x2042 grid points) with GPU Parallel Computing. The reactive parameters, $\beta_r$ and $\Delta X$ are fixed for all the simulations to -$2.10^{-1} \cdot P_s$ and 3, respectively. These parameters correspond to a metamorphic reaction for which $X_s$ decreases as fluid pressure increases (negative Clapeyron slope). $\beta_r$ depends on $P_s$ and corresponds to a sharp reaction taking place over several tenths of MPa for a pressure scale ($p^*$) of 0.1 GPa. $X_{s_{max}}$ is fixed to 5%. The background porosity, $\phi_0$, is fixed to 1%. The initial setup considered two vertically superposed spherical gaussian porosity anomalies with the amplitude of the deeper perturbation equal to 2.5%.

The simulations were conducted in parallel on 80 NVIDIA GeForce GTX TITAN X GPUs (Maxwell GM200 architecture) of the Octopus cluster at the University of Lausanne. The employed numerical algorithm uses a regular Cartesian three-dimensional grid on which spatial derivatives are computed with the smallest possible finite difference stencil. Pseudo-transient iterations are utilized to formulate an explicit algorithm to find the iterative solution for implicit schemes. These simple methods are highly data-local what permits the building of 3-D solvers that run close to hardware peak performance [55,56] and that scale linearly on thousands of GPUs [57]. The MATLAB HPC compiler, *HPC.m* [57], was deployed to generate a high-performance massively parallel multi-GPU solver for the above presented non-dimensional equations governing the evolution of porosity waves induced by metamorphic reactions and viscous deformation. The generated solver includes shared and distributed memory parallelization with CUDA and CUDA-Aware MPI and several performance optimizations. The most important optimisations are 1) on the fly computation of variables that are not degrees of freedom to avoid unnecessary access to the off-chip memory, 2) minimization of required synchronization points in the shared memory parallelization to obtain maximal concurrency and to permit further improvement of the on-chip memory usage (the main calculations are executed in one single CUDA kernel, i.e. no device synchronization is performed in between) and 3) register queue (a technique of explicit management of registers) to further optimize the utilization of the hardware's on-chip memory [56].

## 3 Results

3-D reactive porosity waves were successfully resolved in 3-D with the multi-GPU solver reaching a performance close to hardware limit. The scaling on the 80 GPUs of the Octopus cluster is linear and close to ideal (Figure S1; Supplementary material). Moreover, the solver achieves a speedup over the vectorised MATLAB version of about 257x on one GPU and of about 17 313x on 80 GPUs.

For the two Damköhler numbers used here, porosity distribution through time displays the same features (Figure 1 and Figure 2 and Movie M1 and Movie M2 in Supplementary materials) due to similar evolutions of fluid pressure with initial fast increase at the wave top and decrease at the wave bottom (Figure 1C and Figure 2C). This evolution of fluid pressure results from fluid fluxes conservation between the highly permeable porosity anomalies and the low porous surrounding rock. According to Equation (1), fluxes conservation is performed through an increase of the fluid pressure gradient [9,45]. As fluid pressure changes in the viscous case, the rock compacts and decompacts at the waves' bottom and top, respectively, leading to upward wave propagation (Figure 1 and Figure 2). In the reactive case, pressure changes have the same effects on porosity evolution and wave propagation. However, porosity increase is not due to rock compaction but to devolatilization ($X_s$ increase; Figure 1B and Figure 2B).

The two porosity anomalies propagate independently of their source towards the model top. The bigger anomaly moves at higher speed (Figure 3). As a result, they meet at the middle of the model without segregating but by going through each other for finally recovering their initial shape (Figure 1, Figure 2 and Figure 3). This behaviour of the waves as solitons results from direct porosity dependencies on fluid pressure in Equations (1) and (2) (Figure 1C and Figure 2C).

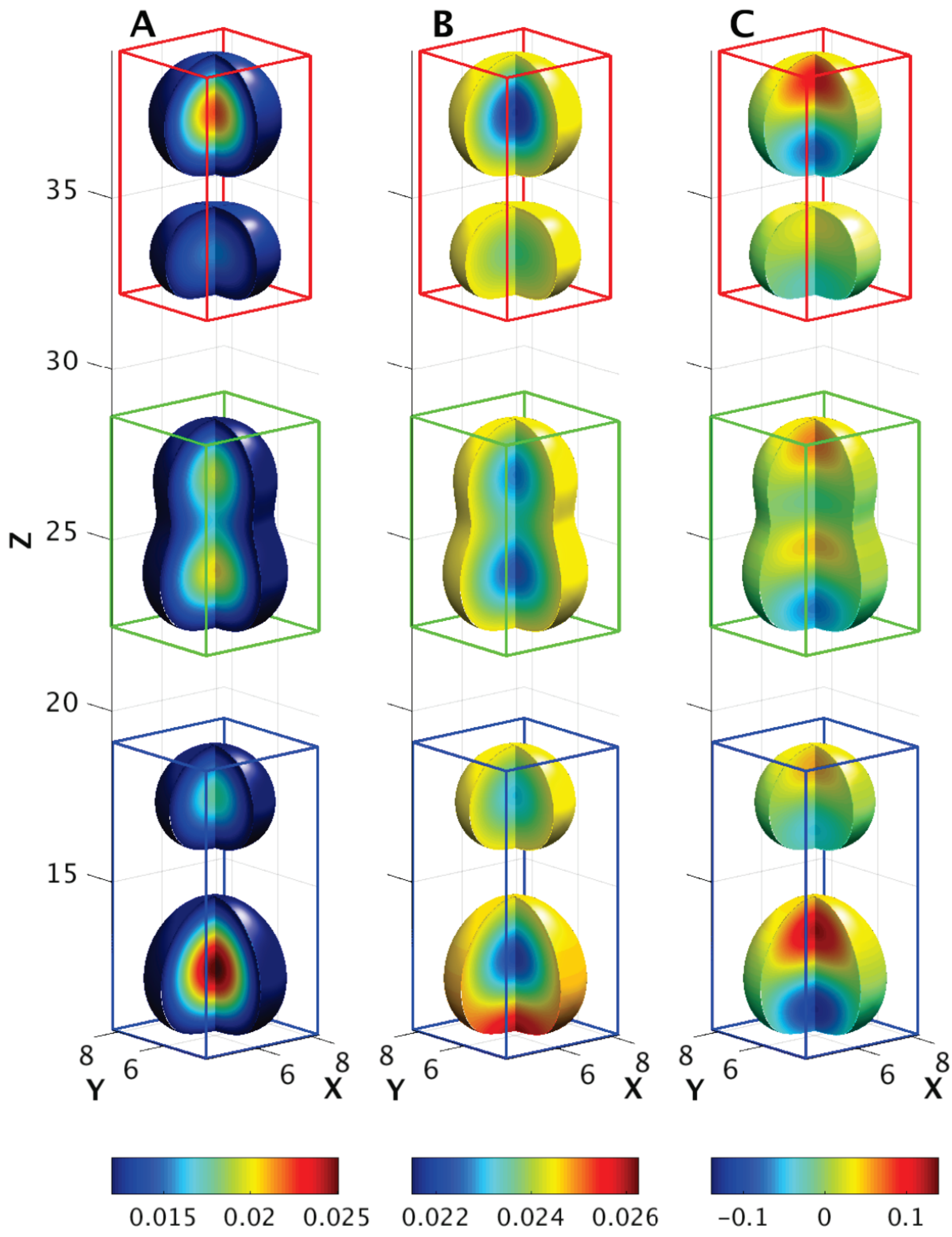

*Figure 1. Evolution in time of 3-D solitary waves when fluid flow is controlled by viscous deformation ($Da = 10^{-2}$). The variables are plotted at 0.15 $t^*$, 0.345 $t^*$ and 0.54 $t^*$ in the blue, green and red boxes, respectively. Variables are displayed on isosurface at $\phi = 0.012$ and on two perpendicular half planes vertically cutting the isosurface in its center. A. Porosity. B. Amount of fluid bound to the rock ($X_s$). C. Fluid pressure ($P_f$).*

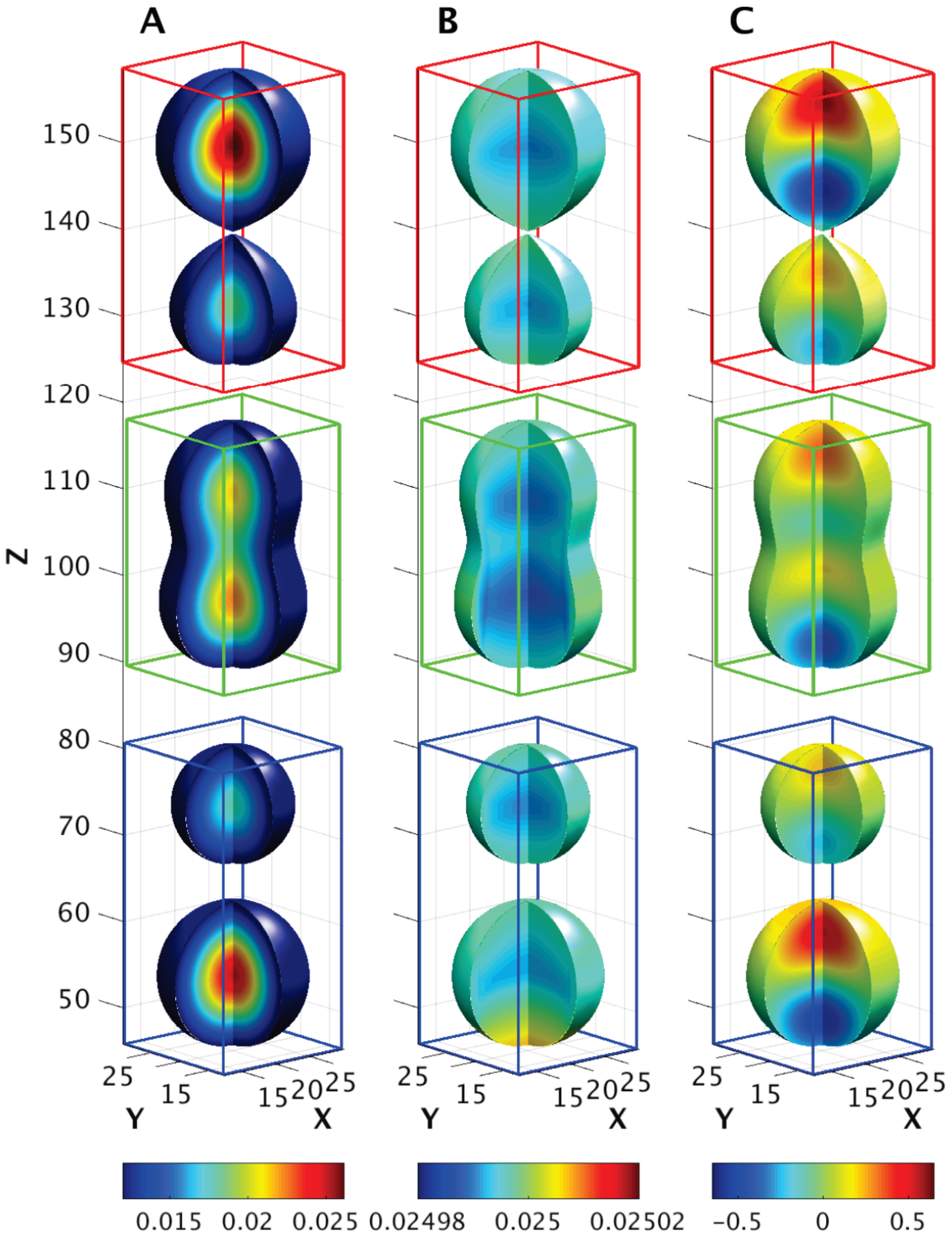

*Figure 2. Evolution in time of 3-D solitary waves when fluid flow is controlled by reaction ($Da = 10^2$). The variables are plotted at 0.01 $t^*$, 0.021 $t^*$ and 0.035 $t^*$ in the blue, green and red boxes, respectively. Variables are displayed on isosurface at $\phi = 0.012$ and on two perpendicular half planes vertically cutting the isosurface in its center. A. Porosity. B. Amount of fluid bound to the rock ($X_s$). C. Fluid pressure ($P_f$).*

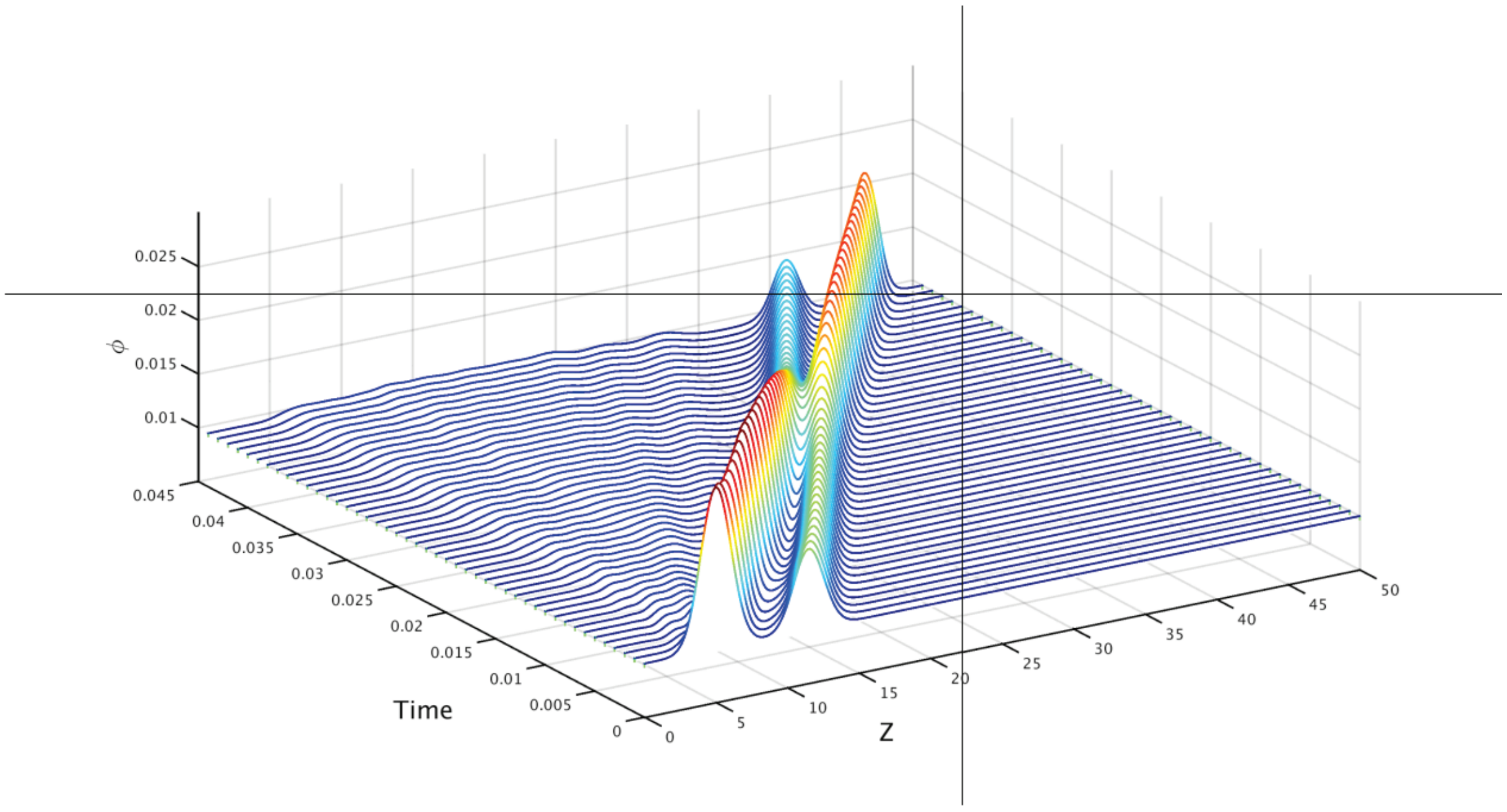

*Figure 3. Evolution of porosity in time along the z-axis and in the middle of the X and Y domains ($X = Y = 6.25$) when fluid flow is controlled by reaction ($Da = 10^2$). The simulation for $Da = 10^{-2}$ displays a similar evolution of fluid pressure with time.*

## 4 Discussion and Conclusion

Couplings between reaction and deformation can lead to reaction infiltration instability with applications to magmatic systems [28,31–37]. Flow organizes into channels allowing for efficient fluid extraction if the system is continuously fed by fluids. We show here that kinetically controlled fluid release during metamorphic reactions also generate travelling waves. These waves can detach from their source in a soliton-like fashion and thus provide a more efficient mechanism for fluid extraction than reaction infiltration instability. The formed waves are identical to the waves developing in purely viscous materials (Figure 1 and Figure 2). However the triggering mechanism is completely different since compaction is either related to viscous creep or to (de-)volatilization-induced porosity changes.

Soliton-like waves have been studied in details for viscous deformation [1,4] but were never described as resulting from reactions. Metamorphic reactions have already been included in models of porosity waves propagation [44,46,47] but the propagation was triggered by deformation rather than by the reaction itself. These latter simulations were all performed for lower crust conditions at high temperature where viscous deformation is expected to occur ($\nu_{\phi}$~$10^{18}$ to $10^{19}$ Pa.s; [10]). Even though viscous deformation occurs, it does not necessary control fluid flow since mineralogical reactions are also thermally activated [53]. As a result, fast reaction kinetics are used in the simulations of [44,46,47] ($\sim 1 < \tau < 10^4$yr) corresponding to high Damköhler numbers ($10 < Da < 10^3$) for which reacting porosity waves should form (Figure 2). Nonetheless, they were not observed due to the used low resolutions compared to the length scale of compaction induced by reaction. Indeed, simulations were performed at the hundreds of meters-scale close from the viscous compaction length ($10 < \delta < 10^3$m) whereas the reaction-induced compaction length ($\delta_r = \frac{k_0(\rho_{s_0} - \rho_f)g\tau}{\mu_f}$) is approximately 100 times smaller for the parameters used in these studies. Below the brittle/ductile transition (below approximately 15 km), reacting porosity waves with a meter-scale wavelength are thus expected to form and can contribute to fluid flow and in particular to metamorphic vein formation.

Above the brittle/ductile transition, lower temperatures induce higher viscosities ($\nu_\phi > 10^{21}$ Pa.s; [10]) and longer reaction durations ($\tau > 10^4$yr; [53]). Therefore, high Damköhler numbers can still be reached leading to reacting porosity waves. In combination, permeabilities are also expected to increase ($k_0 \sim 10^{-16}$; [42]). These parameters imply a reaction-induced compaction length ($\delta_r$) in the order of kilometers with possible implications for fluid transport at large scale in low temperature and permeable environments such as subduction zones.

Reacting porosity waves is thus a new mechanism for fluid flow in the Earth's crust. Fluid-filled porosity evolves as a result of (de-)volatilization reactions resulting from fluid pressure changes. Generated waves are solitary waves going through each other in three dimensions and allowing for efficient fluid transport. Reacting porosity waves are expected to form in the whole Earth's crust at the meter to centimeter-scale in the lower crust, and at the kilometer scale in the upper crust, providing thus an efficient mechanism for fluid extraction in the Earth's crust.

## 5 Acknowledgments

This work was supported by grants and computational resources from the Center for Advanced Modelling Science (CADMOS) and from the Swiss National Supercomputing Centre (CSCS) under project ID #s518. The financial support for CADMOS and the Blue Gene/Q system is provided by the Canton of Geneva, Canton of Vaud, Hans Wilsdorf Foundation, Louis-Jeantet

Foundation, University of Geneva, University of Lausanne, and Ecole Polytechnique Fédérale de Lausanne.

# 6 References

[1] O. V Vasilyev, Y.Y. Podladchikov, D. a. Yuen, Modeling of compaction driven flow in poro-viscoelastic medium using adaptive wavelet collocation method, Geophys. Res. Lett. 25 (1998) 3239–3242. doi:10.1029/98GL52358.

[2] D.R. Scott, D.J. Stevenson, Magma ascent by porous flow, J. Geophys. Res. B. 91 (1986) 9283. doi:10.1029/JB091iB09p09283.

[3] D.R. Scott, D.J. Stevenson, J.A. Whitehead, Observations of solitary waves in a viscously deformable pipe, Nature. 319 (1986) 759–761. doi:10.1038/319759a0.

[4] C. Wiggins, M. Spiegelman, Magma migration and magmatic solitary waves in 3-D, Geophys. Res. Lett. 22 (1995) 1289–1292. doi:10.1029/95GL00269.

[5] D.R. Scott, D.J. Stevenson, Magma solitons, Geophys. Res. Lett. 11 (1984) 1161–1164. doi:10.1029/GL011i011p01161.

[6] J.A.D. Connolly, Y.Y. Podladchikov, Compaction-driven fluid flow in viscoelastic rock, Geodin. Acta. 11 (1998) 55–84. doi:10.1016/S0985-3111(98)80006-5.

[7] M.S. Appold, J.A. Nunn, Numerical models of petroleum migration via buoyancy-driven porosity waves in viscously deformable sediments, Geofluids. 2 (2002) 233–247. doi:10.1046/j.1468-8123.2002.00040.x.

[8] B. Chauveau, E. Kaminski, Porous compaction in transient creep regime and implications for melt , petroleum , and CO 2 circulation, J. Geophys. Res. Solid Earth. 113 (2008) 1–15. doi:10.1029/2007JB005088.

[9] J.A.D. Connolly, Y.Y. Podladchikov, A Hydromechanical Model for Lower Crustal Fluid Flow, Springer-Verlag Heidelberg, 2012. doi:10.1007/978-3-642-28394-914.

[10] R. Bürgmann, G. Dresen, Rheology of the Lower Crust and Upper Mantle: Evidence from Rock Mechanics, Geodesy, and Field Observations, Annu. Rev. Earth Planet. Sci. 36 (2008) 531–567. doi:10.1146/annurev.earth.36.031207.124326.

[11] C. V Putnis, K. Tsukamoto, Y. Nishimura, Direct observations of pseudomorphism: Compositional and textural evolution at a fluid-solid interface, Am. Mineral. 90 (2005) 1909–1912. doi:10.2138/am.2005.1990.

[12] A. Putnis, Mineral replacement reactions: from macroscopic observations to microscopic mechanisms, Mineral. Mag. 66 (2002) 689–708. doi:10.1180/0026461026650056.

[13] A. Putnis, Mineral Replacement Reactions, Rev. Mineral. Geochemistry. 70 (2009) 87–124. doi:10.2138/rmg.2009.70.3.

[14] A. Putnis, H. Austrheim, Fluid-induced processes: metasomatism and metamorphism, Geofluids. 10 (2010) 254–269. doi:10.1111/j.1468-8123.2010.00285.x.

[15] C. Raufaste, B. Jamtveit, T. John, P. Meakin, D.K. Dysthe, The mechanism of porosity formation during solvent-mediated phase transformations, Proc. R. Soc. A Math. Phys. Eng. Sci. 467 (2010) 22. doi:10.1098/rspa.2010.0469.

[16] D.S. Korzhinskii, Theory of metasomatic zoning., Clarendon Press, Oxford., 1970.

[17] J.B. Thompson, Geochemical reaction and open systems, Geochim. Cosmochim. Acta. 34 (1970) 529–551. doi:10.1016/0016-7037(70)90015-3.

[18] A. Hofmann, Chromatographic theory of infiltration metasomatism and its application to feldspars, Am. J. Sci. 272 (1972) 69–90. doi:10.2475/ajs.272.1.69.

[19] J.D. Frantz, H.K. Mao, Bimetasomatism resulting from intergranular diffusion; I, A theoretical model for monomineralic reaction zone sequences, Am. J. Sci. 276 (1976) 817–840. doi:10.2475/ajs.276.7.817.

[20] B. Guy, Mathematical revision of Korzhinskii theory of infiltration metasomatic zoning, Eur. J. Mineral. 5 (1993) 317–339. http://apps.webofknowledge.com.ezproxy.lib.utexas.edu/full{_}record.do?product=WOS{&}search{_}mode=OneClickSearch{&}qid=17{&}SID=3BfD7dfd1@B5ILmoN8p{&}page=18{&}doc=175.

[21] P.C. Lichtner, Q. Kang, Upscaling pore-scale reactive transport equations using a

multiscale continuum formulation, Water Resour. Res. 43 (2007). doi:10.1029/2006WR005664.

[22] P.C. Lichtner, J.W. Carey, Incorporating solid solutions in reactive transport equations using a kinetic discrete-composition approach, Geochim. Cosmochim. Acta. 70 (2006) 1356–1378. doi:10.1016/j.gca.2005.11.028.

[23] T. Xu, E. Sonnenthal, N. Spycher, K. Pruess, TOUGHREACT - A simulation program for non-isothermal multiphase reactive geochemical transport in variably saturated geologic media: Applications to geothermal injectivity and CO2 geological sequestration, Comput. Geosci. 32 (2006) 145–165. doi:10.1016/j.cageo.2005.06.014.

[24] R.C. Fletcher, E. Merino, Mineral growth in rocks: Kinetic-rheological models of replacement, vein formation, and syntectonic crystallization, Geochim. Cosmochim. Acta. 65 (2001) 3733–3748. doi:10.1016/S0016-7037(01)00726-8.

[25] B. Jamtveit, C. V Putnis, A. Malthe-Sørenssen, Reaction induced fracturing during replacement processes, Contrib. to Mineral. Petrol. 157 (2009) 127–133. doi:10.1007/s00410-008-0324-y.

[26] B. Malvoisin, F. Brunet, J. Carlut, S. Rouméjon, M. Cannat, Serpentinization of oceanic peridotites: 2. Kinetics and processes of San Carlos olivine hydrothermal alteration, J. Geophys. Res. 117 (2012) B04102. doi:10.1029/2011JB008842.

[27] O. Plumper, A. Royne, A. Magraso, B. Jamtveit, The interface-scale mechanism of reaction-induced fracturing during serpentinization, Geology. 40 (2012) 1103–1106. doi:10.1130/G33390.1.

[28] M. Spiegelman, P.B. Kelemen, and consequences of flow organization during melt transport : The reaction infiltration instability in compactible media at the Weizmann Science , problem to viscously deformable media , appropriate to the mental Sciences and through a series of theoretic, J. Geophys. Res. 106 (2001) 2061–2077.

[29] G. Daccord, R. Lenormand, O. Liétard, Chemical dissolution of a porous medium by a reactive fluid-I. Model for the “wormholing” phenomenon, Chem. Eng. Sci. 48 (1993) 169–178. doi:10.1016/0009-2509(93)80293-Y.

[30] I.J. Hewitt, Modelling melting rates in upwelling mantle, Earth Planet. Sci. Lett. 300 (2010) 264–274. doi:10.1016/j.epsl.2010.10.010.

[31] M.A. Hesse, A.R. Schiemenz, Y. Liang, E.M. Parmentier, Compaction-dissolution waves in an upwelling mantle column, Geophys. J. Int. 187 (2011) 1057–1075. doi:10.1111/j.1365-246X.2011.05177.x.

[32] T. Keller, R.F. Katz, The role of volatiles in reactive melt transport in the asthenosphere, J. Petrol. 57 (2016) 1073–1108. doi:10.1093/petrology/egw030.

[33] J.S. Jordan, M.A. Hesse, Reactive transport in a partially molten system with binary solid solution, Geochemistry, Geophys. Geosystems. 16 (2015) 4153–4177. doi:10.1002/2015GC005956.

[34] Y. Liang, A. Schiemenz, M.A. Hesse, E.M. Parmentier, Waves, channels, and the preservation of chemical heterogeneities during melt migration in the mantle, Geophys. Res. Lett. 38 (2011) 1–5. doi:10.1029/2011GL049034.

[35] A. Schiemenz, Y. Liang, E.M. Parmentier, A high-order numerical study of reactive dissolution in an upwelling heterogeneous mantle-I. Channelization, channel lithology and channel geometry, Geophys. J. Int. 186 (2011) 641–664. doi:10.1111/j.1365-246X.2011.05065.x.

[36] J.F. Rudge, D. Bercovici, M. Spiegelman, Disequilibrium melting of a two phase multicomponent mantle, Geophys. J. Int. 184 (2011) 699–718. doi:10.1111/j.1365-246X.2010.04870.x.

[37] S.M. Weatherley, R.F. Katz, Melting and channelized magmatic flow in chemically heterogeneous, upwelling mantle, Geochemistry, Geophys. Geosystems. 13 (2012) 1–23. doi:10.1029/2011GC003989.

[38] M. Wangen, Stability of reaction-fronts in porous media, Appl. Math. Model. 37 (2013) 4860–4873. doi:10.1016/j.apm.2012.10.004.

[39] M. Pec, B.K. Holtzman, M. Zimmerman, D.L. Kohlstedt, Reaction infiltration instabilities in experiments on partially molten mantle rocks, Geology. 43 (2015) 575–578. doi:10.1130/G36611.1.

[40] F. Osselin, P. Kondratiuk, A. Budek, O. Cybulski, P. Garstecki, P. Szymczak, Microfluidic observation of the onset of reactive-infitration instability in an analog fracture, Geophys. Res. Lett. 43 (2016) 6907–6915. doi:10.1002/2016GL069261.

[41] B.R. Hacker, Subduction factory 1. Theoretical mineralogy, densities, seismic wave speeds, and H 2 O contents, J. Geophys. Res. 108 (2003) 1–26. doi:10.1029/2001JB001127.

[42] S.E. Ingebritsen, C.E. Manning, Diffuse fluid flux through orogenic belts: implications for the world ocean., Proc. Natl. Acad. Sci. U. S. A. 99 (2002) 9113–9116. doi:10.1073/pnas.132275699.

[43] R.D. Hyndman, P.A. McCrory, A. Wech, H. Kao, J. Ague, Cascadia subducting plate fluids channelled to fore-arc mantle corner: ETS and silica deposition, J. Geophys. Res. Solid Earth. 120 (2015) 4344–4358. doi:10.1002/2015JB011920.

[44] J.A.D. Connolly, Devolatilization-generated fluid pressure and deformation-propagated fluid flow during prograde regional metamorphism, J. Geophys. Res. 102 (1997) 18149. doi:10.1029/97JB00731.

[45] J.A.D. Connolly, The mechanics of metamorphic fluid expulsion, Elements. 6 (2010) 165–172. doi:10.2113/gselements.6.3.165.

[46] M. Tian, J.J. Ague, The impact of porosity waves on crustal reaction progress and CO2 mass transfer, Earth Planet. Sci. Lett. 390 (2014) 80–92. doi:10.1016/j.epsl.2013.12.044.

[47] R.M. Skarbek, A.W. Rempel, Dehydration-induced porosity waves and episodic tremor and slip, Geochemistry, Geophys. Geosystems. 17 (2016) 442–469. doi:10.1002/2015GC006155.

[48] J. Vry, R. Powell, K.M. Golden, K. Petersen, The role of exhumation in metamorphic dehydration and fluid production, Nat. Geosci. 3 (2010) 31–35. doi:10.1038/ngeo699.

[49] O. Plümper, T. John, Y.Y. Podladchikov, J.C. Vrijmoed, M. Scambelluri, Fluid escape from subduction zones controlled by channel-forming reactive porosity, Nat. Geosci. 1 (2016). doi:10.1038/NGEO2865.

[50] B. Malvoisin, Y.Y. Podladchikov, J.C. Vrijmoed, Coupling changes in densities and porosity to fluid pressure variations in reactive porous fluid flow: Local thermodynamic equilibrium, Geochemistry, Geophys. Geosystems. 16 (2015) 4362–4387. doi:10.1002/2015GC006019.

[51] M.J. Holdaway, J.W. Goodge, Rock pressure vs. fluid pressure as a controlling influence on mineral stability: an example from New Mexico, Am. Mineral. 75 (1990) 1043–1058.

[52] S. Llana-Fúnez, J. Wheeler, D.R. Faulkner, Metamorphic reaction rate controlled by fluid pressure not confining pressure: Implications of dehydration experiments with gypsum, Contrib. to Mineral. Petrol. 164 (2012) 69–79. doi:10.1007/s00410-012-0726-8.

[53] J.A. Schramke, D.M. Kerrick, A.C. Lasaga, The reaction muscovite + quartz andalusite + K-feldspar + water; Part 1, Growth kinetics and mechanism, Am. J. Sci. 287 (1987) 517–559. doi:10.2475/ajs.287.6.517.

[54] V. Barcilon, O.M. Lovera, Solitary waves in magma dynamics, J. Fluid Mech. 204 (1989) 121–133. doi:10.1017/S0022112089001680.

[55] M. Krotkiewski, M. Dabrowski, Efficient 3D stencil computations using CUDA, Parallel Comput. 39 (2013) 533–548. doi:10.1016/j.parco.2013.08.002.

[56] P. Micikevicius, 3D Finite Difference Computation on GPUs using CUDA 2701 San Tomas Expressway, Second Work. Gen. Purp. Process. Graph. Process. Units (GPGPU '09). (2009) 79–84. doi:10.1145/1513895.1513905.

[57] S. Omlin, Y.Y. Podladchikov, HPC.m: from MATLAB to HPC on GPU, CPU and MIC, Comput. {&} Geosci. (n.d.).

[58] NVIDIA Corp., NVIDIA nvprof Users Guide, (n.d.). http://docs.nvidia.com/cuda/profiler-users-guide/ (accessed October 26, 2016).

# 7 Supplementary material

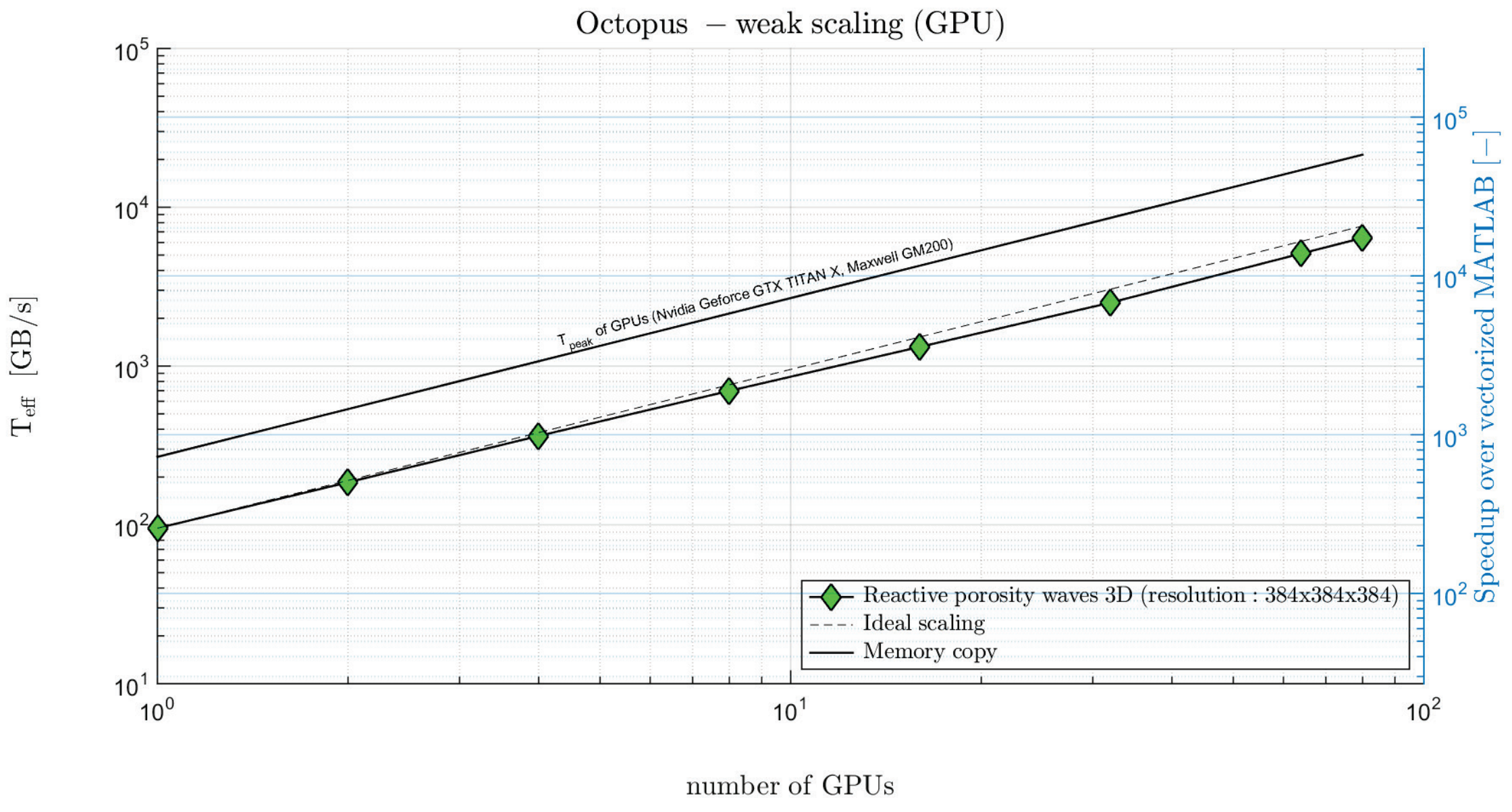

*Figure S1. Performance and scaling of the multi-GPU solver allowing for the simulation of porosity waves induced by metamorphic reactions and viscous deformation on the Octopus cluster. The effective memory throughput $T_{eff}$* [GB/s] [57] *(left axis) and the speedup over the vectorized MATLAB version of the solver (right axis) are shown. $T_{eff}$ is computed as the* effective main memory access per iteration, i.e. the *minimally required main memory access per iteration* [GB], divided by the *execution time per iteration $t_{it}$ [s]. For the here implemented equations, it is every iteration minimally required to update (2 memory accesses, read and write) the 3 unknown degrees of freedom (DOFs) $\phi$, $X_s$ and $p$, and to read (1 memory access) the known DOF $\beta_r$, i.e. $T_{eff} = (2*3+1*1)*N*4\ bytes*10^{-9}/t_{it}\ [GB/s]$, where $N$ is the total number of grid points, being about $384^3$ per GPU (note that $p$ is counted here as unknown DOF, because the solver includes elasticity, which was deactivated for the simulations presented in this paper; note that $\beta_r$ is counted here as known DOF, because it is taken as a potentially spatially heterogeneous input parameter in the solver). $T_{eff}$ can never exceed the total memory throughput (here 218 GB/s on one GPU; measured with NVIDIA nvprof* [58]*) nor the peak memory throughput $T_{peak}$ of the GPUs (here 268 GB/s per GPU; measured with a memory copy benchmark application).*

# Conclusion and Outlook

# 1 Response to the major challenges and conduction of state-of-the-art modelling

We have successfully addressed all of the major challenges in the worldwide efforts to bridge the current gap between hardware and software evolution: (1) we have created the MATLAB HPC compiler HPC.m, which increases the programmer's productivity and permits non-computer scientists to develop parallel high performance software; (2) we have developed new parallel algorithms that are optimally suited for today's hardware, i.e. algorithms that are appropriate for parallelisation and have simple regular data access patterns.

These developments have allowed us to model state-of-the-art earth science problems in three dimensions at high resolution in space and time. We were able to simulate mechanical and reactive porosity waves in fluid-filled porous media, which are both driven by strongly nonlinear physical processes that are very localised in space and time and for which it is a priori unknown where and when they occur.

The simulation of mechanical porosity waves in fluid-filled viscoelastoplastic porous media gives new insights into the dynamics of channel formation. The two key ingredients for high porosity channel generation are: (1) an asymmetry in decompaction versus compaction bulk rheology [1], and (2) an appropriate observation length scale in order to capture the compaction length of the model. The modelled physical processes might be an answer to nonlinear behaviours observed in many sedimentary basins all over the world, and explain the formation of seismic chimneys in loosely consolidated reservoirs. The model could therefore find many applications in the domains of energy and risk assessment [2].

Reacting porosity waves constitute a new mechanism for fluid flow in the Earth's crust. Fluid-filled porosity evolves as a result of (de-)volatilization reactions resulting from fluid pressure changes. Generated waves are solitary waves that pass through each other in three dimensions and allow for efficient fluid transport. Reacting porosity waves are expected to form in the whole Earth's crust at the meter to centimeter-scale in the lower crust and at the kilometer scale in the upper crust, thus, providing an efficient mechanism for fluid extraction in the Earth's crust.

With the generation of multi-GPU solvers for the simulation of mechanical and reactive porosity waves and for eight other applications across the disciplines of earth sciences, we demonstrated the great performance and versatility of HPC.m. All profiled solvers achieve on a single GPU a 250x to 500x higher performance than the fully vectorised MATLAB input script and on one workstation with 4 GPUs, the speedup is about 1000x to 2000x. The generated solvers' performance is close to hardware's peak. All solvers scale linearly on 80 GPUs of the octopus cluster and achieve on 80 GPUs a 17 000x to 35 000x higher performance than the fully vectorised MATLAB input script. Moreover, we have shown linear weak scaling for the solver that enables the simulations of mechanical porosity waves on the full Piz Daint supercomputer[1] at the Swiss National Supercomputing Centre (CSCS, Lugano, Switzerland), i.e. on nearly 5000 GPUs and on the full Lemanicus supercomputer at the Swiss Federal Institute of Technology Lausanne (EFPL, Lausanne, Switzerland), i.e. on up to 1024 CPUs (16 384 CPU cores). On the 4913 GPUs of the Piz Daint supercomputer that we could access, the total speedup over the corresponding vectorised MATLAB solver (executed on Octopus) is over 500 000x. We expect a similar scaling on Piz Daint for all the generated solvers.

HPC.m reduces the time required for transforming a prototype to a near peak performance supercomputing application from typically months or years to a few seconds. It is the first source-to-source translator, to the authors knowledge, that can perform in an automatic fashion all task that are needed for the generation of a near peak performance supercomputing application from a code developed in a classical prototyping environment such as MATLAB.

The present thesis contributes to the spreading of HPC across the disciplines in earth sciences. HPC.m can be a very valuable tool for any earth science modeller and the designed parallel algorithms may be a good basis for the development of numerical solvers in many disciplines.

## 2 Future work

Future work may include the automatic generation of geometric multigrid with HPC.m to improve the convergence speed of iterative solvers (see [3] for a comprehensive book on multigrid methods including theory and practise and [4] for a practical guide explaining how to overcome typical barriers to achieving Textbook-Multigrid Efficiency in Computational Fluid Dynamics). Geometric multigrid is a simple but extremely powerful method that accelerates the convergence of basic iterative solvers through reduction of the fine grid errors by their relaxation on multiple coarser grids. Geometric multigrid can be efficiently implemented in combination with simple variations of the classical Jacobi or Gauss-Seidel iterations, e.g. weighted Jacobi or Red-black Gauss-Seidel iterations [3].

[1] Piz Daint is currently listed as the number 8 on the TOP500 list of the world's top supercomputers [9].

The algorithms developed in this thesis employ classical Piccard or Jacobi iterations. Jacobi iterations are easily obtained by adding local time stepping to Piccard iterations and it is sufficient to multiply the local time steps by an adequate weight to obtain weighted Jacobi iterations. Our algorithms are therefore optimally suited for the acceleration with geometric multigrid. Preliminary results with a developed black box multigrid solver confirm this: we obtained Textbook-Multigrid Efficiency (TME) [3] with a Poisson solver, and near TME with both a Stokes solver and a linear viscous two-phase solver for porous media. Adding nonlinearity to a multigrid solver is feasible; there exist several methods [3,4]. We estimate consequently that our iterative algorithms could converge with geometric multigrid acceleration in one to several orders of magnitude less iterations per physical time step (depending on the importance of material property contrasts).

The classical geometric multigrid method is well suited for shared memory parallelization, i.e. it can be implemented rather easily on one CPU or GPU. To parallelize an application with geometric multigrid for distributed memory is a bigger challenge, especially if the objective is massive parallelization for supercomputers with thousands or hundreds of thousands of processors. In fact, in an ideal multigrid implementation the coarse grid contains only few grid points, which hardly produces an amount of work that would keep thousands of processors busy. For massive parallelization, the classical multigrid method will probably have to be adapted in order to achieve all of the following: near TME, near hardware peak performance and close to ideal scaling. Implementations that scale linearly on thousands of processors have nevertheless already been developed and are used for benchmarking supercomputers [5,6].

Future work may also include the automatic instrumentation of the HPC$^{.m}$ output codes for in-situ visualization with a software like *Visit* [7] or *ParaView* [8]. In-situ visualisation software enables the user to connect directly to a running simulation and permits to visualize intermediate or final simulation results continually, *accessing the data directly in memory*. Simulation data never needs to be written to hard disk in order to pass it to the visualization software, as it is traditionally being done.

# 3 References

[1] J.A.D. Connolly, Y.Y. Podladchikov, Decompaction weakening and channeling instability in ductile porous media: Implications for asthenospheric melt segregation, J. Geophys. Res. Solid Earth. 112 (2007). doi:10.1029/2005JB004213.

[2] L. Räss, V.M. Yarushina, N.S.C. Simon, Y.Y. Podladchikov, Chimneys, channels, pathway flow or water conducting features - an explanation from numerical modelling and implications for CO2 storage, in: T. Dixon, H. Herzog, S. Twinning (Eds.), Energy Procedia, ELSEVIER SCIENCE BV, NETHERLANDS, Austin, TX, 2014: pp. 3761–3774. doi:10.1016/j.egypro.2014.11.405.

[3] A. Brandt, O.E. Livne, Multigrid techniques : 1984 guide with applications to fluid dynamics, Revised Edition, Society for Industrial and Applied Mathematics, Philadelphia, 2011.

[4] A. Brandt, Barriers to Achieving Textbook Multigrid Efficiency (TME) in CFD, 1998. hdl.handle.net/2060/19980201402 (accessed November 10, 2016).

[5] M.F. Adams, J. Brown, J. Shalf, B. Van Straalen, E. Strohmaier, S. Williams, HPGMG 1.0: A Benchmark for Ranking High Performance Computing Systems, 2014. https://escholarship.org/uc/item/00r9w79m (accessed November 10, 2016).

[6] N. Sakharnykh, S. Layton, K. Clark, GPU Implementation of HPGMG-FV, Supercomputing. (2015) 18. http://crd.lbl.gov/assets/pubs_presos/SC15HPGMGBoFGPU.pdf (accessed November 10, 2016).

[7] VisItusers.org, (n.d.). www.visitusers.org (accessed November 14, 2016).

[8] U. Ayachit, A. Bauer, B. Geveci, P. O'Leary, K. Moreland, N. Fabian, J. Mauldin, ParaView Catalyst: Enabling In Situ Data Analysis and Visualization, in: Proc. First Work. Situ Infrastructures Enabling Extrem. Anal. Vis. - ISAV2015, ACM Press, New York, New York, USA, 2015: pp. 25–29. doi:10.1145/2828612.2828624.

[9] E. Strohmaier, J. Dongarra, H. Simon, M. Meuer, Top500 List - June 2016, TOP500.org. (2016). www.top500.org/list/2016/06/ (accessed November 7, 2016).

# APPENDIX

# A MATLAB input scripts for HPC.m (optimized versions)

## A.1 3D heat diffusion

```
clear all
mandatory_params = struct('PRECIS' ,{'4'   }...                     % Mandatory parameters
                         ,'NDIMS'  ,{'3'   }...
                         ,'nx'     ,{'384' }...
                         ,'ny'     ,{'384' }...
                         ,'nz'     ,{'384' }...
                         ,'OVERLAP',{'2'   }...
                         );
gpu_params = struct('NB_PROCESSORS'           ,{'24'  }...          % GPU parameters
                   ,'MAX_NB_THREADS_PER_BLOCK',{'1024'}...
                   ,'BLOCK_X'                 ,{'32'  }...
                   ,'BLOCK_Y'                 ,{'8'   }...
                   ,'BLOCK_Z'                 ,{'1'   }...
                   ,'GRID_X'                  ,{'12'  }...
                   ,'GRID_Y'                  ,{'48'  }...
                   ,'GRID_Z'                  ,{'12'  }...
                   ,'MAX_OVERLENGTH_X'        ,{'0'   }...
                   ,'MAX_OVERLENGTH_Y'        ,{'0'   }...
                   ,'MAX_OVERLENGTH_Z'        ,{'0'   }...
                   );
define_params(mandatory_params);                                    % Define the mandatory parameters
define_params(gpu_params);                                          % Define the GPU parameters
sizes = struct('Te'  ,{'nx  ','ny  ','nz  '}...                     % Declare the array sizes
              ,'Te2' ,{'nx  ','ny  ','nz  '}...
              ,'Ci'  ,{'nx  ','ny  ','nz  '}...
              );
allocate_sizes(sizes);                                              % Allocate the arrays with the given sizes
set_up_gpu();                                                       % Activate the use of GPU
set_up_process_grid(4,4,4);                                         % Activate distributed memory parallelization indicating...
                                                                    % ...the dimensions of the grid of local problems
```

```
% Physics
lam     = 1;                                              % Thermal conductivity
rho     = 1;                                              % Density
c0      = 2;                                              % Heat capacity
lx      = 1;                                              % Length of computational domain in x-dimension
ly      = 1;                                              % Length of computational domain in y-dimension
lz      = 1;                                              % Length of computational domain in z-dimension
te0     = 1;                                              % Background temperature
zc1     = 0.5;                                            % Z-location of 1st temperature anomaly
zc2     = 0.3;                                            % Z-location of 2nd temperature anomaly
teA1    = 100;                                            % Amplitude of 1st temperature anomaly
teA2    = teA1/3;                                         % Amplitude of 2nd temperature anomaly
rv      = 0.05*lz;                                        % Vertical size of the temperature anomaly
rh      = 0.05*lx;                                        % Horizontal size of the temperature anomaly
% Numerics
dx      = lx/(nx_global()-1);                             % Space step in x-dimension
dy      = ly/(ny_global()-1);                             % Space step in y-dimension
dz      = lz/(nz_global()-1);                             % Space step in z-dimension
nt      = 1000001;                                        % Number of time steps
nout    = 10000;                                          % Number of time steps between writing data to disk
% Initial conditions
Ci(:) = 1/c0;                                             % 1/Heat capacity
for iz=1:size(Te,3)
  for iy=1:size(Te,2)
    for ix=1:size(Te,1)
      x_anom  = x_global(ix,dx,size(Te,1)) - 0.5*lx;      % X-distance from the Gaussian temperature anomalies
      y_anom  = y_global(iy,dy,size(Te,2)) - 0.5*ly;      % Y-distance from the Gaussian temperature anomalies
      z_anom1 = z_global(iz,dz,size(Te,3)) - zc1*lz;      % Z-distance from the 1st Gaussian temperature anomaly
      z_anom2 = z_global(iz,dz,size(Te,3)) - zc2*lz;      % Z-distance from the 2nd Gaussian temperature anomaly
      Te(ix,iy,iz) = te0          + teA1*exp(-power(x_anom/rh,2)-power(y_anom/rh,2)-power(z_anom1/rv,2));  % 1st anomaly
      Te(ix,iy,iz) = Te(ix,iy,iz) + teA2*exp(-power(x_anom/rh,2)-power(y_anom/rh,2)-power(z_anom2/rv,2));  % 2nd anomaly
    end
  end
end
Te2 = Te;                                                 % Initialize also Te2 (to have the correct boundary conditions)
time_phys=0; isave=0;                                     % Initialize the physical time and the counter for saving data to disk
```

```
% Action
for it = 1:nt
    if (mod(it,nout)==1); isave=isave+1;                              % Save data and metadata every nout iterations
        save_info();                                                  % Save metadata
        save_coords();                                                % Save the Cartesian coordinates of the process
        save_array(Te,'Te',isave);                                    % Save the array Te
    end
    if (it==3); tic(); end                                            % Start the chronometer after two "warmup iterations" to avoid ...
                                                                      % ... to include any library loading time into the measurement
    start_of_parallel_iteration();                                    % Declare that the parallel iteration starts here
    if (mod(it,100)==1)                                               % Update the size of the time step every 100th iteration
        dt = min(min(dx*dx,dy*dy),dz*dz)*rho/lam/max_global(Ci)/8.1; % Time step for 3D heat diffusion
    end
    on_the_fly('qx    = -lam*d_xi(Te)/dx;');                          % Fourier's law of heat conduction (computation on-the-fly)
    on_the_fly('qy    = -lam*d_yi(Te)/dy;');                          % ...
    on_the_fly('qz    = -lam*d_zi(Te)/dz;');                          % ...
    on_the_fly('dTedt = 1/rho*inn(Ci).*(- d_xa(qx)/dx - d_ya(qy)/dy - d_za(qz)/dz);'); % Conservation of energy (computation on-the-fly)
    Te2(2:end-1,2:end-1,2:end-1) = inn(Te) + dt*dTedt;                % Update of temperature
    [Te2, Te] = swap_pointers(Te2,Te);                                % Swap pointers to permit the on-the-fly computation of dTedt
    update_boundaries(Te);                                            % Indicate that the local boundaries of Te must be updated
    time_phys = time_phys + dt;                                       % Update of physical time
    end_of_parallel_iteration();                                      % Declare that the parallel iteration ends here
end
time_s = toc();                                                       % Stop the chronometer
% Performance
A_eff = (2*1+1)*1/1e9*nx_global()*ny_global()*nz_global()*PRECIS;    % Effective main memory access per iteration [GB]
t_it  = time_s/(nt-2);                                                % Execution time per iteration [s]
T_eff = A_eff/t_it;                                                   % Effective memory throughput [GB/s]
if (me==0); fprintf('\ntime_s=%.4f T_eff=%.4f\n',time_s,T_eff); end  % Print the execution time and the effective memory throughput
```

## A.2 2D glacier flow

```
clear all
mandatory_params = struct('PRECIS' ,{'4'   }...                      % Mandatory parameters
                         ,'NDIMS'  ,{'2'   }...
                         ,'nx'     ,{'2048'}...
                         ,'ny'     ,{'1296'}...
                         ,'nz'     ,{'1'   }...
                         ,'OVERLAP',{'2'   }...
                         );
gpu_params = struct('NB_PROCESSORS'           ,{'24'  }...          % GPU parameters
                   ,'MAX_NB_THREADS_PER_BLOCK',{'1024'}...
                   ,'BLOCK_X'                 ,{'32'  }...
                   ,'BLOCK_Y'                 ,{'8'   }...
                   ,'BLOCK_Z'                 ,{'1'   }...
                   ,'GRID_X'                  ,{'64'  }...
                   ,'GRID_Y'                  ,{'162' }...
                   ,'GRID_Z'                  ,{'1'   }...
                   ,'MAX_OVERLENGTH_X'        ,{'0'   }...
                   ,'MAX_OVERLENGTH_Y'        ,{'0'   }...
                   ,'MAX_OVERLENGTH_Z'        ,{'0'   }...
                   );
define_params(mandatory_params);                                     % Define the mandatory parameters
define_params(gpu_params);                                           % Define the GPU parameters
sizes = struct('H'   ,{'nx  ','ny  ','1'}...                         % Declare the array sizes
              ,'H2'  ,{'nx  ','ny  ','1'}...
              ,'B'   ,{'nx  ','ny  ','1'}...
              );
allocate_sizes(sizes);                                               % Allocate the arrays with the given sizes
set_up_gpu();                                                        % Activate the use of GPU
set_up_process_grid(10,8,1);                                         % Activate distributed memory parallelization indicating...
                                                                     % ...the dimensions of the grid of local problems

% Physics
g_mu   = 1;                                                          % Gravity times shear viscosity
lx     = 2;                                                          % Length of computational domain in x-dimension
ly     = 1;                                                          % Length of computational domain in y-dimension
```

```
% Numerics
dx     = lx/(nx_global()-1);                                    % Space step in x-dimension
dy     = ly/(ny_global()-1);                                    % Space step in y-dimension
nt     = 500000001;                                             % Number of time steps
nout   = 2500000;                                               % Number of time steps between writing data to disk
% Initial conditions
create_data_serially(sizes,mandatory_params,'n','WRITE_DATA_FOR_MATLAB_AND_CONTINUE'); % Create initialization data serially
B  = load_array(B,'B');                                         % Load topography
H  = load_array(H,'H');                                         % Load initial height of glacier
H2 = H;                                                         % Initialize also H2 (to have the correct boundary conditions)
double_precision(time_phys);                                    % Declare time_phys as double precision as its value gets extremely big
time_phys=0; isave=0;                                           % Initialize the physical time and the counter for saving data to disk
% Action
for it = 1:nt
    if (mod(it,nout)==1); isave=isave+1;                        % Save key values, data and metadata every nout iterations
        if (me==0 && isave>=2); fprintf('\nisave=%.0f time_phys=%.4f dt=%.3e it=%d max_H=%.4f\n',isave,time_phys,dt,it,max_H); end % Save key values
        save_info();                                            % Save metadata
        save_coords();                                          % Save the Cartesian coordinates of the process
        save_array(H,'H',isave);                                % Save the array H
    end
    if (it==3); tic(); end                                      % Start the chronometer after two "warmup iterations" to avoid ...
                                                                % ... to include any library loading time into the measurement
    start_of_parallel_iteration();                              % Declare that the parallel iteration starts here
    if (mod(it,100)==1)                                         % Update the size of the time step every 100th iteration
       max_H = max_global(H);                                   % Find global maximum of H
       dt    = min(dx*dx,dy*dy)/g_mu*3.0/power_c(max_H,3)/4.1;  % Time step for 2D glacier flow
    end
    on_the_fly('qx      = av_xa(H.*H.*H)/3*g_mu.*d_xa(H+B)/dx;');   % Flux (height times velocity; computation on-the-fly)
    on_the_fly('qy      = av_ya(H.*H.*H)/3*g_mu.*d_ya(H+B)/dy;');   % ...
    on_the_fly('dHdt    = d_xi(qx)/dx + d_yi(qy)/dy;');         % Change of height in time (computation on-the-fly)
    H2(2:end-1,2:end-1) = max( H(2:end-1,2:end-1) + dt*dHdt, 0);    % Update of height (values <0 are set to 0)
    [H2, H]             = swap_pointers(H2,H);                  % Swap pointers to permit the on-the-fly computation of dHdt
    update_boundaries(H);                                       % Indicate that the local boundaries of H must be updated
    time_phys = time_phys + dt;                                 % Update of physical time
    end_of_parallel_iteration();                                % Declare that the parallel iteration ends here
end
```

```
time_s = toc();                                                         % Stop the chronometer
% Performance
A_eff = (2*1+1)*1/1e9*nx_global()*ny_global()*nz_global()*PRECIS;    % Effective main memory access per iteration [GB]
t_it  = time_s/(nt-2);                                                 % Execution time per iteration [s]
T_eff = A_eff/t_it;                                                    % Effective memory throughput [GB/s]
if (me==0); fprintf('\ntime_s=%.4f T_eff=%.4f\n',time_s,T_eff); end  % Print the execution time and the effective memory throughput
```

## A.3 3D acoustic wave propagation

```
clear all
mandatory_params = struct('PRECIS' ,{'4'   }...                  % Mandatory parameters
                         ,'NDIMS'  ,{'3'   }...
                         ,'nx'     ,{'384' }...
                         ,'ny'     ,{'384' }...
                         ,'nz'     ,{'384' }...
                         ,'OVERLAP',{'2'   }...
                         );
gpu_params = struct('NB_PROCESSORS'           ,{'24'  }...       % GPU parameters
                   ,'MAX_NB_THREADS_PER_BLOCK',{'1024'}...
                   ,'BLOCK_X'                 ,{'32'  }...
                   ,'BLOCK_Y'                 ,{'8'   }...
                   ,'BLOCK_Z'                 ,{'1'   }...
                   ,'GRID_X'                  ,{'12'  }...
                   ,'GRID_Y'                  ,{'48'  }...
                   ,'GRID_Z'                  ,{'12'  }...
                   ,'MAX_OVERLENGTH_X'        ,{'0'   }...
                   ,'MAX_OVERLENGTH_Y'        ,{'0'   }...
                   ,'MAX_OVERLENGTH_Z'        ,{'0'   }...
                   );
define_params(mandatory_params);                                  % Define the mandatory parameters
define_params(gpu_params);                                        % Define the GPU parameters
sizes = struct('P'  ,{'nx  ','ny  ','nz  '}...                    % Declare the array sizes
              ,'P2' ,{'nx  ','ny  ','nz  '}...
              ,'Vp' ,{'nx  ','ny  ','nz  '}...
              );
allocate_sizes(sizes);                                            % Allocate the arrays with the given sizes
set_up_gpu();                                                     % Activate the use of GPU
set_up_process_grid(4,4,4);                                       % Activate distributed memory parallelization indicating...
                                                                  % ...the dimensions of the grid of local problems
```

```
% Physics
vp0    = 2;                                                    % Propagation velocity value
rad    = 0.1;                                                  % Radius of pressure anomaly
p_in   = 100.0;                                                % Pressure inside of sphere (anomaly)
p_out  = 1.0;                                                  % Pressure outside of sphere (background)
lx     = 1;                                                    % Length of computational domain in x-dimension
ly     = 1;                                                    % Length of computational domain in y-dimension
lz     = 1;                                                    % Length of computational domain in z-dimension
% Numerics
dx     = lx/(nx_global()-1);                                   % Space step in x-dimension
dy     = ly/(ny_global()-1);                                   % Space step in y-dimension
dz     = lz/(nz_global()-1);                                   % Space step in z-dimension
dt     = power(min(min(dx,dy),dz),2.0)/vp0/8.1;                % Time step
nt     = 1000001;                                              % Number of time steps
nout   = 10000;                                                % Number of time steps between writing data to disk
% Initial conditions
Vp(:) = vp0;                                                   % Propagation velocity
P(:)  = p_out;                                                 % Pressure (intialized to background value)
for iz = 1:size(P,3)
  for iy = 1:size(P,2)
    for ix = 1:size(P,1)
      x_sph = x_global(ix,dx,size(P,1)) - 0.5*lx;              % X-distance from the spherical pressure anomaly center
      y_sph = y_global(iy,dy,size(P,2)) - 0.5*ly;              % Y-distance from the spherical pressure anomaly center
      z_sph = z_global(iz,dz,size(P,3)) - 0.5*lz;              % Z-distance from the spherical pressure anomaly center
      if (x_sph*x_sph + y_sph*y_sph + z_sph*z_sph < rad*rad); P(ix,iy,iz) = p_in; end  % Spherical pressure anomaly
    end
  end
end
P2 = P;                                                        % Initialize also P2 (to have the correct boundary conditions)
time_phys=0; isave=0;                                          % Initialize the physical time and the counter for saving data to disk
```

```
% Action
for it = 1:nt
    if (mod(it,nout)==1); isave=isave+1;                                % Save data and metadata every nout iterations
        save_info();                                                    % Save metadata
        save_coords();                                                  % Save the Cartesian coordinates of the process
        save_array(P,'P',isave);                                        % Save the array P
    end
    if (it==3); tic(); end                                              % Start the chronometer after two "warmup iterations" to avoid ...
                                                                        % ... to include any library loading time into the measurement
    start_of_parallel_iteration();                                      % Declare that the parallel iteration starts here
    on_the_fly('dqxdt = - d_xi(P)/dx;');                                % Change of flux int time (computation on-the-fly)
    on_the_fly('dqydt = - d_yi(P)/dy;');                                % ...
    on_the_fly('dqzdt = - d_zi(P)/dz;');                                % ...
    on_the_fly('dPdt  = inn(Vp).*(- d_xa(dqxdt)/dx - d_ya(dqydt)/dy - d_za(dqzdt)/dz);'); % Change of pressure in time (computation on-the-fly)
    P2(2:end-1,2:end-1,2:end-1) = 2.0*inn(P) - inn(P2) + dt*dt*dPdt;    % Update of pressure
    [P2, P] = swap_pointers(P2,P);                                      % Swap pointers to permit the on-the-fly computation of dPdt
    update_boundaries(P);                                               % Indicate that the local boundaries of P must be updated
    time_phys = time_phys + dt;                                         % Update of physical time
    end_of_parallel_iteration();                                        % Declare that the parallel iteration ends here
end
time_s = toc();                                                         % Stop the chronometer
% Performance
A_eff = (2*1+1)*1/1e9*nx_global()*ny_global()*nz_global()*PRECIS;       % Effective main memory access per iteration [GB]
t_it  = time_s/(nt-2);                                                  % Execution time per iteration [s]
T_eff = A_eff/t_it;                                                     % Effective memory throughput [GB/s]
if (me==0); fprintf('\ntime_s=%.4f T_eff=%.4f\n',time_s,T_eff); end     % Print the execution time and the effective memory throughput
```

## A.4 2D shallow water

```
clear all
mandatory_params = struct('PRECIS' ,{'4'   }...                  % Mandatory parameters
                         ,'NDIMS'  ,{'2'   }...
                         ,'nx'     ,{'2048'}...
                         ,'ny'     ,{'1296'}...
                         ,'nz'     ,{'1'   }...
                         ,'OVERLAP',{'2'   }...
                         );
gpu_params = struct('NB_PROCESSORS'          ,{'24'  }...           % GPU parameters
                   ,'MAX_NB_THREADS_PER_BLOCK',{'1024'}...
                   ,'BLOCK_X'                ,{'32'  }...
                   ,'BLOCK_Y'                ,{'8'   }...
                   ,'BLOCK_Z'                ,{'1'   }...
                   ,'GRID_X'                 ,{'64'  }...
                   ,'GRID_Y'                 ,{'162' }...
                   ,'GRID_Z'                 ,{'1'   }...
                   ,'MAX_OVERLENGTH_X'       ,{'0'   }...
                   ,'MAX_OVERLENGTH_Y'       ,{'0'   }...
                   ,'MAX_OVERLENGTH_Z'       ,{'0'   }...
                   );
define_params(mandatory_params);                                    % Define the mandatory parameters
define_params(gpu_params);                                          % Define the GPU parameters
sizes = struct('H'   ,{'nx  ','ny  ','1'}...                        % Declare the array sizes
              ,'H2'  ,{'nx  ','ny  ','1'}...
              ,'B'   ,{'nx  ','ny  ','1'}...
              ,'HVx' ,{'nx  ','ny  ','1'}...
              ,'HVx2',{'nx  ','ny  ','1'}...
              ,'HVy' ,{'nx  ','ny  ','1'}...
              ,'HVy2',{'nx  ','ny  ','1'}...
              );
allocate_sizes(sizes);                                              % Allocate the arrays with the given sizes
set_up_gpu();                                                       % Activate the use of GPU
set_up_process_grid(10,8,1);                                        % Activate distributed memory parallelization indicating...
                                                                    % ...the dimensions of the grid of local problems
```

```
% Physics
g      = 1;                                                        % Gravity
lx     = 2;                                                        % Length of computational domain in x-dimension
ly     = 1;                                                        % Length of computational domain in y-dimension
% Numerics
lf     = 1;                                                        % Switch for Lax-Friedrich (LF) stabilisation (1 - on, 0 - off)
dx     = lx/(nx_global()-1);                                       % Space step in x-dimension
dy     = ly/(ny_global()-1);                                       % Space step in y-dimension
nt     = 4000001;                                                  % Number of time steps
nout   = 20000;                                                    % Number of time steps between writing data to disk
% Initial conditions
create_data_serially(sizes,mandatory_params,'n','WRITE_DATA_FOR_MATLAB_AND_CONTINUE'); % Create initialization data serially
B  = load_array(B,'B');                                            % Load topography
H  = load_array(H,'H');                                            % Load initial height of water
H2 = H;                                                            % Initialize also H2 (to have the correct boundary conditions)
time_phys=0; isave=0;                                              % Initialize the physical time and the counter for saving data to disk
% Action
for it = 1:nt
    if (mod(it,nout)==1); isave=isave+1;                           % Save key values, data and metadata every nout iterations
        if (me==0 && isave>=2); fprintf('\nisave=%.0f time_phys=%.4f dt=%.3e it=%d max_H=%.4f\n',isave,time_phys,dt,it,max_H); end % Save key values
        save_info();                                               % Save metadata
        save_coords();                                             % Save the Cartesian coordinates of the process
        save_array(H,'H',isave);                                   % Save the array H
    end
    if (it==3); tic(); end                                         % Start the chronometer after two "warmup iterations" to avoid ...
                                                                   % ... to include any library loading time into the measurement
    start_of_parallel_iteration();                                 % Declare that the parallel iteration starts here
    if (mod(it,100)==1)                                            % Update the size of the time step every 100th iteration
       max_H = max_global(H);                                      % Find global maximum of H
       dt = min(dx,dy)/sqrt_c(g*max_H)/2.1;                        % Time step for 2D shallow water equations
    end
    on_the_fly('Vx       = 1./H.*HVx;');                           % Velocity (computation on-the-fly)
    on_the_fly('Vy       = 1./H.*HVy;');                           % ...
    on_the_fly('qxx      = HVx.*Vx + 0.5*g*H.*H;');                % Flux of the flux (computation on-the-fly)
    on_the_fly('qyy      = HVy.*Vy + 0.5*g*H.*H;');                % ...
    on_the_fly('qxy      = HVy.*Vx;');                             % ...
    on_the_fly('qyx      = HVx.*Vy;');                             % ...
```

```
    on_the_fly('Sx        = -inn(H)*g.*d_xa(av_xi(B))/dx;');         % Source (computation on-the-fly)
    on_the_fly('Sy        = -inn(H)*g.*d_ya(av_yi(B))/dy;');         % ...
    on_the_fly('qxx_av    =  av_xa(qxx) - lf*dx/4/dt*d_xa(HVx);');   % Average of flux of the flux with LF stabilization (computation on-the-fly)
    on_the_fly('qyy_av    =  av_ya(qyy) - lf*dy/4/dt*d_ya(HVy);');   % ...
    on_the_fly('qxy_av    =  av_ya(qxy) - lf*dy/4/dt*d_ya(HVx);');   % ...
    on_the_fly('qyx_av    =  av_xa(qyx) - lf*dx/4/dt*d_xa(HVy);');   % ...
    on_the_fly('HVx_av    =  (av_xa(HVx) - lf*dx/4/dt*d_xa(B+H))     % Average of the flux (height times velocity) with LF stabilization and ...
       .*(x_global(ix,dx,nx)>dx   || x_global(ix,dx,nx)<lx-dx        % ... applying zero-flux boundary conditions (the last expression that is ...
       || y_global(iy,dy,ny)>dy/2 || y_global(iy,dy,ny)<ly-dy/2);'); % ... multiplied evaluates to 0 at the the global boundaries and to 1 ...
    on_the_fly('HVy_av    =  (av_ya(HVy) - lf*dy/4/dt*d_ya(B+H))     % ... everywhere else) (computation on-the-fly)
       .*(x_global(ix,dx,nx)>dx/2 || x_global(ix,dx,nx)<lx-dx/2      % ...
       || y_global(iy,dy,ny)>dy   || y_global(iy,dy,ny)<ly-dy  );'); % ...
    on_the_fly('dHVxdt    = -d_xi(qxx_av)/dx - d_yi(qxy_av)/dy + Sx;'); % Change of the flux (height times velocity) in time ...
    on_the_fly('dHVydt    = -d_yi(qyy_av)/dy - d_xi(qyx_av)/dx + Sy;'); % ... (computation on-the-fly)
    on_the_fly('dHdt      = -d_xi(HVx_av)/dx - d_yi(HVy_av)/dy;');   % Change of the height in time (computation on-the-fly)
    HVx2(2:end-1,2:end-1) = inn(HVx) + dt*dHVxdt;                    % Update of the flux (height times velocity)
    HVy2(2:end-1,2:end-1) = inn(HVy) + dt*dHVydt;                    % ...
    H2(  2:end-1,2:end-1) = inn(H)   + dt*dHdt;                      % Update of the height
    [HVx2, HVx]           = swap_pointers(HVx2,HVx);                 % Swap pointers to permit the on-the-fly computation of dHVxdt
    [HVy2, HVy]           = swap_pointers(HVy2,HVy);                 % Swap pointers to permit the on-the-fly computation of dHVydt
    [H2,   H]             = swap_pointers(H2,H);                     % Swap pointers to permit the on-the-fly computation of dHdt
    H(1,:)=H(2,:); H(end,:)=H(end-1,:); H(:,1)=H(:,2); H(:,end)=H(:,end-1); % Boundary conditions on height
    update_boundaries(H,HVx,HVy);                                    % Indicate that the local boundaries of H, HVx and HVy must be updated
    time_phys = time_phys + dt;                                      % Update of physical time
    end_of_parallel_iteration();                                     % Declare that the parallel iteration ends here
end
time_s = toc();                                                      % Stop the chronometer
% Performance
A_eff = (2*3+1)*1/1e9*nx_global()*ny_global()*nz_global()*PRECIS;    % Effective main memory access per iteration [GB]
t_it  = time_s/(nt-2);                                               % Execution time per iteration [s]
T_eff = A_eff/t_it;                                                  % Effective memory throughput [GB/s]
if (me==0); fprintf('\ntime_s=%.4f T_eff=%.4f\n',time_s,T_eff); end  % Print the execution time and the effective memory throughput
```

## A.5 2D thermal convection

```
clear all
mandatory_params = struct('PRECIS' ,{'4'   }...                         % Mandatory parameters
                         ,'NDIMS'  ,{'2'   }...
                         ,'nx'     ,{'4094'}...
                         ,'ny'     ,{'2046'}...
                         ,'nz'     ,{'1'   }...
                         ,'OVERLAP',{'2'   }...
                         );
gpu_params = struct('NB_PROCESSORS'           ,{'24'  }...              % GPU parameters
                   ,'MAX_NB_THREADS_PER_BLOCK',{'1024'}...
                   ,'BLOCK_X'                 ,{'32'  }...
                   ,'BLOCK_Y'                 ,{'8'   }...
                   ,'BLOCK_Z'                 ,{'1'   }...
                   ,'GRID_X'                  ,{'128' }...
                   ,'GRID_Y'                  ,{'256' }...
                   ,'GRID_Z'                  ,{'1'   }...
                   ,'MAX_OVERLENGTH_X'        ,{'1'   }...
                   ,'MAX_OVERLENGTH_Y'        ,{'1'   }...
                   ,'MAX_OVERLENGTH_Z'        ,{'0'   }...
                   );
define_params(mandatory_params);                                        % Define the mandatory parameters
define_params(gpu_params);                                              % Define the GPU parameters
sizes = struct('P'   ,{'nx  ','ny  ','1'}...                            % Declare the array sizes
              ,'P2'  ,{'nx  ','ny  ','1'}...
              ,'T'   ,{'nx  ','ny  ','1'}...
              ,'T2'  ,{'nx  ','ny  ','1'}...
              ,'Vx'  ,{'nx+1','ny  ','1'}...
              ,'Vx2' ,{'nx+1','ny  ','1'}...
              ,'Vy'  ,{'nx  ','ny+1','1'}...
              ,'Vy2' ,{'nx  ','ny+1','1'}...
              ,'Rho' ,{'nx  ','ny  ','1'}...
              );
allocate_sizes(sizes);                                                  % Allocate the arrays with the given sizes
set_up_gpu();                                                           % Activate the use of GPU
set_up_process_grid(8,8,1);                                             % Activate distributed memory parallelization indicating...
```

```
                                                                        % ...the dimensions of the grid of local problems
% Physics
lam     = 1;                                                            % Thermal conductivity
mu      = 1;                                                            % Shear viscosity
g       = 1e10;                                                         % Gravity
ly      = 1;                                                            % Length of computational domain in y-dimension
lx      = ly*2;                                                         % Length of computational domain in x-dimension
rho0    = 1;                                                            % Density, background value
alf     = 1e-2;                                                         % Thermal expansion coefficient
rh      = 0.1*ly;                                                       % Size of the temperature anomaly
t0      = 0;                                                            % Temperature, background value
ta      = 1;                                                            % Temperature, value at anomaly
% Numerics
dx      = lx/(nx_global()-1);                                           % Space step in x-dimension
dy      = ly/(ny_global()-1);                                           % Space step in y-dimension
dt      = dx*dx/4/2;                                                    % Time step
beta    = 5e-9*dt/dx/dx;                                                % Numerical compressibility
nt      = 20000001;                                                     % Number of time steps
nout    = 100000;                                                       % Number of time steps between writing data to disk
% Initial conditions
for iy=1:size(T,2)
    for ix=1:size(T,1)
        x_anom = x_global(ix,dx,size(T,1))-0.51*lx;                     % X-distance from the Gaussian temperature anomaly
        y_anom = y_global(iy,dy,size(T,2))-0.51*ly;                     % Y-distance from the Gaussian temperature anomaly
        T(ix,iy) = t0 + ta*exp_c(-power_c(x_anom/rh,2) - power_c(y_anom/rh,2)); % Gaussian temperature anomaly
        if (y_global(iy,dy,size(T,2)) == ly); T(ix,iy) = t0; end        % Boundary condition on temperature (t0 on top)
        if (y_global(iy,dy,size(T,2)) == 0 ); T(ix,iy) = ta; end        % Boundary condition on temperature (ta on bottom)
    end
end
for iy=1:size(P,2)
    for ix=1:size(P,1)
        P(ix,iy) = (rho0-rho0*alf*(t0 + ta/2))*g*(ly-y_global(iy,dy,size(P,2))); % Initial pressure (lithostatic)
    end
end
T2   = T;                                                               % Initialize also T2 (to have the correct boundary conditions)
P2   = P;                                                               % Initialize also P2 (to have the correct boundary conditions)
Rho = rho0-rho0*alf*T;                                                  % Density
```

```
time_phys=0; isave=0;                                        % Initialize the physical time and the counter for saving data to disk
% Action
for it = 1:nt
    if (mod(it,nout)==1); isave=isave+1;                     % Save data and metadata every nout iterations
        save_info();                                         % Save metadata
        save_coords();                                       % Save the Cartesian coordinates of the process
        save_array(P,  'P',  isave);                         % Save the array P
        save_array(T,  'T',  isave);                         % Save the array T
        save_array(Rho,'Rho',isave);                         % Save the array Rho
        save_array(Vx, 'Vx', isave);                         % Save the array Vx
        save_array(Vy, 'Vy', isave);                         % Save the array Vy
    end
    if (it==3); tic(); end                                   % Start the chronometer after two "warmup iterations" to avoid ...
                                                             % ... to include any library loading time into the measurement
    start_of_parallel_iteration();                           % Declare that the parallel iteration starts here
    on_the_fly('divV        = d_xa(Vx)/dx + d_ya(Vy)/dy;');        % Divergence of velocity
    on_the_fly('qxx         = P - 2*mu*(d_xa(Vx)/dx - divV/2);');    % Flux (equal normal stress with opposite sign)
    on_the_fly('qyy         = P - 2*mu*(d_ya(Vy)/dy - divV/2);');    % ...
    on_the_fly('qxy         =   -   mu*(d_yi(Vx)/dy + d_xi(Vy)/dx);'); % Flux (equal shear stress with opposite sign)
    on_the_fly('dVxdt       = -1./av_xi(Rho).*(d_xi(qxx)/dx + d_ya(qxy)/dy);'); % Change of velocity in time
    on_the_fly('dVydt       = -1./av_yi(Rho).*(d_yi(qyy)/dy + d_xa(qxy)/dx + av_yi(Rho)*g);'); % ...
    on_the_fly('Rho         = rho0-rho0*alf*T;','mod(it,nout)==0');  % Density (with indication to write it every nout iterations to memory)
    on_the_fly('dPdt        = -divV/beta;');                     % Change of pressure in time
    on_the_fly('qxT         = -lam*d_xi(T)/dx;');                % Heat flux
    on_the_fly('qyT         = -lam*d_yi(T)/dy;');                % ...
    on_the_fly('dT_diff     = - (d_xa(qxT)/dx + d_ya(qyT)/dy);');    % Diffusion of temperature
    on_the_fly('dT_advx_neg = - (Vx(3:end-1,2:end-1)<0).*Vx(3:end-1,2:end-1).*d_xa(T(2:end  ,2:end-1))/dx;'); % Advection of temperature
    on_the_fly('dT_advx_pos = - (Vx(2:end-2,2:end-1)>0).*Vx(2:end-2,2:end-1).*d_xa(T(1:end-1,2:end-1))/dx;'); % ...
    on_the_fly('dT_advy_neg = - (Vy(2:end-1,3:end-1)<0).*Vy(2:end-1,3:end-1).*d_ya(T(2:end-1,2:end  ))/dy;'); % ...
    on_the_fly('dT_advy_pos = - (Vy(2:end-1,2:end-2)>0).*Vy(2:end-1,2:end-2).*d_ya(T(2:end-1,1:end-1))/dy;'); % ...
    on_the_fly('dTdt        = dT_diff + dT_advx_neg + dT_advx_pos + dT_advy_neg + dT_advy_pos;');  % Change of temperature in time
    T2(2:end-1,2:end-1)     = T(2:end-1,2:end-1)   + dt*dTdt;        % Update of temperature
    P2                      = P                    + dt*dPdt;        % Update of pressure
    Vx2(2:end-1,2:end-1)    = Vx(2:end-1,2:end-1)  + dt*dVxdt;       % Update of velocity
    Vy2(2:end-1,2:end-1)    = Vy(2:end-1,2:end-1)  + dt*dVydt;       % ...
    [T2, T]                 = swap_pointers(T2,T);                   % Swap pointers to permit the on-the-fly computation of dTedt
    [P2, P]                 = swap_pointers(P2,P);                   % Swap pointers to permit the on-the-fly computation of dPdt
```

```
    [Vx2, Vx]                = swap_pointers(Vx2,Vx);               % Swap pointers to permit the on-the-fly computation of dVxdt
    [Vy2, Vy]                = swap_pointers(Vy2,Vy);               % Swap pointers to permit the on-the-fly computation of dVydt
    T(1  ,:)                 = T(2    ,:);                          % Boundary conditions for the temperature in the x-dimension
    T(end,:)                 = T(end-1,:);                          % ...
    update_boundaries(T,Vx,Vy);                                     % Indicate that the local boundaries of T, Vx and Vy must be updated
    time_phys = time_phys + dt;                                     % Update of physical time
    end_of_parallel_iteration();                                    % Declare that the parallel iteration ends here
end
time_s = toc();                                                     % Stop the chronometer
% Performance
A_eff = (2*4+0)*1/1e9*nx_global()*ny_global()*nz_global()*PRECIS;   % Effective main memory access per iteration [GB]
t_it  = time_s/(nt-2);                                              % Execution time per iteration [s]
T_eff = A_eff/t_it;                                                 % Effective memory throughput [GB/s]
if (me==0); fprintf('\ntime_s=%.4f T_eff=%.4f\n',time_s,T_eff); end  % Print the execution time and the effective memory throughput
```

## A.6 3D reactive porosity waves

```
clear all
mandatory_params = struct('PRECIS' ,{'4'   }...                       % Mandatory parameters
                         ,'NDIMS'  ,{'3'   }...
                         ,'nx'     ,{'128' }...
                         ,'ny'     ,{'128' }...
                         ,'nz'     ,{'408' }...
                         ,'OVERLAP',{'2'   }...
                         );
gpu_params = struct('NB_PROCESSORS'          ,{'24'  }...            % GPU parameters
                   ,'MAX_NB_THREADS_PER_BLOCK',{'1024'}...
                   ,'BLOCK_X'                ,{'32'  }...
                   ,'BLOCK_Y'                ,{'8'   }...
                   ,'BLOCK_Z'                ,{'1'   }...
                   ,'GRID_X'                 ,{'4'   }...
                   ,'GRID_Y'                 ,{'16'  }...
                   ,'GRID_Z'                 ,{'51'  }...
                   ,'MAX_OVERLENGTH_X'       ,{'0'   }...
                   ,'MAX_OVERLENGTH_Y'       ,{'0'   }...
                   ,'MAX_OVERLENGTH_Z'       ,{'0'   }...
                   );
define_params(mandatory_params);                                      % Define the mandatory parameters
define_params(gpu_params);                                            % Define the GPU parameters
sizes = struct('Xs'          ,{'nx  ','ny  ','nz  '}...               % Declare the array sizes
              ,'Xs2'         ,{'nx  ','ny  ','nz  '}...
              ,'Xs_old'      ,{'nx  ','ny  ','nz  '}...
              ,'Pf'          ,{'nx  ','ny  ','nz  '}...
              ,'Pf2'         ,{'nx  ','ny  ','nz  '}...
              ,'Pf_old'      ,{'nx  ','ny  ','nz  '}...
              ,'Phi'         ,{'nx  ','ny  ','nz  '}...
              ,'Phi2'        ,{'nx  ','ny  ','nz  '}...
              ,'Phi_old'     ,{'nx  ','ny  ','nz  '}...
              ,'Rhofg'       ,{'nx  ','ny  ','nz  '}...
              ,'Rhosg'       ,{'nx  ','ny  ','nz  '}...
              ,'Beta_X'      ,{'nx  ','ny  ','nz  '}...
              ,'Xseq'        ,{'nx  ','ny  ','nz  '}...
```

```
               ,'K_muf_Beta_ef',{'nx  ','ny  ','nz  '}...
               ,'Ei'           ,{'nx  ','ny  ','nz  '}...
               ,'Ef'           ,{'nx  ','ny  ','nz  '}...
               ,'Es'           ,{'nx  ','ny  ','nz  '}...
               ,'E_max'        ,{'nx  ','ny  ','nz  '}...
               ,'dXsdt'        ,{'nx  ','ny  ','nz  '}...
               ,'dPfdt'        ,{'nx  ','ny  ','nz  '}...
               ,'dPfdt2'       ,{'nx  ','ny  ','nz  '}...
               );
allocate_sizes(sizes);                                                  % Allocate the arrays with the given sizes
set_up_gpu();                                                           % Activate the use of GPU
set_up_process_grid(4,4,4);                                             % Activate distributed memory parallelization indicating...
                                                                        % ...the dimensions of the grid of local problems

% Physics
eta      = 1;                                                           % Bulk viscosity
k_muf0   = 1;                                                           % Permeability over fluid shear viscosity, reference value
drho     = 1;                                                           % Density
phi0     = 0.01;                                                        % Initial porosity
g        = 1;                                                           % Gravity
% Dependent scales
ls       = sqrt(k_muf0*eta);                                            % Length scale
ps       = drho*ls;                                                     % Pressure scale
ts       = eta/ps;                                                      % Time scale
% Nondimensionalization
Da       = 100;                                                         % Damköhler number (viscous time over reaction time);
tau      = 1/Da*ts;                                                     % Introduction of the Damköhler number (viscous time over chemical time)
De       = 0.001;                                                       % Deborah number (viscous over elastic time scale); small to study viscous case
beta_d   = De/ps;                                                       % Introduction of the Deborah number for viscous vs. elastic deformation
rhos     = drho+drho/2;                                                 % Solid density
rhof     = drho/2;                                                      % Fluid density
lz       = ls*50;                                                       % Length of computational domain in z-dimension
lx       = lz/4;                                                        % Length of computational domain in x-dimension
ly       = lz/4;                                                        % Length of computational domain in y-dimension
zc1      = 0.1;                                                         % Z-location of 1st porosity anomaly
zc2      = 0.23;                                                        % Z-location of 2nd porosity anomaly
phia1    = 1.5*phi0;                                                    % Amplitude of 1st porosity anomaly
phia2    = 0.75*phi0;                                                   % Amplitude of 2nd porosity anomaly
```

```
rv        = 0.03*lz;                                                % Vertical size of the porosity anomaly
rh        = 0.12*lx;                                                % Horizontal size of the porosity anomaly
beta_s    = beta_d*11/12;                                           % Solid compressibility without reaction
beta_f    = 6*beta_d;                                               % Fluid compressibility
cs        = -1;                                                     % Clapeyron slope of the reaction
beta_xx   = 2e-1*cs;                                                % Xseq variation with pressure (sign determines the Clapeyron slope)
delta_X   = 3e0;                                                    % Solid density variation with Xs
maxXs     = 0.05;                                                   % Maximum mass fraction of volatile species in solid
% Numerics
dx        = lx/(nx_global()-1);                                     % Space step in x-dimension
dy        = ly/(ny_global()-1);                                     % Space step in y-dimension
dz        = lz/(nz_global()-1);                                     % Space step in z-dimension
eps       = 2e-8;                                                   % Convergence limit
dt_phys   = 1e-6;                                                   % Physical time step
nt        = 61001;                                                  % Number of time steps
nout      = 1000;                                                   % Number of time steps between writing data to disk
nup       = 10;                                                     % Number of iterations between updates of time step and errors
iterMin   = 1;                                                      % Minimum number of iterations
iterMax   = 10000;                                                  % Maximum number of iterations
% Initial conditions
Xs(:) = maxXs/2;                                                    % Mass fraction of volatile species in solid
for iz=1:size(Phi,3)
    for iy=1:size(Phi,2)
        for ix=1:size(Phi,1)
            x_anom  = x_global(ix,dx,size(Phi,1)) - 0.5*lx;         % X-distance from the Gaussian porosity anomalies
            y_anom  = y_global(iy,dy,size(Phi,2)) - 0.5*ly;         % Y-distance from the Gaussian porosity anomalies
            z_anom1 = z_global(iz,dz,size(Phi,3)) - zc1*lz;         % Z-distance from the 1st Gaussian porosity anomaly
            z_anom2 = z_global(iz,dz,size(Phi,3)) - zc2*lz;         % Z-distance from the 2nd Gaussian porosity anomaly
            Phi(ix,iy,iz) = phi0          + phia1*exp_c(-power_c(x_anom/rh,2) - power_c(y_anom/rh,2) - power_c(z_anom1/rv,2)); % 1st anomaly
            Phi(ix,iy,iz) = Phi(ix,iy,iz) + phia2*exp_c(-power_c(x_anom/rh,2) - power_c(y_anom/rh,2) - power_c(z_anom2/rv,2)); % 2nd anomaly
        end
    end
end
time_phys=0; isave=0; niter=0; dtPf=0;                             % Initialize different variables for the time loop / iterations
```

```
% Action
for it = 1:nt
    if (mod(it,nout)==1); isave=isave+1;                                   % Save data and metadata every nout iterations
        save_info();                                                       % Save metadata
        save_coords();                                                     % Save the Cartesian coordinates of the process
        save_array(Xs,  'Xs',   isave);                                    % Save the array Xs
        save_array(Pf,  'Pf',   isave);                                    % Save the array Pf
        save_array(Phi, 'Phi',  isave);                                    % Save the array Phi
        save_array(Xseq,'Xseq', isave);                                    % Save the array Xseq
    end
    if (it==3); tic(); niter=0; end                                        % Start the chronometer after two "warmup iterations" to avoid ...
                                                                           % ... to include any library loading time into the measurement
    start_of_parallel_iteration();                                         % Declare that the parallel iteration starts here
    Xs_old     = Xs;                                                       % Store Xs from the previous time step
    Pf_old     = Pf;                                                       % Store Pf from the previous time step
    Phi_old    = Phi;                                                      % Store Phi from the previous time step
    maxabs_err = 2*eps;                                                    % Initialize the maximum absolute error
    iter       = 0;                                                        % Initialize the iteration counter
    % Pseudo transient iterations
    while ( maxabs_err>eps &&  iter<iterMax )  ||  (iter<iterMin)
        iter = iter + 1;
        % Useful parameters
        on_the_fly('Pe              = -Pf;');
        on_the_fly('Xseq            = min(max(beta_xx*Pf + maxXs/2,0),maxXs);','mod(it,nout)==0');
        on_the_fly('Beta_P          = Phi.*beta_f + beta_d - (1 + Phi).*beta_s;');
        on_the_fly('Beta_Phi        = beta_d - 1./(1-Phi)*beta_s;');
        on_the_fly('K_muf           = k_muf0.*1/(phi0*phi0*phi0).*Phi.*Phi.*Phi;');
        % Darcy flux
        on_the_fly('qDx             = -av_xa(K_muf).*(d_xa(Pf)/dx         );');
        on_the_fly('qDy             = -av_ya(K_muf).*(d_ya(Pf)/dy         );');
        on_the_fly('qDz             = -av_za(K_muf).*(d_za(Pf)/dz - rhof*g);');
        if (mod(iter,nup)==1);  K_muf_Beta_P=K_muf./Beta_P;  dtPf = min(min(dx*dx,dy*dy),dz*dz)/max_global(K_muf_Beta_P)/16.1;  end  % Pf time step
        if (mod(iter,nup)==0);  Ei=Phi;  Ef=Pf/10;  Es=Xs;  end      % Store all unkown DOFs before their update
```

```
        % Updates of the unkown DOFs (Xs, Pf, Phi)
        on_the_fly('dPfdt_phys          = (Pf  -  Pf_old)/dt_phys;');
        on_the_fly('dXsdt_phys          = (Xs  -  Xs_old)/dt_phys;');
        on_the_fly('dPhidt_phys         = (Phi - Phi_old)/dt_phys;');
        Xs2                             =          Xs_old + dt_phys*dXsdt;
        Pf2                             =          Pf     + dtPf*dPfdt - dtPf*dPfdt_phys;
        Phi2                            = min(max(Phi_old + dt_phys*dPhidt,1e-4),1);
        % Time derivatives
        dPfdt2(2:end-1,2:end-1,2:end-1) = 1.0./inn(Beta_P).*( -d_xi(qDx)/dx - d_yi(qDy)/dy - d_zi(qDz)/dz + inn(Pe)./(eta*(1 - inn(Phi))) ...
                                          + (1-rhos./(rhof*(1+delta_X))).*(1-inn(Phi)).*(1+delta_X)./(1-inn(Xs)).*inn(dXsdt) );
        on_the_fly('dXsdt               = (Xseq-Xs)/tau;');
        on_the_fly('dPhidt              = (1 - Phi).*(Beta_Phi.*dPfdt  -  Pe./(eta.*(1 - Phi)) - (1 + delta_X)./(1-Xs).*dXsdt);');
        % Swap all pointers (permits to do all computations within an iterations without synchronization in between)
        [Xs2,    Xs]    = swap_pointers(Xs2,    Xs);
        [Pf2,    Pf]    = swap_pointers(Pf2,    Pf);
        [Phi2,   Phi]   = swap_pointers(Phi2,   Phi);
        [dPfdt2, dPfdt] = swap_pointers(dPfdt2, dPfdt);
        % Boundary conditions, followed by local boundary update.
        dPfdt(1,:,:)=dPfdt(2,:,:); dPfdt(end,:,:)=dPfdt(end-1,:,:);
        dPfdt(:,1,:)=dPfdt(:,2,:); dPfdt(:,end,:)=dPfdt(:,end-1,:);
        dPfdt(:,:,1)=dPfdt(:,:,2); dPfdt(:,:,end)=dPfdt(:,:,end-1);
        update_boundaries(dPfdt);
        % Compute the errors (errors = variables before update - variables after update)
        if (mod(iter,nup)==0)
            Ei          = abs(Phi    - Ei);                      % Errors for Phi
            Ef          = abs(Pf/10 - Ef);                       % Errors for Pf
            Es          = abs(Xs     - Es);                      % Errors for Xs
            E_max       = max(max(Ei,Ef),Es);                    % Maximum error (locally)
            maxabs_err = max_global(E_max);                      % Global maximum of error
        end
    end
    niter     = niter + iter;
    time_phys = time_phys + dt_phys;                             % Update of physical time
    end_of_parallel_iteration();                                 % Declare that the parallel iteration ends here
end
time_s = toc();                                                  % Stop the chronometer
```

```
% Performance
A_eff = (2*3+1)*1/1e9*nx_global()*ny_global()*nz_global()*PRECIS;    % Effective main memory access per iteration [GB]
t_it  = time_s/(nt-2);                                               % Execution time per iteration [s]
T_eff = A_eff/t_it;                                                  % Effective memory throughput [GB/s]
if (me==0); fprintf('\ntime_s=%.4f T_eff=%.4f\n',time_s,T_eff); end  % Print the execution time and the effective memory throughput
```

## A.7 3D poroviscoelastic two-phase flow

```
clear all
mandatory_params = struct('PRECIS' ,{'4'   }...                    % Mandatory parameters
                         ,'NDIMS'  ,{'3'   }...
                         ,'nx'     ,{'254' }...
                         ,'ny'     ,{'254' }...
                         ,'nz'     ,{'406' }...
                         ,'OVERLAP',{'2'   }...
                         );
gpu_params = struct('NB_PROCESSORS'           ,{'24'  }...          % GPU parameters
                   ,'MAX_NB_THREADS_PER_BLOCK',{'1024'}...
                   ,'BLOCK_X'                 ,{'32'  }...
                   ,'BLOCK_Y'                 ,{'8'   }...
                   ,'BLOCK_Z'                 ,{'1'   }...
                   ,'GRID_X'                  ,{'8'   }...
                   ,'GRID_Y'                  ,{'32'  }...
                   ,'GRID_Z'                  ,{'408' }...
                   ,'MAX_OVERLENGTH_X'        ,{'1'   }...
                   ,'MAX_OVERLENGTH_Y'        ,{'1'   }...
                   ,'MAX_OVERLENGTH_Z'        ,{'1'   }...
                   );
define_params(mandatory_params);                                    % Define the mandatory parameters
define_params(gpu_params);                                          % Define the GPU parameters
sizes = struct('Phi'     ,{'nx  ','ny  ','nz  '}...                 % Declare the array sizes
              ,'Phi2'    ,{'nx  ','ny  ','nz  '}...
              ,'Pt'      ,{'nx  ','ny  ','nz  '}...
              ,'Pt2'     ,{'nx  ','ny  ','nz  '}...
              ,'Pto'     ,{'nx  ','ny  ','nz  '}...
              ,'Pte'     ,{'nx  ','ny  ','nz  '}...
              ,'Pf'      ,{'nx  ','ny  ','nz  '}...
              ,'Pf2'     ,{'nx  ','ny  ','nz  '}...
              ,'Pfo'     ,{'nx  ','ny  ','nz  '}...
              ,'Pfe'     ,{'nx  ','ny  ','nz  '}...
              ,'Vx'      ,{'nx+1','ny  ','nz  '}...
              ,'Vy'      ,{'nx  ','ny+1','nz  '}...
              ,'Vz'      ,{'nx  ','ny  ','nz+1'}...
```

```
              ,'Vze'      ,{'nx  ','ny  ','nz+1'}...
              ,'Sxx'      ,{'nx  ','ny-2','nz-2'}...
              ,'Sxxo'     ,{'nx  ','ny-2','nz-2'}...
              ,'Syy'      ,{'nx-2','ny  ','nz-2'}...
              ,'Syyo'     ,{'nx-2','ny  ','nz-2'}...
              ,'Szz'      ,{'nx-2','ny-2','nz  '}...
              ,'Szzo'     ,{'nx-2','ny-2','nz  '}...
              ,'Sxy'      ,{'nx-1','ny-1','nz-2'}...
              ,'Sxyo'     ,{'nx-1','ny-1','nz-2'}...
              ,'Sxz'      ,{'nx-1','ny-2','nz-1'}...
              ,'Sxzo'     ,{'nx-1','ny-2','nz-1'}...
              ,'Syz'      ,{'nx-2','ny-1','nz-1'}...
              ,'Syzo'     ,{'nx-2','ny-1','nz-1'}...
              ,'Mn'       ,{'nx  ','ny  ','nz  '}...
              ,'Etas'     ,{'nx  ','ny  ','nz  '}...
              ,'divqD'    ,{'nx  ','ny  ','nz  '}...
              ,'divVn'    ,{'nx  ','ny  ','nz  '}...
              ,'LPhiSo'   ,{'nx-2','ny-2','nz-2'}...
              ,'LPhiSe'   ,{'nx-2','ny-2','nz-2'}...
              ,'Rx'       ,{'nx-1','ny-2','nz-2'}...
              ,'Ry'       ,{'nx-2','ny-1','nz-2'}...
              ,'Rz'       ,{'nx-2','ny-2','nz-1'}...
              ,'divV'     ,{'nx  ','ny  ','nz  '}...
              ,'Vx_BG'    ,{'nx+1','ny  ','nz  '}...
              ,'E_max'    ,{'nx-2','ny-2','nz-2'}...
              ,'K_muf'    ,{'nx  ','ny  ','nz  '}...
              ,'qDz'      ,{'nx-2','ny-2','nz-1'}...
              ,'Mus'      ,{'nx  ','ny  ','nz  '}...
              ,'Rog'      ,{'nx  ','ny  ','nz  '}...
              ,'Tau2'     ,{'nx-2','ny-2','nz-2'}...
              );
allocate_sizes(sizes);                                   % Allocate the arrays with the given sizes
set_up_gpu();                                            % Activate the use of GPU
set_up_process_grid(4,4,5);                              % Activate distributed memory parallelization indicating...
                                                         % ...the dimensions of the grid of local problems
```

```
% Physics
mu0      = 1.0;                                          % Solid shear viscosity, background value
rhofg    = 1.0;                                          % Fluid density times gravity
k_muf0   = 1.0;                                          % Permeability divided by fluid viscosity, reference value
etas0    = 1.0;                                          % Bulk viscosity, reference value
xsi_w    = 1.0;                                          % Wanted xsi (coefficient in the updates of the stresses)
% Nondimensionalization
xy_BG    = 1.0;                                          % Background shear stress
rhosg    = 2.0*rhofg;                                    % Solid density times gravity
ra       = 3.0;                                          % Height - width ratio
lz       = ra*10.0;                                      % Length of computational domain in z-dimension
lx       = (lz*nx_global())/nz_global();                 % Length of computational domain in x-dimension
ly       = (lz*ny_global())/nz_global();                 % Length of computational domain in y-dimension
cMu      = 1e-1;                                         % Geometry Coefficient in viscosity - porosity relation
bet_phi  = 5e-3;                                         % Effective bulk compressibility
bet_dry  = 1.3*bet_phi;                                  % Drained compressibility
b        = 0.89;                                         % Skempton's coefficient
alpha    = 0.9;                                          % Biot-Willis coefficient
r        = 1e3;                                          % Decompaction weakening factor for Pf>Pt
mod_G    = 0.5/bet_phi;                                  % Shear modulus
np       = 3.0;                                          % Exponent in nonlinear solid shear viscosity relation
tau_b    = 0.05;                                         % Characteristic stress at the transition between linear ...
                                                         % ... and power law viscous behaviour
phi0     = 1e-2;                                         % Porosity, background value
rogBG    = rhofg.*phi0 + (1.0-phi0).*rhosg;              % Density, background value
phia1    = 1.5*phi0;                                     % Porosity, value at 1st anomaly
phia2    = 2.0*phi0;                                     % Porosity, value at 2nd anomaly
zc1      = 0.25;                                         % Z position of 1st porosity anomaly
zc2      = 0.1;                                          % Z position of 2nd porosity anomaly
rh       = 1.0;                                          % Horizontal size of the porosity anomaly
rv       = 1.0;                                          % Vertical size of the porosity anomaly
% Numerics
dampX    = 1.0-1.0/nx_global();                          % Damping coefficient
dampY    = 1.0-1.0/ny_global();                          % ...
dampZ    = 1.0-1.0/nz_global();                          % ...
rel      = 1e-1;                                         % Relaxation coefficient for shear viscosity
rels     = 5e-1;                                         % Relaxation coefficient for bulk viscosity
```

```
eps       = 5e-4;                                                    % Convergence limit
roi       = 1.0;                                                     % Inertial density
dx        = lx/(nx_global()-1.0);                                    % Space step in x-dimension
dy        = ly/(ny_global()-1.0);                                    % Space step in y-dimension
dz        = lz/(nz_global()-1.0);                                    % Space step in z-dimension
bet_n     = 10.0;                                                    % Numerical compressibility
dtPt_sc   = (1.0+bet_n)/nz_global();                                 % Scale for iterative total pressure time step
dtV_sc    = 1.0/(1.0+bet_n);                                         % Scale for iterative velocity time step
dt        = 3.5e-06;                                                 % Physical time step
nt        = 40000;                                                   % Number of time steps
nout      = 2000;                                                    % Number of time steps between writing data to disk
nup       = 15;                                                      % Number of iterations between updates of time step and errors
iterMin   = 1;                                                       % Minimum number of iterations
iterMax   = 10000;                                                   % Maximum number of iterations
% Initial conditions
Mn(:)      = mu0;                                                    % Initialize the shear viscosity with the background value
Phi(:)     = phi0;                                                   % Initialize the porosity with the background value
for iz=1:size(Phi,3)
  for iy=1:size(Phi,2)
    for ix=1:size(Phi,1)
      x_anom  = x_global(ix,dx,size(Phi,1)) - 0.5*lx;               % X-distance from the Gaussian porosity anomalies
      y_anom  = y_global(iy,dy,size(Phi,2)) - 0.5*ly;               % Y-distance from the Gaussian porosity anomalies
      z_anom1 = z_global(iz,dz,size(Phi,3)) - zc1*lz;               % Z-distance from the 1st Gaussian porosity anomaly
      z_anom2 = z_global(iz,dz,size(Phi,3)) - zc2*lz;               % Z-distance from the 2nd Gaussian porosity anomaly
      Phi(ix,iy,iz) = phi0          + phia1*exp_c(-power_c(x_anom/rh,2) - power_c(y_anom/rh,2) - power_c(z_anom1/rv,2)); % 1st anomaly
      Phi(ix,iy,iz) = Phi(ix,iy,iz) + phia2*exp_c(-power_c(x_anom/rh,2) - power_c(y_anom/rh,2) - power_c(z_anom2/rv,2)); % 2nd anomaly
    end
  end
end
Phi2 = Phi;                                                          % Initialize also Phi2 (to have the correct boundary conditions)
for iz=1:size(Vx_BG,3)
  for iy=1:size(Vx_BG,2)
    for ix=1:size(Vx_BG,1)
      Vx_BG(ix,iy,iz) = xy_BG*(y_global(iy,dy,size(Vx_BG,2)) - 0.5*ly); % Initialize background velocity
    end
  end
end
```

```
LPhiSo = inn(log_c(1.0-Phi));                                       % Initialize the solid porosity in the log scale
time_phys=0; isave=0; niter=0; iter=0; maxabs_err=2.0*eps;          % Initialize different variables for the time loop / iterations
dtS=dt; dt_cum=dt;                                                  % ...
% Action
for it = 1:nt
    if (mod(it,nout)==1); isave=isave+1;                            % Save data and metadata every nout iterations
        save_info();                                                % Save metadata
        save_coords();                                              % Save the Cartesian coordinates of the process
        save_array(Phi,   'Phi',isave);                             % Save the array Phi
        save_array(K_muf, 'K_muf',isave);                           % Save the array K_muf
        save_array(qDz,   'qDz',isave);                             % Save the array qDz
        save_array(divqD, 'divqD',isave);                           % Save the array divqD
        save_array(Mn,    'Mn',isave);                              % Save the array Mn
        save_array(Etas,  'Etas',isave);                            % Save the array Etas
        save_array(Mus,   'Mus',isave);                             % Save the array Mus
        save_array(Sxx,   'Sxx',isave);                             % Save the array Sxx
        save_array(Syy,   'Syy',isave);                             % Save the array Syy
        save_array(Szz,   'Szz',isave);                             % Save the array Szz
        save_array(Sxy,   'Sxy',isave);                             % Save the array Sxy
        save_array(Sxz,   'Sxz',isave);                             % Save the array Sxz
        save_array(Syz,   'Syz',isave);                             % Save the array Syz
        save_array(Tau2,  'Tau2',isave);                            % Save the array Tau2
        save_array(Rog,   'Rog',isave);                             % Save the array Rog
        save_array(Vx,    'Vx',isave);                              % Save the array Vx
        save_array(Vy,    'Vy',isave);                              % Save the array Vy
        save_array(Vz,    'Vz',isave);                              % Save the array Vz
        save_array(Pt,    'Pt',isave);                              % Save the array Pt
        save_array(Pf,    'Pf',isave);                              % Save the array Pf
        save_array(Rx,    'Rx',isave);                              % Save the array Rx
        save_array(Ry,    'Ry',isave);                              % Save the array Ry
        save_array(Rz,    'Rz',isave);                              % Save the array Rz
        save_array(divV,  'divV',isave);                            % Save the array divV
    end
    if (it==3); tic(); niter=0; end                                 % Start the chronometer after two "warmup iterations" to avoid...
                                                                    % ... to include any library loading time into the measurement
    start_of_parallel_iteration();                                  % Declare that the parallel iteration starts here
```

```
    if (it>1);  LPhiSo = LPhiS;  end                              % Store the solid porosity (log) from the previous time step
    Pto     = Pt;                                                 % Store the total pressure from the previous time step
    Pfo     = Pf;                                                 % Store the fluid pressure from the previous time step
    dtS     = 1.0/mod_G/xsi_w*max_global(Mn)*phi0*cMu;            % Physical time step for stresses
    if (dt_cum >= dtS || it==1)                                   % When the cumulated time steps (dt_cum) have reached the ...
        Sxxo = Sxx; Syyo = Syy; Szzo = Szz;                       % ... value of dtS, store all the stresses from the previous ...
        Sxyo = Sxy; Sxzo = Sxz; Syzo = Syz;                       % ... time step ...
        dt_cum = 0;                                               % ... and reinitialize dt_cum.
    end
    iter       = 0;                                               % Initialize the iteration counter
    maxabs_err = 2.0*eps;                                         % Initialize the maximum absolute error
    % Pseudo transient iterations
    while (maxabs_err > eps && iter < iterMax)  ||  (iter<iterMin)
        iter = iter + 1;
        if (mod(iter,nup)==0);  LPhiSe=LPhiS; Pte=Pt; Pfe=Pf; Vze=Vz;  end  % Store all variables at iteration start.
        % Nonlinear parameters
        on_the_fly('Rog   = rhofg.*Phi + (1.0-Phi).*rhosg - rogBG;','mod(it,nout)==0'); % Density times gravity
        on_the_fly('K_muf = k_muf0*1.0/(phi0*phi0*phi0)*Phi.*Phi.*Phi;','mod(it,nout)==0'); % Permeability over fluid shear viscosity
        on_the_fly('Mus   = Mn.*phi0*cMu;','mod(it,nout)==0');       % Solid shear viscosity
        on_the_fly('Rfact = (Pf>Pt)*(1.0/r-1.0) + 1.0;');            % Decompaction weakening factor
        on_the_fly('Etasi = etas0.*Mn.*(1.0./(Phi./phi0)).*Rfact;'); % Bulk viscosity
        if (mod(iter,nup)==1);
            Etas          = exp_c(rels*log_c(Etasi) + (1.0-rels)*log_c(Etas)); % Apply relaxation on bulk viscosity
        end
        if (it==1); Etas  = Etasi.*1.0./Rfact; end                   % Initialization of bulk viscosity at first time step
        % Timesteps
        on_the_fly('dtV   = dz*dz*(1.0./Mus).*roi./6.1*dtV_sc;');    % Iterative velocity time step
        on_the_fly('dtPt  = Mus.*dtPt_sc;');                         % Iterative total pressure time step
        on_the_fly('dtPf  = dz*dz*(1.0./(K_muf./(1.0 + bet_dry*b/alpha)))./6.1;'); % Iterative fluid pressure time step
        % Divergence of velocities
        divV              = d_xa(Vx)/dx + d_ya(Vy)/dy + d_za(Vz)/dz; % Divergence of velocity
        % Pressures (implicit update) and numerical divV
        divVn             = ( divV + bet_dry.*((Pt-Pto) - alpha*(Pf-Pfo))/dt + (Pt-Pf).*1.0./Etas.*(1.0./(1.0-Phi)) );
        Pt2               = ( Pt -dtPt.*(divV  + bet_dry.*(         -Pto - alpha.*(Pf-Pfo))/dt - Pf.*(1.0./(Etas.*(1.0-Phi))))) ...
                            ./(1.0 + dtPt.*(bet_dry/dt         + (1.0./(Etas.*(1.0-Phi)))));
        Pf2               = ( Pf -dtPf.*(divqD + bet_dry.*(-alpha/b.*Pfo - alpha.*(Pt-Pto))/dt - Pt.*(1.0./(Etas.*(1.0-Phi))))) ...
                            ./(1.0 + dtPf.*(bet_dry*alpha/b/dt + (1.0./(Etas.*(1.0-Phi)))));
```

```
[Pt2,Pt] = swap_pointers(Pt2,Pt);                          % Swap pointers to prevent synch. between the pressure updates
[Pf2,Pf] = swap_pointers(Pf2,Pf);                          % ...
% Strain rates deviators
on_the_fly('Exx   = ( d_xi(Vx)/dx - in_yz(divV)/3.0   );');
on_the_fly('Eyy   = ( d_yi(Vy)/dy - in_xz(divV)/3.0   );');
on_the_fly('Ezz   = ( d_zi(Vz)/dz - in_xy(divV)/3.0   );');
on_the_fly('Exy   = ((d_yi(Vx + Vx_BG)/dy + d_xi(Vy)/dx).*0.5);');
on_the_fly('Exz   = ((d_zi(Vx        )/dz + d_xi(Vz)/dx).*0.5);');
on_the_fly('Eyz   = ((d_yi(Vz        )/dy + d_zi(Vy)/dz).*0.5);');
% Stresses deviators Visco-Elastic and 2nd Stress invariant
on_the_fly('Xsi   = (1.0./(mod_G*dtS)).*Mus;');              % Coefficient for updates of stresses
Sxx   = Sxxo.* in_yz(Xsi).*(1.0./(1.0+ in_yz(Xsi))) + 2.0*(Exx + bet_n*in_yz(divVn)).* in_yz(Mus).*(1.0./(1.0+ in_yz(Xsi)));
Syy   = Syyo.* in_xz(Xsi).*(1.0./(1.0+ in_xz(Xsi))) + 2.0*(Eyy + bet_n*in_xz(divVn)).* in_xz(Mus).*(1.0./(1.0+ in_xz(Xsi)));
Szz   = Szzo.* in_xy(Xsi).*(1.0./(1.0+ in_xy(Xsi))) + 2.0*(Ezz + bet_n*in_xy(divVn)).* in_xy(Mus).*(1.0./(1.0+ in_xy(Xsi)));
Sxy   = Sxyo.*av_xyi(Xsi).*(1.0./(1.0+av_xyi(Xsi))) + 2.0*Exy.*av_xyi(Mus).*(1.0./(1.0+av_xyi(Xsi)));
Sxz   = Sxzo.*av_xzi(Xsi).*(1.0./(1.0+av_xzi(Xsi))) + 2.0*Exz.*av_xzi(Mus).*(1.0./(1.0+av_xzi(Xsi)));
Syz   = Syzo.*av_yzi(Xsi).*(1.0./(1.0+av_yzi(Xsi))) + 2.0*Eyz.*av_yzi(Mus).*(1.0./(1.0+av_yzi(Xsi)));
on_the_fly('Tau2 = sqrt_c(0.5*(in_x(Sxx).*in_x(Sxx) + in_y(Syy).*in_y(Syy) + in_z(Szz).*in_z(Szz) + av_xya(Sxy).*av_xya(Sxy)*2.0 +
            av_xza(Sxz).*av_xza(Sxz)*2.0 + av_yza(Syz).*av_yza(Syz)*2.0 ) );','mod(it,nout)==0');
% Residuals
Rx    = 1.0/roi*(d_xa(Sxx)/dx + d_ya(Sxy)/dy + d_za(Sxz)/dz - d_xi(Pt)/dx             ) +  Rx*dampX;
Ry    = 1.0/roi*(d_xa(Sxy)/dx + d_ya(Syy)/dy + d_za(Syz)/dz - d_yi(Pt)/dy             ) +  Ry*dampY;
Rz    = 1.0/roi*(d_xa(Sxz)/dx + d_ya(Syz)/dy + d_za(Szz)/dz - d_zi(Pt)/dz - av_zi(Rog)) +  Rz*dampZ;
% Darcy flux and its divergence
on_the_fly('qDx   = -av_xi(K_muf).* d_xi(Pf)/dx;');
on_the_fly('qDy   = -av_yi(K_muf).* d_yi(Pf)/dy;');
on_the_fly('qDz   = -av_zi(K_muf).*(d_zi(Pf)/dz + (rhofg-rogBG));','mod(it,nout)==0');
divqD(2:end-1,2:end-1,2:end-1) = d_xa(qDx)/dx + d_ya(qDy)/dy + d_za(qDz)/dz;
% Porosity update - including advection
on_the_fly('dPhidt = inn(bet_phi.*((Pf-Pfo)/dt-(Pt-Pto)/dt) + 1.0./Etas.*(Pf-Pt)) + av_xa(inn(Vx).*d_xi(1.0-Phi)/dx) +
            av_ya(inn(Vy).*d_yi(1.0-Phi)/dy) + av_za(inn(Vz).*d_zi(1.0-Phi)/dz);');
on_the_fly('LPhiS  = LPhiSo + dt*(-dPhidt);');
Phi2(2:end-1,2:end-1,2:end-1) = 1.0-exp_c(LPhiS);
[Phi2,Phi] = swap_pointers(Phi2,Phi);                   % Swap pointers to permit Phi update including its spatial derivatives
% Velocity updates and boundary conditions
Vx(2:end-1,2:end-1,2:end-1) = inn(Vx) + av_xi(dtV).*Rx;
Vy(2:end-1,2:end-1,2:end-1) = inn(Vy) + av_yi(dtV).*Ry;
```

```
        Vz(2:end-1,2:end-1,2:end-1) = inn(Vz) + av_zi(dtV).*Rz;
        Vx(:,1,:) = Vx(:,2,:); Vx(:,end,:) = Vx(:,end-1,:);   Vx(:,:,1) = Vx(:,:,2); Vx(:,:,end) = Vx(:,:,end-1);
        Vy(1,:,:) = Vy(2,:,:); Vy(end,:,:) = Vy(end-1,:,:);   Vy(:,:,1) = Vy(:,:,2); Vy(:,:,end) = Vy(:,:,end-1);
        Vz(:,1,:) = Vz(:,2,:); Vz(:,end,:) = Vz(:,end-1,:);   Vz(1,:,:) = Vz(2,:,:); Vz(end,:,:) = Vz(end-1,:,:);
        % Shear viscosity update (with relaxation) and boundary conditions
        if (mod(iter,nup)==1);
            on_the_fly('Mni = mu0*(1.0./(1.0 + power_c((Tau2./tau_b),(np - 1.0))));');
            Mn(2:end-1,2:end-1,2:end-1) = exp_c(rel*log_c(Mni) + (1.0-rel)*log_c(inn(Mn)));
            Mn(1,:,:) = Mn(2,:,:); Mn(end,:,:) = Mn(end-1,:,:);
            Mn(:,1,:) = Mn(:,2,:); Mn(:,end,:) = Mn(:,end-1,:);
            Mn(:,:,1) = Mn(:,:,2); Mn(:,:,end) = Mn(:,:,end-1);
            update_boundaries(Mn);                            % Indicate to update local boundaries of Mn
        end
        % Local boundary updates
        update_boundaries(Vx,Vy,Vz,divqD,Phi);
        % Error check
        if (mod(iter,nup)==0)
            LPhiSe = abs(LPhiSe-LPhiS);
            Pte    = abs(Pte-Pt);
            Pfe    = abs(Pfe-Pf);
            Vze    = abs(Vze-Vz);
            E_max  = max(max(max(LPhiSe,inn(Pte)),inn(Pfe)),inn(Vze(:,:,1:end-1))); % Maximum error (locally)
            maxabs_err = max_global(E_max);                   % Global maximum of error
        end
    end %(while loop)
    niter     = niter     + iter;
    dt_cum    = dt_cum    + dt;
    time_phys = time_phys + dt;
    end_of_parallel_iteration();
end %(time loop)
time_s = toc();                                               % Stop the chronometer
% Performance
A_eff = (2*12+1)*1/1e9*nx_global()*ny_global()*nz_global()*PRECIS;   % Effective main memory access per iteration [GB]
t_it  = time_s/(niter);                                       % Execution time per iteration [s]
T_eff = A_eff/t_it;                                           % Effective memory throughput [GB/s]
if (me==0); fprintf('\ntime_s=%.4f T_eff=%.4f\n',time_s,T_eff); end  % Print the execution time and the effective memory throughput
```

## *Acknowledgement*

Firstly, I would like to express my sincere gratitude to my supervisor, Prof. Yuri Podladchikov. Yuri opened my mind for numerical modelling and supercomputing. He showed me a new way to look at informatics and science. Thanks to him, I realized that scientific computing must not be approached the same way as, let us say, webpage development. He made me revise all the traditional informatics conventions, which are often considered untouchable by computers scientists. I am most happy that with this PhD thesis, I could contribute to spread Yuri's unique vision about how science, and in particular numerical modelling, should be done. Also, I am most grateful for all the freedom that he gave me during this PhD to develop what I am passionate about and of what I am convinced. Finally, I am very grateful that I could be a part in the design, test and assembly of the Octopus supercomputer. This was a highly instructional and unique experience.

Besides my supervisor, I would like to thank the rest of my thesis committee, Prof. Taras Gerya and Prof. Stefan Schmalholz, who kindly accepted to examine my thesis during the hectic days before Christmas. I would like to thank Stefan also for his constant support in organisational and administrative questions. I would like to thank Prof. Michel Jaboyedoff for presiding the thesis defence.

I would like to express my sincere gratitude to Prof. Dave Yuen, who reviewed my article about HPC.m and who made it possible for me to give invited talks at Cray Inc. (Minnesota, USA) and at the Minnesota Supercomputing Institute (Minnesota, USA). Thanks to him, I could also give a talk in China and visit Tianhe-2, the fastest supercomputer in the world (www.top500.org) at this time.

I would like to thank particularly to Philippe Logean, from whom I could learn an enormous amount of practical Informatics knowledge during my years at the institute. Moreover, he helped our research group in the assembling of the Octopus supercomputer and he is now doing an incredible job administrating it. I would like to thank Philippe as well as Dr. Hamid Hussain-Khan, Dr. Vincent Keller and Simon Hiscox for the stimulating discussions in the design phase of the Octopus supercomputer.

I thank particularly Ludovic Räss, Dr. Vangelis Moulas, Dr. Benjamin Malvoisin and Igor Podladchikov for all their help and stimulating discussions. Working with them was a professionally enormously enriching opportunity and a great pleasure. I thank Joshua

Vaughan-Hammon for reviewing the English of the introduction and the conclusion with me. I thank my fellow labmates for the stimulating discussions and for the good moments apart from work.

Lastly, I would like to thank my family for their endless support during my entire life. And most of all I would like to thank my loving, supportive, encouraging, and always cheerful wife Rossi, whose faithful support during all the stages of this PhD is so appreciated. Thank you.